\documentclass[prb,twocolumn,superscriptaddress,floatfix]{revtex4-2}
\usepackage{color}
\usepackage{amsmath}
\usepackage{amssymb}
\usepackage{bm}      	% bold math
\usepackage{graphicx}
\usepackage[unicode=true,pdfusetitle,
 bookmarks=false,bookmarksnumbered=false,bookmarksopen=false,
 breaklinks=false,pdfborder={0 0 1},backref=false,colorlinks=true,urlcolor=blue,citecolor=blue,linkcolor=black]
 {hyperref}
\usepackage{subfigure}
\usepackage{textcomp}

\newcommand{\ord}[1]{{\mathit{O}}\left(#1\right)}

\newcommand{\bsub}{\begin{subequations}}
\newcommand{\esub}{\end{subequations}}

\newcommand{\vex}[1]{\bm{\mathrm{#1}}}

\newcommand{\tauh}{\hat{\tau}}
\newcommand{\sh}{\hat{s}}
\newcommand{\shb}{\hat{\vex{s}}}
\numberwithin{equation}{section}

\begin{document}
\title{Magnetic instability and spin-glass order beyond the Anderson-Mott transition in interacting power-law random banded matrix fermions}
\def\rice{Department of Physics and Astronomy, Rice University, Houston, Texas
77005, USA}
\def\rcqm{Rice Center for Quantum Materials, Rice University, Houston, Texas
77005, USA}
\author{Xinghai Zhang}\affiliation{\rice}
\author{Matthew S. Foster}\affiliation{\rice}\affiliation{\rcqm}
\date{\today}

\begin{abstract}
In the presence of quenched disorder, the 
interplay
between local magnetic-moment formation and Anderson localization for electrons at a 
zero-temperature, metal-insulator transition (MIT) remains a long unresolved problem. 
Here, we study the 
emergence
of these 
phenomena
in a power-law random banded matrix model of spin-1/2 fermions with repulsive Hubbard interactions. 
Focusing on the regime of \emph{weak} interactions, we perform  both analytical field theory and numerical self-consistent Hartree-Fock 
calculations. 
We show that interference-mediated effects strongly enhance the density of states and magnetic fluctuations upon approaching the MIT from the metallic side.
These are consistent with results due to Finkel'stein obtained four decades ago.
Our numerics further show that local moments nucleate from 
typical states 
as we cross
the MIT, 
with a density that grows continuously into the insulating phase. 
We identify spin-glass order in the insulator by computing the overlap distribution between converged Hartree-Fock mean-field moment profiles. 
Our results indicate that  itinerant interference effects can morph smoothly into moment formation and magnetic frustration within a single model,
revealing a common origin for these disparate phenomena.
\end{abstract}
\maketitle
\tableofcontents

%%%%%%%%%%%%%%%%%%%%%%%%%%%%%%%%%%%%%%%%%%%%%%%%%%%%%%%%%%%%%%%%%%%%%%%%%%%%%%%%%%%%%%%
%%%%%%%%%%%%%%%%%%%%%%%%%%%%%%%%%%%%%%%%%%%%%%%%%%%%%%%%%%%%%%%%%%%%%%%%%%%%%%%%%%%%%%%
%%%%%%%%%%%%%%%%%%%%%%%%%%%%%%%%%%%%%%%%%%%%%%%%%%%%%%%%%%%%%%%%%%%%%%%%%%%%%%%%%%%%%%%
%%%%%%%%%%%%%%%%%%%%%%%%%%%%%%%%%%%%%%%%%%%%%%%%%%%%%%%%%%%%%%%%%%%%%%%%%%%%%%%%%%%%%%%
%%%%%%%%%%%%%%%%%%%%%%%%%%%%%%%%%%%%%%%%%%%%%%%%%%%%%%%%%%%%%%%%%%%%%%%%%%%%%%%%%%%%%%%
%%%%%%%%%%%%%%%%%%%%%%%%%%%%%%%%%%%%%%%%%%%%%%%%%%%%%%%%%%%%%%%%%%%%%%%%%%%%%%%%%%%%%%%
%%%%%%%%%%%%%%%%%%%%%%%%%%%%%%%%%%%%%%%%%%%%%%%%%%%%%%%%%%%%%%%%%%%%%%%%%%%%%%%%%%%%%%%
%%%%%%%%%%%%%%%%%%%%%%%%%%%%%%%%%%%%%%%%%%%%%%%%%%%%%%%%%%%%%%%%%%%%%%%%%%%%%%%%%%%%%%%
%%%%%%%%%%%%%%%%%%%%%%%%%%%%%%%%%%%%%%%%%%%%%%%%%%%%%%%%%%%%%%%%%%%%%%%%%%%%%%%%%%%%%%%
%%%%%%%%%%%%%%%%%%%%%%%%%%%%%%%%%%%%%%%%%%%%%%%%%%%%%%%%%%%%%%%%%%%%%%%%%%%%%%%%%%%%%%%
%%%%%%%%%%%%%%%%%%%%%%%%%%%%%%%%%%%%%%%%%%%%%%%%%%%%%%%%%%%%%%%%%%%%%%%%%%%%%%%%%%%%%%%

\section{Introduction}

Metal-insulator transitions (MITs) in impure quantum materials present a many-faceted puzzle, with important
roles played by Coulomb interactions, magnetic fluctuations, and Anderson localization. 
Thermodynamic observables such as the spin susceptibility can be enhanced upon approaching the MIT in doped semiconductors 
\cite{Paalanen1988};
this has been interpreted as evidence for the proliferation of unscreened local magnetic moments \cite{Milo1989,Bhatt1992,BK1994,Feng1999,Miranda2012}. 
These can nucleate even on the metallic side for sufficiently strong interactions \cite{Milo1989,Bhatt1992}. At the same time,
long wavelength quantum interference effects give rise to Anderson localization 
of states at the Fermi energy \cite{Lee1985},
which 
determine
transport properties. The advent of both components 
seems to suggest
a ``two-fluid'' picture 
for 
weakly localizing
itinerant electrons 
and 
local magnetism, 
similar to heavy-fermion physics \cite{Paalanen1988,Miranda2012,Coleman2015}. 
However, it remains unclear under what conditions this two-fluid
picture should apply.

A strong interplay between magnetism and localization was theoretically suggested four decades ago in calculations performed
by Finkel'stein \cite{Finkelstein1983}.
Using a field theory framework that incorporates interference effects between hydrodynamic modes of the interacting electron liquid,
Finkel'stein identified an apparent magnetic instability upon approaching the MIT from the metallic side \cite{Castellani1984,BK1994,Finkelstein2010,Finkelstein2023}.
Similar to the predicted enhancement of the pairing amplitude near the superconductor-insulator transition
\cite{Feigelman2007,Feigelman2010,Burmistrov2012},
itinerant spin-exchange interactions get boosted near the MIT because quantum-critical eigenstates \cite{Evers2008}
exhibit large spatial overlap and strong state-to-state correlations (``Chalker scaling'' \cite{Chalker1988,Chalker1990,Cuevas2007}).
However, the precise nature of the resulting instability and any incipient order has remained controversial  \cite{BK1994}.
Possible interpretations included nucleation of local moments,
itinerant ferromagnetic order \cite{KB1996,Andreev1998,Chamon2000},
or magnetic droplet formation \cite{NAL2000}. 

Experimental evidence for anomalous magnetic behavior near the interacting MIT was
already observed in the 1980s in phosphorous-doped silicon. See e.g.\ Ref.~\cite{BK1994} for a review.
In the 1990s, these issues were reignited by the observation of a 2D MIT \cite{Kravchenko1994, Kravchenko1995, Abrahams2001, Kravchenko2010, Shashkin2017}.
Competing theories proposed to explain the phenomenon are still under debate \cite{DS2005, Punn2001, Punn2005, Ani2007, DS2013, DS2014}. 
Again magnetic anomalies near the critical point were observed \cite{Shashkin2001, Shashkin2006, Ani2006}, possibly linked to the formation of 
local moments or spin droplets \cite{Pudalov2012,Pudalov2018,Hossain2020}.

In this paper, we seek to clarify the \emph{theoretical} situation, at least in a particular concrete model. We study spin-1/2 fermions 
with repulsive Hubbard interactions and power-law random-banded matrix (PRBM) hopping. 
Noninteracting PRBM models describe fermions with long-ranged random hopping in 1D, and exhibit a well-understood Anderson MIT as a function of the hopping power law \cite{Mirlin1996}.
The Anderson MIT in this 1D model has similar properties to the MIT in higher-dimensional systems \cite{Mirlin2000b}, and for this reason generated significant interest 
(for reviews, see Refs.~\cite{Mirlin2000a,Evers2008}).
Unlike higher-dimensional models, the 1D nature of the PRBM model allows us to perform large-scale self-consistent numerical calculations to incorporate virtual effects of interactions.
Meanwhile, the interplay of the Anderson MIT and interaction effects can be studied analytically via the Finkel'stein nonlinear sigma model (NLsM)
effective field theory.
% and perturbative calculation is feasible.
This enables us to employ complementary analytical and numerical tools to study the interacting model. 
Using a Keldysh version of the Finkel'stein 
NLsM
%field theory 
\cite{Kamenev2009,Kamenev2011,Liao2017}, we calculate quantum corrections to observables on the metallic side and at the transition.
We find an enhanced low-energy density of states, antilocalizing corrections to transport, and a boost to the magnetic susceptibility; these diverge logarithmically 
with decreasing temperature at the MIT. The obtained results are qualitatively identical to those of Finkel'stein in $d = 2 + \epsilon$ dimensions \cite{Finkelstein1983,BK1994},
indicating some type of magnetic instability near the transition. 
The main purpose of the analytical calculations is to ground and provide qualitative benchmarks for
our subsequent numerical study.

We also perform self-consistent Hartree-Fock numerics \cite{DH1993,Tusch1993}, for which  we are able to reach large system sizes (8000 sites). 
Hartree-Fock was previously employed in studies of dirty, interacting spinless fermions \cite{Epperlein1997,Vojta1998,Amini2014}.
For the spinful PRBM with Hubbard $U$,
we confirm the analytical results on the metallic side. In addition, the numerics reveals the proliferation of local magnetic moments that begins near the MIT. 
The moment density grows monotonically as we tune into the insulator, quickly converging with system size. 
We find that the local moments nucleate from \emph{typical} states at the Fermi energy as we cross the 
MIT.
The moments 
form irregular frozen patterns, with no net ferromagnetic or antiferromagnetic character. 
We calculate overlaps between different mean-field converged spin configurations. 
The distribution function that we obtain is 
\emph{reminiscent} of full replica symmetry breaking (RSB) in infinite-range spin glasses \cite{Binder1986,MezardBook}.
In this work, we do not attempt to resolve the question of whether true RSB occurs in the ground-state manifold in the thermodynamic limit.

Thus we find that both long-wavelength interference effects and local-moment physics play crucial roles in the same model, albeit
on opposite sides of the MIT. Our most interesting finding is that local moments appear to smoothly emerge from the quantum-critical, multifractal-enhanced state that characterizes the MIT, suggesting a \emph{single-fluid} model whose character changes from itinerant to localized at the MIT. 
This should be contrasted with a picture of anomalously localized moments that can arise from the combination of strong interactions and rare disorder fluctuations.
A key question for future research is whether this dual character of the interacting, disordered fluid survives correlation effects beyond the Hartree-Fock framework employed here. 

This paper is organized as follows. In Sec.~\ref{sec:summary}, we summarize the main analytical and numerical results. 
Analytical results for the density of states (DOS), MIT, and spin susceptibility are presented in Sec.~\ref{sec:Keldysh-summary}. 
Numerical results for the DOS, moment formation, spin susceptibility, and evidence for spin-glass order are presented in Sec.~\ref{sec:MF-summary}. 
Additional technical details on the self-consistent mean-field numerics for the PRBM-Hubbard model are given in Sec.~\ref{sec:MF-Details}.
We set up the long-wavelength field theory description of the interacting PRBM model and derive quantum corrections in Sec.~\ref{sec:Keldysh}.
In Sec.~\ref{sec:Conclusion}, we conclude and discuss the outlook for future research.

%%%%%%%%%%%%%%%%%%%%%%%%%%%%%%%%%%%%%%%%%%%%%%%%%%%%%%%%%%%%%%%%%%%%%%%%%%%%%%%%%%%%%%%
%%%%%%%%%%%%%%%%%%%%%%%%%%%%%%%%%%%%%%%%%%%%%%%%%%%%%%%%%%%%%%%%%%%%%%%%%%%%%%%%%%%%%%%
%%%%%%%%%%%%%%%%%%%%%%%%%%%%%%%%%%%%%%%%%%%%%%%%%%%%%%%%%%%%%%%%%%%%%%%%%%%%%%%%%%%%%%%
%%%%%%%%%%%%%%%%%%%%%%%%%%%%%%%%%%%%%%%%%%%%%%%%%%%%%%%%%%%%%%%%%%%%%%%%%%%%%%%%%%%%%%%
%%%%%%%%%%%%%%%%%%%%%%%%%%%%%%%%%%%%%%%%%%%%%%%%%%%%%%%%%%%%%%%%%%%%%%%%%%%%%%%%%%%%%%%
%%%%%%%%%%%%%%%%%%%%%%%%%%%%%%%%%%%%%%%%%%%%%%%%%%%%%%%%%%%%%%%%%%%%%%%%%%%%%%%%%%%%%%%
%%%%%%%%%%%%%%%%%%%%%%%%%%%%%%%%%%%%%%%%%%%%%%%%%%%%%%%%%%%%%%%%%%%%%%%%%%%%%%%%%%%%%%%
%%%%%%%%%%%%%%%%%%%%%%%%%%%%%%%%%%%%%%%%%%%%%%%%%%%%%%%%%%%%%%%%%%%%%%%%%%%%%%%%%%%%%%%
%%%%%%%%%%%%%%%%%%%%%%%%%%%%%%%%%%%%%%%%%%%%%%%%%%%%%%%%%%%%%%%%%%%%%%%%%%%%%%%%%%%%%%%
%%%%%%%%%%%%%%%%%%%%%%%%%%%%%%%%%%%%%%%%%%%%%%%%%%%%%%%%%%%%%%%%%%%%%%%%%%%%%%%%%%%%%%%
%%%%%%%%%%%%%%%%%%%%%%%%%%%%%%%%%%%%%%%%%%%%%%%%%%%%%%%%%%%%%%%%%%%%%%%%%%%%%%%%%%%%%%%

\section{Summary of main results}
\label{sec:summary}

\begin{figure*}[t!]
  \centering
  \includegraphics[width=0.97\textwidth]{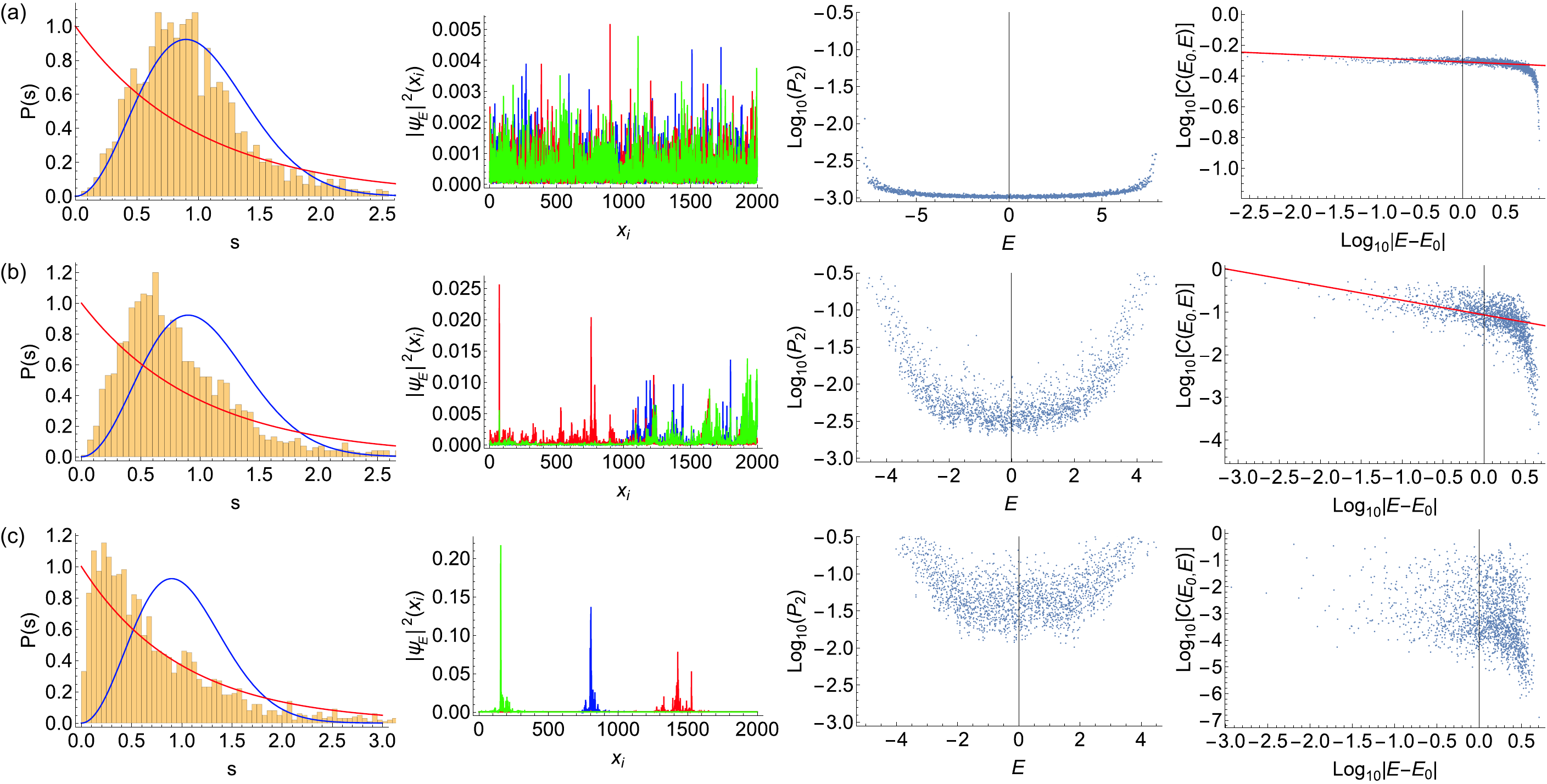}
  \caption{Key features of the non-interacting GUE-PRBM model [Eq.~(\ref{eq:HPRBM+U}) with $U$ = 0], from the ergodic to Anderson-localized regimes. 
Numerical results are shown from exact diagonalization of an $L = 2000$ site system (single disorder realization), with 
(a) $\alpha = 0.5$  (ergodic phase),
(b) $\alpha = 1.0$ (Anderson MIT),
and 
(c) $\alpha = 1.3$ (Anderson insulator).
From left to right each row depicts 
(i) the level statistics, 
(ii) probability profiles in position space for 3 consecutive mid-spectrum eigenstates, 
(iii) the IPR $P_2$ [Eq.~(\ref{IPRDef})] versus energy $E$, and 
(iv) 
the Chalker correlator $C(E_0,E)$ [Eq.~(\ref{CDef})] for the mid-spectrum state with $E_0\approx 0$ versus the energy difference $E-E_0$. 
The blue (red) curve for the level statistics is the GUE Wigner surmise \cite{Livan2017} (Poisson distribution).
The IPR $P_2$ is small [$\ord{1/L}$] for almost all states in the ergodic phase, but becomes substantial in the Anderson insulator. 
The Chalker correlator takes a flat nonzero value for ergodic states, shows power-law dependence in state number at the MIT 
[Chalker scaling, Eq.~(\ref{CScaling})], 
but is strongly suppressed in the insulator (except for discrete states residing in the same localization volume as the selected state). 
The red curve gives the power-law-fit for the Chalker scaling of states not too close to the band-edge Lifshitz tails.
For all results presented here, we have chosen $b = 1$ [Eq.~(\ref{eq:aij})].
}
  \label{fig:PRBM-GUE}
\end{figure*}

We consider an SU(2)-invariant spin-$1/2$ PRBM model with repulsive Hubbard interactions,
\begin{equation}
	  H = \sum_{ij,s} \mathsf{H}_{ij} c^{\dagger}_{i s} c_{j s} + U \sum_i n_{i\uparrow} n_{i\downarrow} -\mu \sum_i n_i.
  	\label{eq:HPRBM+U}
\end{equation}
Here $c_{j s}$ annihilates a spin-$s \in \{\uparrow,\downarrow\}$ fermion at site $x_j$,
and 
$\mathsf{H}_{ij} = \mathsf{G}_{ij} \, a\left( \left|i-j\right| \right)$, 
where $\mathsf{G}_{i j}$ denotes a random matrix in the Gaussian unitary ensemble.
We set the energy variance of the Gaussian random matrix to one,  
\begin{align}
	\overline{\mathsf{G}_{i j} \, \mathsf{G}_{k l}} 
	=
	\left(\Delta \mathsf{G}\right)^2 
	\delta_{i l} \, \delta_{k j},
\qquad
	\Delta \mathsf{G} = 1,
\end{align}
where the overline denotes disorder averaging. 
The Hubbard interaction strength $U$ is then dimensionless, measured in units of the standard deviation $\Delta \mathsf{G}$.
In Eq.~(\ref{eq:HPRBM+U}), $a\left( \left|i-j\right| \right)$ is an envelope function modulating the hopping 
that decays as $|i-j|^{-\alpha}$ at large distances,
\begin{equation}
  a\left(\left|i-j\right|\right)= 
  \begin{cases}
	1 & |i-j|<b\,,\\
	\left|\frac{i-j}{b}\right|^{-\alpha} & |i-j|\ge b\,.
  \end{cases}
  \label{eq:aij}
\end{equation}
We have set the lattice constant equal to one; the length parameter $b$ 
can be used to tune the character of the Anderson MIT. 
For $b\gg 1$, the Anderson MIT in the PRBM model is \emph{weakly multifractal} (see below). 
This is similar to the weak-localization regime in 2D as well as the MIT predicted in $d=2+\epsilon$ dimensions \cite{Evers2008}.
In this regime,  the analytical NLsM description of the model is weakly coupled and perturbation theory is justified.
In this paper we consider $b \gg 1$ ($b = 1$) for the analytical (numerical) calculations; qualitatively similar
results obtain near the MIT via both approaches. We do not consider the regime $b \ll 1$, which is dominated by rare regions and 
strong fractality. The properties of the  $b \ll 1$ model are characteristic of purely 1D systems that evade localization \cite{Evers2008}. 

Throughout this paper, we limit our study to the case of \emph{weak} interactions with $U \leq 1$. 
As in Finkel'stein \cite{Finkelstein1983}, the magnetic instability that we study is predicted to occur 
for \emph{arbitrarily weak} interactions near the MIT, owing to the fractal wave-function enhancement of interactions there. 
Because wave-function fractality occurs near the MIT due to disorder-induced interference phenomena, 
the physics that we study here is completely different from Stoner ferromagnetism in a clean (e.g.) Hubbard model.
Moreover, our restrictions to weak interactions and $b \geq 1$ are chosen to mimic a diffusive higher-dimensional Fermi liquid.
Mott physics and Hubbard bands are also irrelevant here.

In the rest of this
section, we summarize the main results of this paper. 
In Sec.~\ref{sec:PRBM} we first review key aspects of the noninteracting PRBM model, Eq.~(\ref{eq:HPRBM+U}) with $U = 0$. 
In Sec.~\ref{sec:Keldysh-summary}, we present analytical results regarding the interacting MIT.
We employ the non-linear sigma model \cite{Finkelstein1983,BK1994,Finkelstein2010} description of Eq.~(\ref{eq:HPRBM+U}) that 
incorporates density-density and spin-triplet fermion interactions, which descend from the Hubbard $U$. 
Perturbation theory in the sigma model is controlled for large $b \gg 1$ \cite{Mirlin1996, Mirlin2000a, Evers2008}.
We obtain very similar results to those in $d = 2 + \epsilon$ dimensions \cite{Finkelstein1983,BK1994}:
Altshuler-Aronov effects \emph{increase} the density of states, transport, and spin susceptibility,
due to enhanced spin-triplet interactions.  

In Sec.~\ref{sec:MF-summary}, we follow up with mean-field (Hartree-Fock) exact diagonalization numerics for Eq.~(\ref{eq:HPRBM+U})
with $b = 1$ [Eq.~(\ref{eq:aij})].
The results on the metallic side and at the MIT are consistent with the analytical picture.
With numerics, however, we are able to penetrate the interacting Anderson insulating regime. There we observe the nucleation of a small density of local magnetic moments 
(Figs.~\ref{fig:Sz} and \ref{fig:Sz-N}). 
The nucleation proceeds smoothly as we tune away from the MIT, with unpaired spins forming from typical localized states near the Fermi energy,
see Fig.~\ref{fig:P2-rE}. 
We present evidence that these moments form a spin glass
(Fig.~\ref{fig:Pq-a}).
This is interesting because it suggests that \emph{replica symmetry breaking} \cite{Binder1986,MezardBook}
could be
necessary to capture the physics on the insulating side of the interacting MIT.  
By contrast, NLsM studies typically presume replica-symmetric 
saddle points
in both the metallic and insulating regimes 
\cite{bibnote1}.

\subsection{Review of the non-interacting PRBM model \label{sec:PRBM}}

\subsubsection{Basic properties of the PRBM model}

The noninteracting PRBM model exhibits three different regimes, depending upon the exponent $\alpha$. 
For $0 \leq \alpha < 1/2$, the physics is similar to the random-matrix limit $\alpha = 0$. Level statistics are
determined by the Wigner surmise \cite{Livan2017,Mehta}, and wave functions are extended and featureless.  
For $1/2 < \alpha < 1$, the wave functions are ergodic, but exhibit strong fluctuations in finite size. 
The Anderson MIT at $\alpha = 1$ is ``spectrum-wide critical,'' in that almost all wave functions 
except those in the Lifshitz tails 
become critical (multifractal \cite{Mirlin1996, Mirlin2000a, Evers2008}). For $\alpha > 1$, all states Anderson localize and level statistics cross over to the Poisson distribution. 
 
Extended and localized states can be distinguished by the inverse participation ratio (IPR) \cite{Evers2008},
\begin{align}\label{IPRDef}
	P_2(\psi_E) \equiv \sum_{x} \left|\psi_E(x)\right|^4. 
\end{align}
For an extended (critical) eigenfunction $\psi_E$, $P_2 \sim (1/L)^{\tau_2}$ with $\tau_2 = 1$ ($0 < \tau_2 < 1$); here $L$ denotes the system size. 
A localized state with a localization length smaller than $L$ instead has $\tau_2 = 0$ and $P_2 \sim \ord{1}$. 
We note that there are small finite-size power-law \emph{corrections} to $P_2$ even in the insulator, due to L\'evy flights \cite{Evers2008,Kravtsov2012}. 

Another way to characterize the system is via the \emph{Chalker correlator}
\begin{align}\label{CDef}
	C(E,E') 
	\equiv
	2
	\,
	\frac{\sum_{x} \left|\psi_{E}(x)\right|^2 \left|\psi_{E'}(x)\right|^2 }{P_2(\psi_E) + P_2(\psi_{E'})}.
\end{align}
This is a normalized, ``energy-split'' IPR that measures the degree of spatial overlap between the position-space distribution functions of two different eigenstates. In the ergodic phase $C(E,E')$ is a constant for $E \neq E'$. In the Anderson insulator $C(E,E')$ is strongly suppressed except for a discrete set of eigenenergies, because localized states that are nearby in energy are typically far-separated in position space. At the Anderson MIT, one expects power-law scaling 
\cite{Chalker1988,Chalker1990,Cuevas2007,Feigelman2007}
\begin{align}\label{CScaling}
	C(E,E')
	\sim
	|E - E'|^{\tau_2/d - 1},
\end{align}
where $d$ is the number of spatial dimensions.
Although individual critical eigenstates consist of self-similar ensembles of rare, well-separated peaks, these peaks strongly overlap for states with nearby energies, in sharp contrast to the Anderson insulator. See the second and fourth panels of Fig.~\ref{fig:PRBM-GUE}(b). 
This critical scaling can enhance matrix elements of interactions \cite{Feigelman2007,Feigelman2010,Burmistrov2012,Foster2012,Foster2014}. It has been predicted to boost the pairing amplitude near the superconductor-insulator transition, in the absence of long-ranged Coulomb interactions \cite{Feigelman2007,Feigelman2010,Burmistrov2012,Zhang2022}.

In Fig.~\ref{fig:PRBM-GUE}, we exhibit numerical results for  a PRBM model in Gaussian unitary ensemble (GUE) and three different values of $\alpha$. Results are presented for level statistics, representative wave functions, and spectrum-wide statistics for the IPR [Eq.~(\ref{IPRDef})] and Chalker correlator [Eq.~(\ref{CDef})]. 

In the first column of Fig.~\ref{fig:PRBM-GUE}, the distribution of the level spacings in the GUE-PRBM model is shown for ergodic phase [(a), $\alpha=0.5$], 
the transition point [(b), $\alpha=1$],
and 
the localized phase [(c), $\alpha=1.3$]. 
In the ergodic phase, the 
distribution of the level spacings follows the Wigner surmise \cite{Livan2017} (blue curve) that exhibits strong level repulsion, similar to the random matrix limit ($\alpha=0$). 
In the localized phase, there is no level repulsion and the level spacings follow the Poisson distribution (red curve). 
A remnant of level repulsion survives near the transition point, but the distribution deviates from the Wigner surmise. 

The second column of Fig.~\ref{fig:PRBM-GUE} shows the probability density of typical wave functions at different $\alpha$. 
In the ergodic phase, the wave functions are extended, spreading over the whole system. 
The IPR $P_2$ [Eq.~\eqref{IPRDef}] of the extended wave functions is inversely proportional to the system size $L$ in 1D, which is shown in the third column of row (a) in Fig.~\ref{fig:PRBM-GUE}. 
In the localized phase [Fig.~\ref{fig:PRBM-GUE} (c)], the wave functions are spatially localized and the IPR takes values of order of unity, independent of the system size.
At the transition point  [Fig.~\ref{fig:PRBM-GUE} (b)], the wave functions are multifractal with spatially rarified peaks. The IPR has power-law dependence on system size, $P_2\sim L^{-\tau_2}$ with $0<\tau_2<1$.

The last column of Fig.~\ref{fig:PRBM-GUE} shows the Chalker correlator [Eq.~\eqref{CDef}] of the GUE-PRBM model. 
In the ergodic phase, the wave functions at different energies all have similar spatial overlap and the Chalker correlator has weak energy dependence. 
In the localized phase, the wave functions at different energies localize in disparate small regions and thus exhibit negligible spatial overlap.
At the critical point, the spatial overlap of wave functions has power-law dependence on energy difference and exhibits the Chalker scaling behavior [Eq.~\eqref{CScaling}].

\subsubsection{Nonlinear sigma model for the PRBM model}

For $1/2 < \alpha < 3/2$, the PRBM model can be mapped to a 1D version of the standard field theory of noninteracting localization, the non-linear sigma model \cite{Mirlin1996, Mirlin2000a, Evers2008}. The theory is identical to the usual one, except that the gradient-squared term is replaced by a nonlocal power of momentum $|k|^\sigma$, where $\sigma = 2 \alpha - 1$. This leads to superballistic transport throughout the ergodic phase $0 \leq \sigma < 1$. 
The theory becomes renormalizable at the MIT, where $\sigma = 1$.

The DOS per spin of the GUE-PRBM model is well-approximated by the saddle-point of the sigma model, 
which gives a generalized Wigner semicircle law
\begin{equation}
	  \nu\left( E \right) = \frac{1}{\pi J_0} \sqrt{J_0 -\frac{E^2}{4}} \,,
  		\label{eq:DOS0}
\end{equation}
with $J_0 = \sum_j J_{ij}$ and $J_{ij} = a^2\left( \left|i-j\right| \right)$. With the form of $a\left( \left|i-j\right| \right)$ given by Eq.~\eqref{eq:aij}, we have 
\begin{equation}
  	J_0 = 2b^{2\alpha} \left[ \zeta\left( 2\alpha, b \right) - \zeta \left( 2\alpha, L/2 \right) \right] + 2b -1\,.
  	\label{eq:J0}
\end{equation}
Here $\zeta(a, x)$ is the Hurwitz zeta function and $L$ is the system size. The DOS in Eq.~\eqref{eq:DOS0} is compared to numerics for variable $\alpha$ 
in Fig.~\ref{fig:DOS}(a). In the random-matrix regime $-1 \leq \sigma < 0$, the range of the energy spectrum increases with system size as 
$\sim L^{-\sigma/2}$ and the zero-energy DOS (at the spectrum center) decreases with system size $\sim L^{\sigma/2}$. For $\alpha > 1/2$ ($\sigma > 0$), the DOS becomes system-size independent, up to finite-size corrections $\sim L^{-\sigma/2}$. The range of the energy spectrum (peak of the DOS) decreases (increases) with increasing $\alpha$.

%%%%%%%%%%%%%%%%%%%%%%%%%%%%%%%%%%%%%%%%%%%%%%%%%%%%%%%%%%%%%%%%%%%%%%%%%%%%%%%%%%%%%%%
%%%%%%%%%%%%%%%%%%%%%%%%%%%%%%%%%%%%%%%%%%%%%%%%%%%%%%%%%%%%%%%%%%%%%%%%%%%%%%%%%%%%%%%
%%%%%%%%%%%%%%%%%%%%%%%%%%%%%%%%%%%%%%%%%%%%%%%%%%%%%%%%%%%%%%%%%%%%%%%%%%%%%%%%%%%%%%%
%%%%%%%%%%%%%%%%%%%%%%%%%%%%%%%%%%%%%%%%%%%%%%%%%%%%%%%%%%%%%%%%%%%%%%%%%%%%%%%%%%%%%%%
%%%%%%%%%%%%%%%%%%%%%%%%%%%%%%%%%%%%%%%%%%%%%%%%%%%%%%%%%%%%%%%%%%%%%%%%%%%%%%%%%%%%%%%
%%%%%%%%%%%%%%%%%%%%%%%%%%%%%%%%%%%%%%%%%%%%%%%%%%%%%%%%%%%%%%%%%%%%%%%%%%%%%%%%%%%%%%%

\subsection{Analytical results: Magnetic instability}
\label{sec:Keldysh-summary}

To treat the effects of the interactions in the PRBM-Hubbard model
in Eq.~(\ref{eq:HPRBM+U}), we derive 
the interacting (Finkel'stein) version of the NLsM \cite{Finkelstein1983,BK1994}. 
We work at half-filling throughout this paper.
Using the Keldysh formalism \cite{Kamenev2009,Kamenev2011,Liao2017},
we incorporate two effective dimensionless interaction parameters: 
a density-density coupling $\gamma_s > 0$, and a spin-triplet interaction $\gamma_t < 0$. 
For $U > 0$, the density-density (spin-triplet) interaction is repulsive (attractive),
as are the associated Landau parameters in a higher-dimensional diffusive Fermi liquid
\cite{BK1994,Coleman2015}.

The dimensionless effective parameters $\gamma_{s,t}$ incorporate Fermi-liquid renormalization
\cite{Finkelstein1983}. 
The limit $\gamma_t \rightarrow - \infty$ would correspond to the Stoner instability to 
ferromagnetism in a clean system. Calculations performed four decades ago for the diffusive, interacting
Fermi liquid in $d = 2 + \epsilon$ dimensions predicted
some kind of magnetic instability, signaled by the runaway RG flow of $\gamma_t$ towards negative infinity
\cite{Finkelstein1983,Castellani1984,BK1994}. Due to the \emph{antilocalizing} effects of $\gamma_t$-mediated
Altshuler-Aronov corrections, this instability was predicted to pre-empt the expected disorder-driven MIT. 
Despite enormous efforts, the nature of the instability in the field theoretic framework was 
never fully clarified. Possible interpretations include an interference-enhanced ferromagnetic instability \cite{KB1996,Chamon2000,NAL2000},
or the formation of localized spin moments \cite{Milo1989,Bhatt1992,Miranda2012,Pezzoli2009}.
A large-$N$ version of the sigma model was formulated to suppress the intervening 
magnetic phase, in an attempt to explain the 2D MIT \cite{Punn2005,Kravchenko2010}.

In this subsection we show that the sigma model predicts a phenomenology for the PRBM-Hubbard model
that is very similar to the finite-dimensional results. In particular, the triplet channel enhances
the DOS, transport, and the spin susceptibility at the interacting MIT. 
As with the results in $d = 2 + \epsilon$ dimensions, the nature of any magnetic order that onsets near the MIT is not
determined by the standard perturbative field-theory treatment. 
The detailed calculation of the results summarized in this subsection can be found in Sec.~\ref{sec:Keldysh}.

In Sec.~\ref{sec:MF-summary}, below, we instead employ 
self-consistent numerics to show that local moments form a spin glass in the insulating phase.

\subsubsection{Density of states}

In disordered system with interactions, it is known that 
the DOS receives interference-mediated interaction (Altshuler-Aronov) corrections \cite{Lee1985}. 
The DOS corrections can be evaluated via the Finkel'stein nonlinear sigma model with perturbation theory [See Sec.~\ref{sec:DosTech} for more details].
Using the NLsM, at the MIT ($\alpha = 1$) we obtain 
\begin{equation}
  	\delta \nu 
	= 
	\frac{\nu_0 \lambda}{2\pi} \left[ \ln{\left( 1-\gamma_s \right)} + 3 \ln{\left( 1-\gamma_t \right)} \right] \ln\left({\frac{\Lambda}{T}}\right)\,.
  \label{eq:dnu-MIT}
\end{equation}
Here $\nu_0 = 1/\pi\sqrt{J_0}$ is the bare DOS per spin at the Fermi energy, and $\lambda$ is the inverse of the superdiffusive ``conductivity'' 
[see Eq.~\eqref{eq:super-cond}], up to a constant. 
For the PRBM-Hubbard model, $\lambda  \propto b^{-1}$ at the MIT [Eq.~\eqref{eq:lambda}], 
and serves as the control parameter for perturbation theory when $b \gg 1$.
In Eq.~(\ref{eq:dnu-MIT}),
$\Lambda$ is the ultraviolet (UV) cutoff in energy and the temperature $T$ serves as the infrared (IR) cutoff.
Because the repulsive Hubbard $U$ gives $\gamma_s >0$ and $\gamma_t < 0$, 
the singlet channel interaction suppresses the DOS whilst the triplet one enhances it.
Near the MIT, the triplet channel interaction dominates (see below) and the DOS is predicted to be enhanced.

\begin{figure}[t!]
  \centering
  \includegraphics[width=0.3\textwidth]{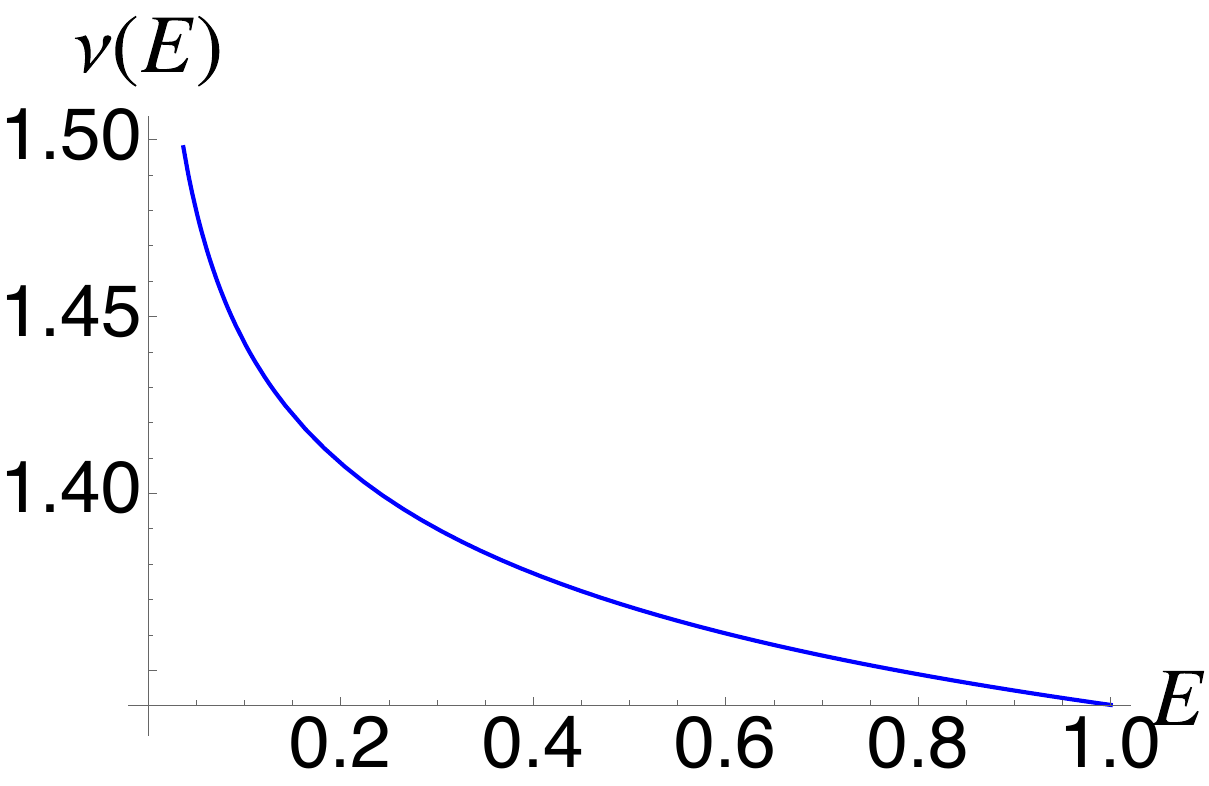}
  \caption{The RG-improved prediction for the DOS of the PRBM-Hubbard model near zero energy [Eq.~\eqref{eq:nu-RG}]. 
	The enhancement is due to the Altshuler-Aronov correction in the triplet interaction channel. 
	Here we set $\lambda=0.1$, $\gamma_t^0=0.1$ and $\Delta=1$.}
  \label{fig:DOS-RG}
\end{figure}

RG flow equations for $\lambda$, $\nu$ and the interactions can be obtained from the corresponding response functions. 
Ignoring the singlet coupling $\gamma_s$ and the logarithmically slow flow of $\lambda$, 
near the MIT we obtain the approximate dependence of the DOS close to the Fermi energy,
\begin{equation}
  \nu\left(E\right)
	\sim 
	e^{
		-3\mathsf{Li}_{2}\left[
		  \gamma_{t}^{0} \left(\frac{\Delta}{E}\right)^{\lambda}
					\right]
	}\,.
\label{eq:nu-RG}
\end{equation}
Here $E$ is the energy relative to the Fermi level,
$\Delta$ is the UV energy scale ($\sim$ bandwidth),
and $\mathsf{Li}_2(x)$ is the dilogarithm function. 
Eq.~(\ref{eq:nu-RG}) applies over an intermediate range of energies, but cannot be trusted in the limit $E \rightarrow 0$;
the latter corresponds to the strong coupling regime wherein the perturbative RG must break down. 
The interaction-enhanced DOS is shown in Fig.~\ref{fig:DOS-RG}. The prediction is consistent with the numerical results that 
we summarize in Sec.~\ref{sec:MF-summary}, see Fig.~\ref{fig:DOS}.

\subsubsection{Density dynamics}

Quantum corrections to transport 
can be obtained by evaluating the density-density response function
via the NLsM. Here we summarize the main results for density response in PRBM-Hubbard model obtained in Sec.~\ref{sec:densresptech}.
The semiclassical density-density response function takes the form
\begin{equation}
  \Pi^0(\omega,k) = 
  -\kappa \frac{D_c |k|^\sigma}{D_c |k|^\sigma -i\omega}\,.
  \label{eq:nn}
\end{equation}
Here $\sigma = 2\alpha-1$ and $\alpha$ is the hopping exponent for the PRBM, 
$D_c$ is the charge superdiffusion constant, and $\kappa$ the charge compressibility,
\begin{equation}
  	D_c \equiv \frac{D}{1-\gamma_s}\,,
	\qquad 
	\kappa \equiv 2 \nu_0 (1-\gamma_s)\,.
  \label{eq:Dc-kappa}
\end{equation}
The energy superdiffusion constant $D=2/(\lambda \pi \nu_0)$. 
Near the MIT 
($0<\sigma<2$, $1/2<\alpha<3/2$), 
the semiclassical 
electron motion is superdiffusive and the conductivity diverges.
Thus, we define the following superdiffusive ``conductivity'' to characterize transport,
\begin{equation}
  	\mathfrak{g} 
	\equiv 
	\lim_{k\to0} \frac{i\omega}{|k|^\sigma} \Pi(\omega, k) 
	= 
	\frac{4}{\pi}\lambda^{-1}.
  	\label{eq:super-cond}
\end{equation}

In the ergodic phase ($\sigma<1$), the Altshuler-Aronov correction to transport is given by
\begin{equation}
  	\delta\mathfrak{g}  
	= 
	\frac{2\Lambda_{k}^{1-\sigma}}{\pi^{2}\left(1-\sigma\right)}
	\left[ \ln\left(1-\gamma_{s}\right) + 3\ln\left(1-\gamma_{t}\right) \right]. 
 \label{eq:deltasigma}
\end{equation}
Here $\Lambda_k$ is the UV cutoff in momentum.
The correction is UV divergent (IR convergent),
indicating the RG-irrelevance of $\lambda$ in the ergodic regime
(similar to the resistivity of a diffusive metal in 3D).

\begin{figure*}[t!]
  \centering
  \includegraphics[width=0.9\textwidth]{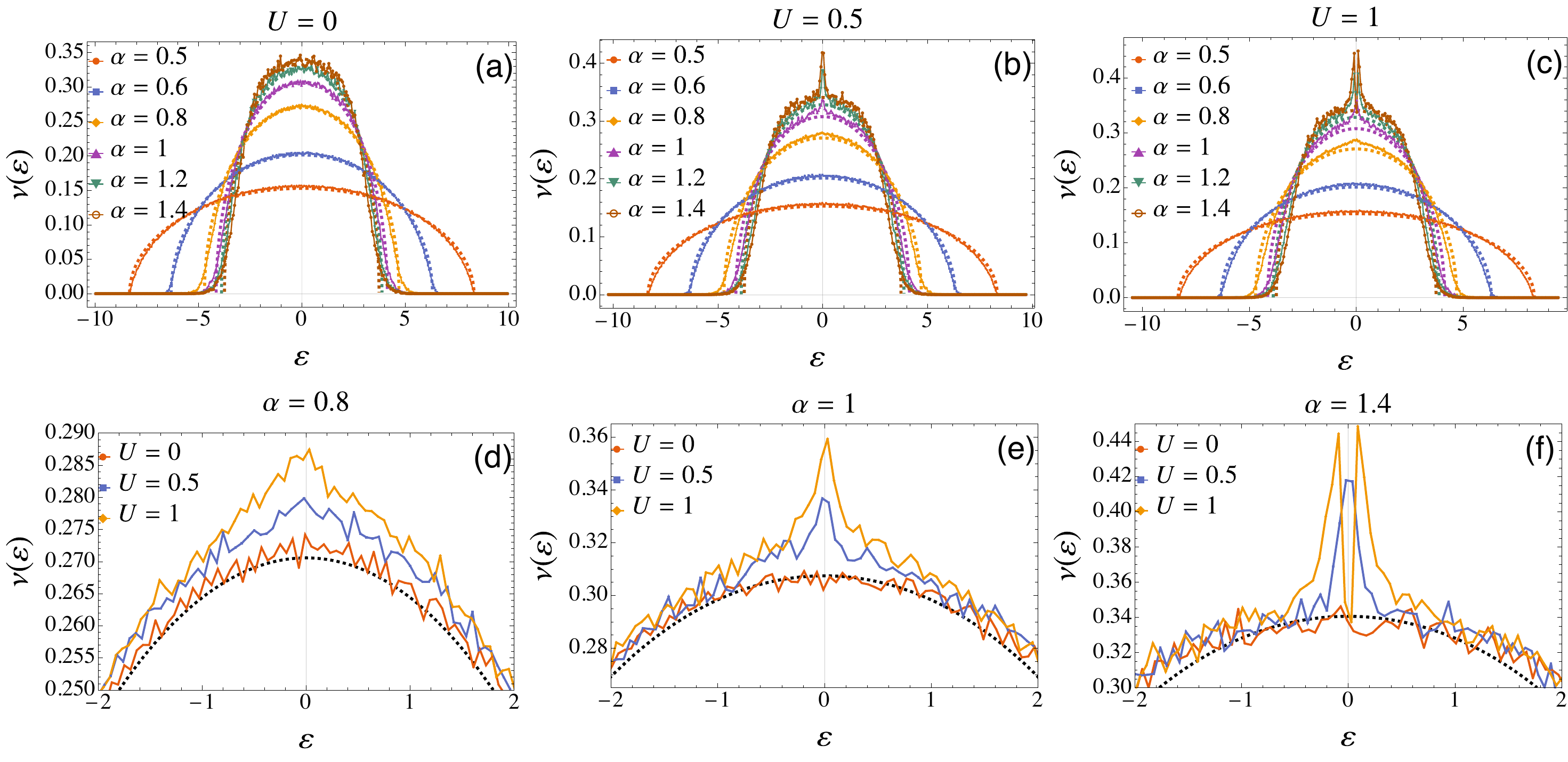}
  \caption{The density of states for PRBM-Hubbard model with repulsive Hubbard interactions. The results are taken for systems of size $L=4000$ and averaged over $N_\mathsf{ens}=50$ disorder realizations. The dashed lines indicate the DOS for non-interacting systems obtained via the NLsM [Eq.~\eqref{eq:DOS0}]. The curve at $\alpha=0.5$ is taken to be slightly away to avoid the divergence of Eq.~\eqref{eq:DOS0} at $\alpha=1/2$. Panels (a), (b) and (c) show the DOS for different hopping powers $\alpha$ with $U=0$, $0.5$ and $1$, respectively. 
Panels (d), (e) and (f) show the DOS for different interaction strength with $\alpha=0.8$, $1$ and $1.4$, respectively. 
The DOS is not changed by the interaction for systems with $\alpha$ close to $0.5$. 
When $\alpha$ approaches the MIT, the DOS receives quantum corrections due to the interplay of interactions and critical wave functions. 
The observed enhancement is consistent with the analytical results due to the Altshuler-Aronov correction in the spin-triplet channel, Eq.~(\ref{eq:nu-RG}) and Fig.~\ref{fig:DOS-RG}. 
In the localize phase, the DOS near the Fermi energy is still enhanced, but begins to show a dip for stronger interactions $U \gtrsim 1$ due to the crossover towards Mott localization. 
Here the energy $E$ is measured relative to the Hartree shift $U/2$, so that $E=0$ corresponds to the Fermi energy.}
  \label{fig:DOS}
\end{figure*}

At the noninteracting critical point ($\sigma=1$), the dependence on the UV cutoff becomes logarithmic and the correction can be cut off in the infrared by the temperature,
\begin{equation}
  	\delta\mathfrak{g}
	=
	\frac{2}{\pi^2}\left[ \ln\left(1-\gamma_{s}\right) + 3\ln\left(1-\gamma_{t}\right) \right]\ln\left(\frac{\Lambda}{T}\right)\,.
  \label{eq:dsigma-critical}
\end{equation}
Different from the usual case \cite{BK1994}, the form of the Altshuler-Aronov corrections in Eqs.~(\ref{eq:dnu-MIT}) and (\ref{eq:dsigma-critical})
is identical. This is due to the nonanalytic character of the density fluctuations in the PRBM [Eq.~(\ref{eq:nn})]. 
As a result, only a subset of the usual diagrams that renormalize transport contribute to Eq.~(\ref{eq:dsigma-critical}),
and these are precisely those responsible for DOS renormalization, see Sec.~\ref{sec:densresptech}.

Eq.~(\ref{eq:dsigma-critical}) implies that the singlet (triplet) interaction is localizing (antilocalizing). 
We find numerically that the interacting MIT is slightly shifted to $\alpha > 1$ by moderate Hubbard interactions, see Fig.~\ref{fig:transition}.  
This is consistent with magnetic fluctuation-dominated \emph{antilocalization}, which we attribute to the same Altshuler-Aronov effect responsible
for the DOS enhancement in Figs.~\ref{fig:DOS-RG} and \ref{fig:DOS}.

\subsubsection{Magnetic fluctuations boosted by multifractality \label{sec:magMFC}}

\begin{figure}
  \centering
  \includegraphics[width=0.35\textwidth]{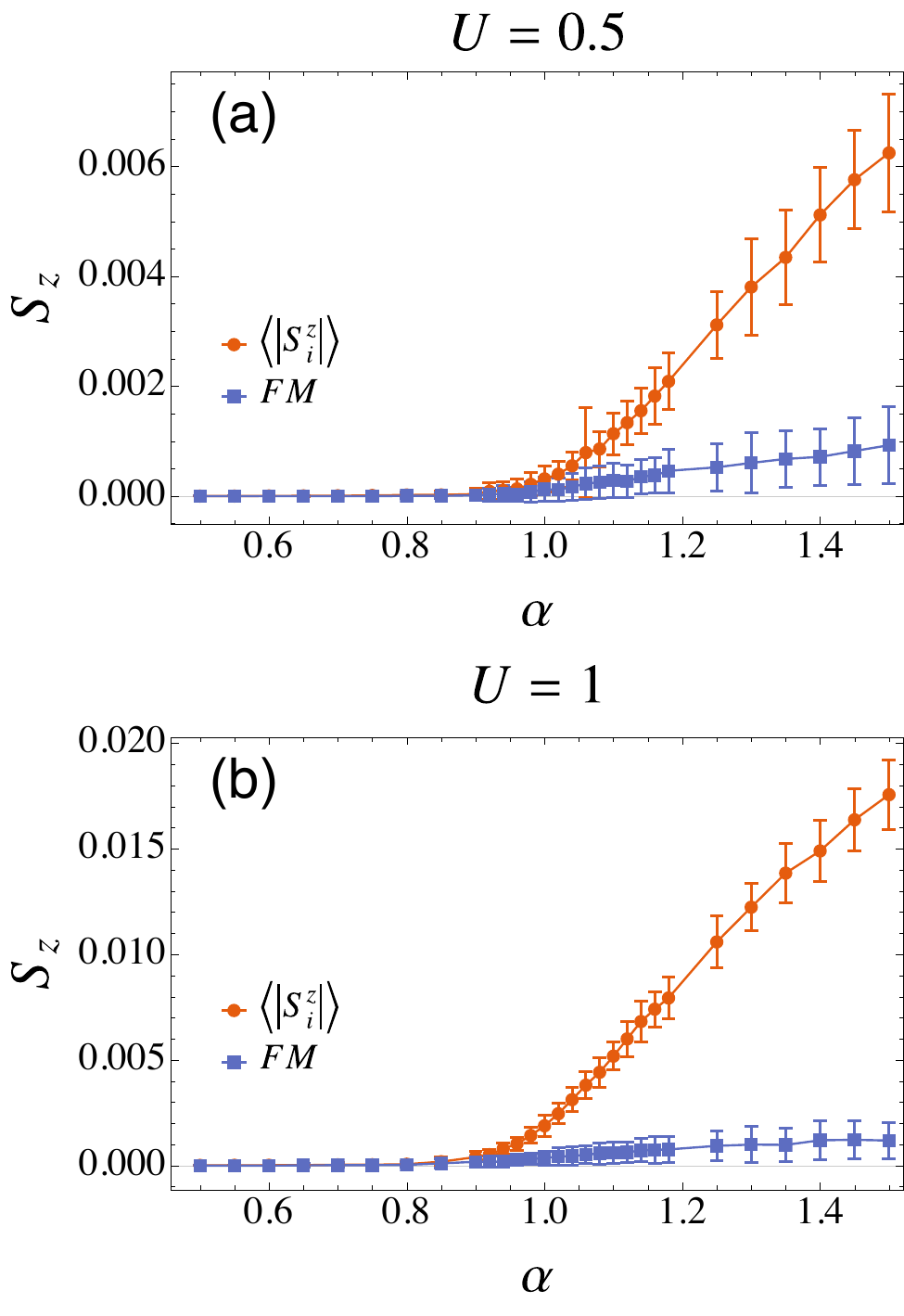}
  \caption{The local moment density $\left\langle\left|S_i^z\right|\right\rangle$ [Eq.~\eqref{eq:SzSG}] 
	and ferromagnetic order parameter $S^z_{\scriptstyle\mathsf{FM}}$ [Eq.~\eqref{eq:SzFM}] for the PRBM-Hubbard model. 
	The data are taken for system of size $L=2000$ and averaged over $N_\mathsf{ens}=100$ disorder realizations. 
	Both vanish in the ergodic regime with $0.5 \lesssim \alpha \ll 1$, and the mean field ground state is spin unpolarized. 
	Both arise close to the critical point near $\alpha=1$ and increase with $\alpha$ into the localized phase. 
	However, the weak net magnetization does not indicate ferromagnetism. 
  	The FM order parameter is much smaller than the local moment density and its variance due to ensemble averaging 
	(error bars in the plot) is comparable to its mean. }
  \label{fig:Sz}
\end{figure}

The magnetic properties can be shown by the spin response function, which is evaluated in Sec.~\ref{sec:spintech}.
The semiclassical spin response function is determined by the dynamical susceptibility
\begin{equation}\label{chiSemiClassical}
  	\chi_{ij}^{0}\left(\omega,k\right)
	=
	\delta_{ij} 
	\chi_0\frac{D_{t}|k|^{\sigma}}{D_{t}|k|^{\sigma}-i\omega}\,.
\end{equation}
Here $D_{t}$ is the spin superdiffusion constant 
and $\chi_0$ is the bare static spin susceptibility,
\begin{equation}
	D_{t}
	\equiv
	\frac{{D}}{1-\gamma_{t}}\,,
	\qquad 
	\chi_0
	\equiv
	2 \nu_0
	\left(1-\gamma_{t}\right)\,.
\end{equation}
The static spin susceptibility can be extracted from the spin response function in the static limit $\omega\to0$.

In the ergodic phase with $\sigma<1$, the static spin susceptibility receives a quantum correction,
\begin{equation}
 	 \chi_{ij}
	=
	\delta_{ij}
	\chi_{0}
	\left[1-\frac{\lambda\gamma_{t}\Lambda_{k}^{1-\sigma}}{\pi\left(1-\sigma\right)}\right]\,.
  \label{eq:chi-ij-0}
\end{equation}
The quantum correction is proportional to $\lambda$, the strength of the triplet-channel interaction $\gamma_t$,  
and the UV cutoff $\Lambda_k^{1-\sigma}$. 
At the critical point, this becomes
\begin{equation}
 	 \chi_{ij}
	=
	\delta_{ij}
	\chi_{0}
	\left[
		1-\frac{\lambda\gamma_{t}}{\pi}\ln\left(\frac{\Lambda}{T}\right)
	\right]\,.
\label{eq:chi-MIT}
\end{equation}
Repulsive Hubbard $U$ gives $\gamma_t < 0$, so that the spin susceptibility is enhanced by quantum interference. 

The RG equations for the singlet and triplet interaction strengths near the MIT take the same form as in the original 
Finkel'stein calculation \cite{Finkelstein1983,DellAnna2017} in $d = 2 + \epsilon$ dimensions, see Eq.~(\ref{eq:RG-gt-E}).
These equations possess a single relevant direction that leads to the runaway RG flow of $\gamma_t \rightarrow -\infty$. 
The mechanism that drives the flow away from weak coupling is the multifractal enhancement of matrix elements
\cite{Feigelman2007,Feigelman2010,Burmistrov2012,Foster2012,Foster2014,Zhang2022},
due to the Chalker scaling of quantum-critical wave functions [Figs.~\ref{fig:PRBM-GUE}(b,d) and Eq.~(\ref{CScaling})].
The divergence of $|\gamma_t|$ drives a simultaneous divergence of the spin susceptibility in Eq.~(\ref{eq:chi-MIT}). 
A susceptibility peak near the MIT can also be discerned in the numerical results presented in the next section,
see Fig.~\ref{fig:Sus-N}. These results signal the onset of magnetic phenomena at the MIT, but the standard field-theoretic approach is
ill-suited to determine its character.

\begin{figure}[b]
  \centering
  \includegraphics[width=0.35\textwidth]{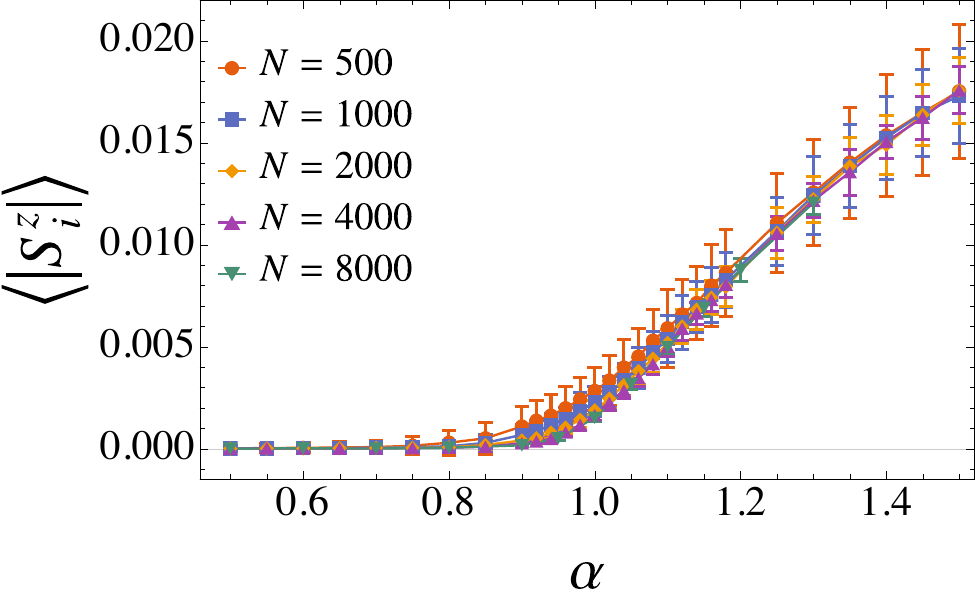}
  \caption{
  	The local moment density $\left\langle\left|S^z_i\right|\right\rangle$ [Eq.~\eqref{eq:SzSG}] for different system sizes with $U = 1$. 
	The $L=500$, $1000$, $2000$ and $4000$ data are averaged over $1000$, $500$, $100$, $50$ and $40$ disorder realizations, respectively. 
	Near the MIT, the onset of $\left\langle\left|S^z_i\right|\right\rangle > 0$ slightly sharpens with system size. 
	The moment density becomes system-size independent in the localized phase and its variance decreases with increasing $L$.}
  \label{fig:Sz-N}
\end{figure}

%%%%%%%%%%%%%%%%%%%%%%%%%%%%%%%%%%%%%%%%%%%%%%%%%%%%%%%%%%%%%%%%%%%%%%%%%%%%%%%%%%%%%%%
%%%%%%%%%%%%%%%%%%%%%%%%%%%%%%%%%%%%%%%%%%%%%%%%%%%%%%%%%%%%%%%%%%%%%%%%%%%%%%%%%%%%%%%
%%%%%%%%%%%%%%%%%%%%%%%%%%%%%%%%%%%%%%%%%%%%%%%%%%%%%%%%%%%%%%%%%%%%%%%%%%%%%%%%%%%%%%%
%%%%%%%%%%%%%%%%%%%%%%%%%%%%%%%%%%%%%%%%%%%%%%%%%%%%%%%%%%%%%%%%%%%%%%%%%%%%%%%%%%%%%%%
%%%%%%%%%%%%%%%%%%%%%%%%%%%%%%%%%%%%%%%%%%%%%%%%%%%%%%%%%%%%%%%%%%%%%%%%%%%%%%%%%%%%%%%
%%%%%%%%%%%%%%%%%%%%%%%%%%%%%%%%%%%%%%%%%%%%%%%%%%%%%%%%%%%%%%%%%%%%%%%%%%%%%%%%%%%%%%%

\subsection{Self-consistent Hartree-Fock numerics \label{sec:MF-summary}}

\subsubsection{Local moment formation}

The analytical results presented in Sec.~\ref{sec:Keldysh-summary} demonstrate the onset of strong magnetic fluctuations near the MIT in the PRBM-Hubbard model.
To elucidate the nature of these fluctuations and any incipient order that develops, we turn to Hartree-Fock numerical calculations. 
We use static mean-field theory to decouple the Hubbard interactions in terms of the onsite magnetization, and perform self-consistent exact diagonalization of 
the resulting quadratic fermion Hamiltonian. 
We restrict our attention to mean-field states with collinear magnetic order.
Details of the procedure and convergence criteria are discussed in Sec.~\ref{sec:MF-Details}. 
All results presented in this section take $b = 1$ in Eq.~(\ref{eq:aij}). Because perturbation theory formally requires $b \gg 1$, in general 
we anticipate only qualitative agreement between analytics and numerics.

Fig.~\ref{fig:DOS} shows the density of states in the PRBM model with and without the repulsive Hubbard-$U$ interaction. 
In the non-interacting case, the numerical DOS [solid lines in Fig.~\ref{fig:DOS}(a)] matches well the Wigner semicircle prediction 
[Eq.~\eqref{eq:DOS0}, dashed lines in Figs.~\ref{fig:DOS}(a)--(c)], except in the Lifshitz tails near the band edges. 

The repulsive Hubbard $U > 0$ interaction is found to enhance the numerical DOS compared to the non-interacting case near the Fermi surface at zero energy [Fig.~\ref{fig:DOS}(d)-(e)], when the system is tuned through the MIT. 
Deep in the ergodic phase $\alpha \sim 0.5$, the effect is negligible.
As shown in Fig.~\ref{fig:DOS}, the DOS at the Fermi energy increases with interaction strength and a peak starts to arise at $E = 0$ for $U=0.5$ and $U=1$ as $\alpha$ approaches 1 from below.
Near the interaction-dressed MIT [Fig.~\ref{fig:DOS}(e)], a sharp peak appears in the DOS, indicating strong enhancement of the DOS by the interaction.
The peak becomes sharper for moderate interactions ($U\sim0.5$) as $\alpha$ increases and the system enters the localized phase, see Fig.~\ref{fig:DOS}(b).
The analytical prediction in Eqs.~(\ref{eq:dnu-MIT}) and (\ref{eq:nu-RG}) and in Fig.~\ref{fig:DOS-RG} shows that the DOS is diminished by interaction in the spin-singlet channel and enhanced by interaction in the spin-triplet channel, due to the Altshuler-Aronov corrections to the DOS.
The enhancement of the DOS indicates that the spin-triplet channel interaction dominates, suggesting the onset of strong magnetic fluctuations. The indication of magnetic fluctuation is consistent with the emergence of spin glass order discussed in later sections.
%This is consistent with the analytical prediction in Eq.~(\ref{eq:nu-RG}) and Fig.~\ref{fig:DOS-RG} due to Altshuler-Aronov corrections in the spin-triplet channel,
%suggesting the onset of strong magnetic fluctuations.  

For stronger interactions ($U \gtrsim 1$), the enhancement of the DOS persists for states near the Fermi energy in the localized phase; 
however, a dip appears in the localized system exactly at the Fermi energy [Fig.~\ref{fig:DOS}(f)], indicating the crossover towards Mott localization 
expected with strong Hubbard $U$.
For small $\alpha$, the energy range of the PRBM model is large 
[see Fig.~\ref{fig:DOS}(a)]
and weak interactions are insufficient to drive Mott physics. 
For large $\alpha$ [Fig.~\ref{fig:DOS}(f)], wave functions become strongly localized in space, 
which promotes the interactions for electrons occupying the same localization volume.
The joint effects of Anderson and Mott localization leads to the emergence of Mott physics for relatively weak interactions 
(compared to the clean Hubbard model) in the \emph{strongly} localized regime $\alpha \gtrsim 1.5$. 
%, but this is not our interest here.
The interplay of Anderson localization and strong-correlated Mott physics is however beyond the scope of the present paper, and we focus here instead upon the relatively weakly interacting regime.

\begin{figure*}[ht]
  \centering
  \includegraphics[width=0.98\textwidth]{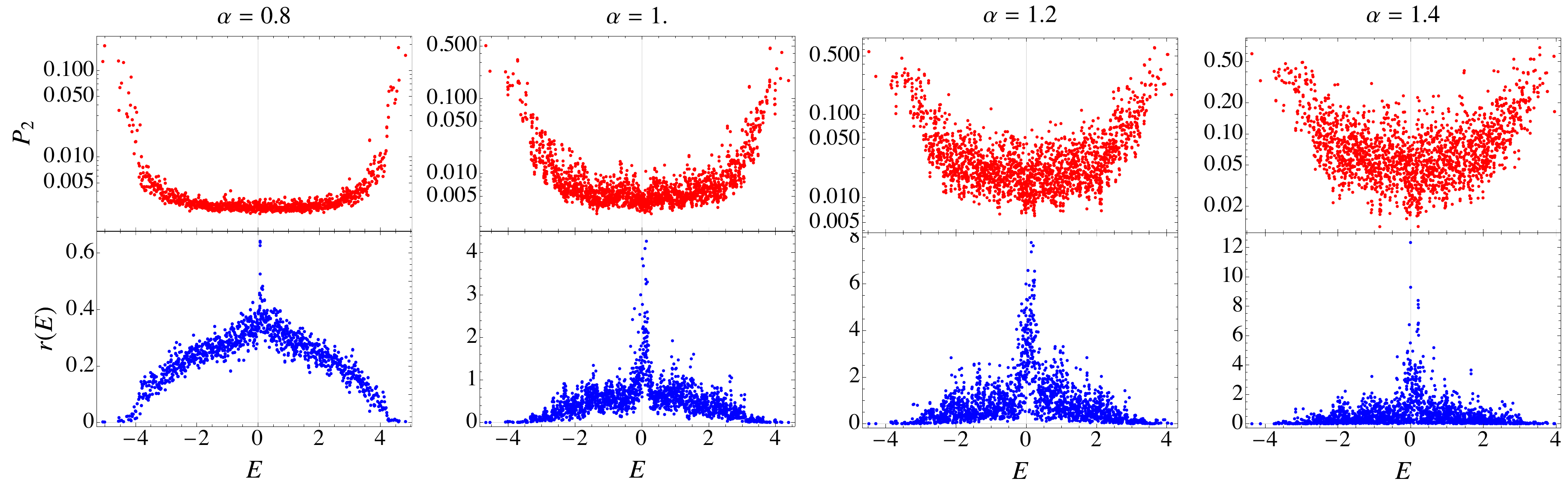}
  \caption{
	The inverse participation ratio $P_2(E)$ [Eq.~(\ref{eq:P2tau2})]
	and the magnetization ratio $r(E)$  [Eq.~(\ref{eq:rE})]
	for the Hartree-Fock single-particle eigenstates in the metallic ($\alpha=0.8$), localized ($\alpha=1.2$ and $\alpha=1.4$) phases and near the critical point ($\alpha=1$).
	The behavior of $P_2(E)$ indicates that only band-edge Liftshitz tail states are excessively localized compared to generic states in the bulk spectrum, including those
	at the Fermi energy. The peak in $r(E)$ near $E = 0$ (the Fermi energy) shows that sites with nonzero magnetization are dominated by eigenstates near the Fermi energy at the band center. Here $U=1$ and $L=1000$.
  }
	\label{fig:P2-rE}
\end{figure*}

Fig.~\ref{fig:Sz} shows the ferromagnetic (FM) and local magnetic moment density order parameters in the PRBM-Hubbard model.
The FM order parameter is evaluated in the Hartree-Fock ground state, 
\begin{equation}
	 S^z_\mathsf{FM} 
  	=
 	\left\langle 
  	\frac{1}{2L}\sum_i \left(n_{i\uparrow} - n_{i\downarrow}\right)
  	\right\rangle \,.
  \label{eq:SzFM}
\end{equation}
The local moment density is defined via
\begin{equation}
  \left\langle\left|S^z\right|\right\rangle 
  =
  \left\langle
  \frac{1}{2L}\sum_i \left|
  n_{i\uparrow} - n_{i\downarrow}
  \right| 
  \right\rangle \,.
  \label{eq:SzSG}
\end{equation}
Here $\left\langle\cdots\right\rangle$ denotes ensemble averaging over different disorder realizations.
The local moment density is conceptually similar to the Edwards-Anderson order parameter \cite{Edwards1975} for spin glasses. 
Deep in the ergodic phase both $S^z_\mathsf{FM}$ and  $\left\langle\left|S_i^z\right|\right\rangle$  are negligible.
Near the MIT, local moments start to form and  $\left\langle\left|S_i^z\right|\right\rangle$   increases with $\alpha$ in the localized phase.
Concomittantly ferromagnetism appears to nucleate and increase with $\alpha \gtrsim 1$.
However, the FM magnetization is extremely small, corresponding to a few majority spins, 
and its variance induced by disorder averaging is comparable to its mean. The net magnetization decreases with increasing system size $L$. 
For the studied values of $U$, antiferromagnetic order is even weaker, and also suppressed with increasing $L$.
By contrast, the local moment density is well-converged with $L$ in the insulating phase, see Fig.~\ref{fig:Sz-N}. 
The onset $\left\langle\left|S_i^z\right|\right\rangle >0$ sharpens slightly with increasing $L$ near the MIT $\alpha \sim 1$. 
Representative moment profiles for a small system and different $\alpha$ are exhibited in Appendix~\ref{app:spinprofiles}, Fig.~\ref{fig:Szi-alpha}.
Different initial conditions in the self-consistent Hartree-Fock procedure produce similar spatial profiles for $\left\langle\left|S_i^z\right|\right\rangle$,
but different signs of the localized moments, see  Fig.~\ref{fig:Szi-alpha=1.1} in Appendix~\ref{app:spinprofiles}. This is suggestive of spin-glass order,
discussed further below.

\begin{figure}[b]
  \centering
  \includegraphics[width=0.35\textwidth]{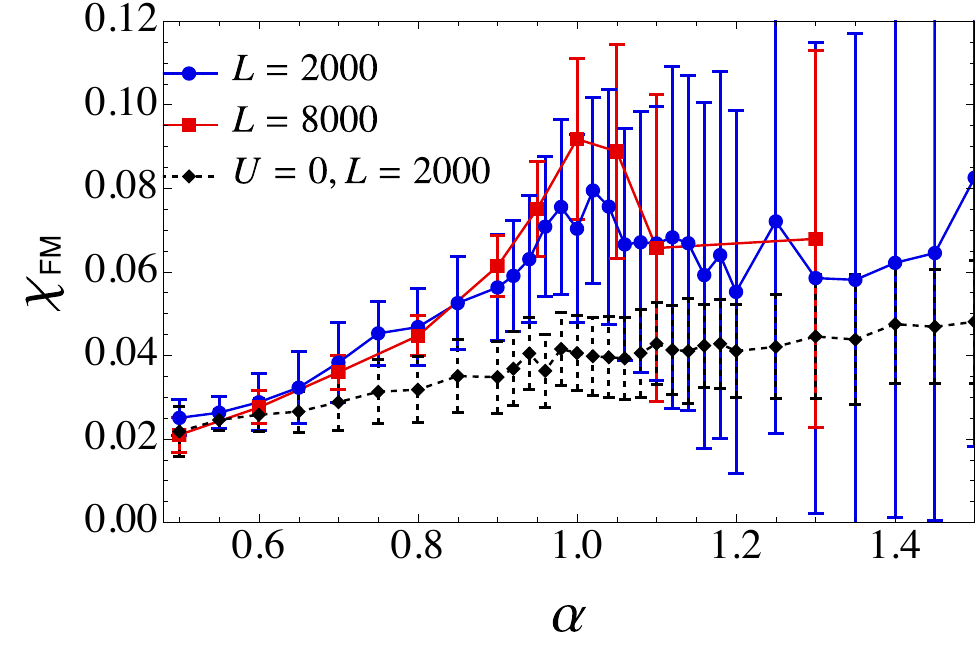}
  \caption{The spin susceptibility in the PRBM-Hubbard model for different system sizes with $U=1$. 
		The $L=2000$ (blue circle) and $L=8000$ (red square) data are averaged with $100$ and $40$ disorder realizations, respectively. 
		The incremental applied field is $\Delta h=0.02$ for $L=2000$ and $\Delta h=0.01$ for $L=8000$. 
		The spin susceptibility can be boosted by the interplay of spin-triplet interactions and multifractality near the MIT, see 
		Eq.~(\ref{eq:chi-MIT}) and the discussion in Sec.~\ref{sec:magMFC}. 
		The interacting susceptibility is indeed enhanced compared to the non-interacting system (black dashed  curve).
		Near the MIT, the interacting $\chi_{\scriptstyle \mathsf{FM}}$ shows a non-diverging peak.
		At the same time, the fluctuations induced by ensemble averaging are very large for all $\alpha \gtrsim 1$ in the interacting case,
		consistent with the absence of ferromagnetism (Fig.~\ref{fig:Sz}).
}
  \label{fig:Sus-N}
\end{figure}

The question remains whether the formation local moments follows the ``single-fluid'' or the ``two-fluid'' picture. In the two-fluid picture, local moments are formed by excessively localized electrons whilst most of the electrons remain itinerant. 
To answer this question, we define the magnetic inverse participation ratio (mag-IPR),
  \begin{equation}
	P_2^{\scriptscriptstyle \mathsf{mag}}
	(E)
	 \equiv 
	 \sum_i
   \left| \psi_{E}(x_i)\right|^2
 \left|S_i^z\right|
 \, .
	\label{eq:mag-IPR}
  \end{equation}
Here $\psi_{E}$ denotes the single-particle eigenstate wave function with energy $E$, 
while  $S_i^z$ is the spin polarization of the half-filled Hartree-Fock ground state.
We then define the magnetization ratio of a state,
  \begin{equation}
	r(E)
	\equiv 
	2 P_2^{\scriptscriptstyle \mathsf{mag}}
	(E)
	/
	P_2
	(E)\, .
	\label{eq:rE}
  \end{equation}
Here $P_2(E)$ is the inverse participation ratio of $\psi_E$
[see Eq.~\eqref{eq:P2tau2}]. 
The magnetization ratio can be used to characterize the contribution of a state to the spin polarization. 
If the spin polarization were contributed by a single state $\psi_{E_0}$, then the 
magnetization ratio would be determined by the Chalker correlation function of the states
$r(E) \sim C(E - E_0)$ [see Eqs.~(\ref{CDef}) and (\ref{CScaling})]. 
More generally, $r(E)$ is a summation of Chalker-like correlations between $\psi_E$ and the states dominating the magnetization profile.  
In Fig.~\ref{fig:P2-rE}, we show both the IPR $P_2(E)$ (red) and the magnetization ratio $r(E)$ (blue) for the whole energy spectrum of the Hartree-Fock Hamiltonian. It can be seen that $r(E)$ always peaks around the Fermi energy, in both the extended and localized phases. This indicates that the overall spin polarization is mainly contributed by the states near the Fermi energy, and that the spin polarization results from mismatch of the number of spin-up and -down electrons below the Fermi energy. On the other hand, the IPR of states near the Fermi energy are similar to generic states away from the band-edge Lifshitz tails.
This implies that the wave functions relevant to the magnetization profile localize or delocalize in the same fashion and that the local moments are not excessively localized, 
consistent with the single-fluid picture.

The ferromagnetic spin susceptibility can be obtained via
\begin{equation}
  \chi_{\scriptstyle \mathsf{FM}}\left(\Delta h\right)
  \equiv
  \lim_{\Delta h\to0}\frac{S_{\scriptstyle \mathsf{FM}}^{z}\left(\Delta h\right)-S_{\scriptstyle \mathsf{FM}}^{z}\left(-\Delta h\right)}{2\Delta h}\,.
  \label{eq:chi}
\end{equation}
In a ferromagnet, the susceptibility should diverge when $\Delta h$ goes to $0$.
On the other hand, the spin susceptibility shows a non-diverging cusp near the spin glass transition \cite{Binder1986}.
In the PRBM-Hubbard model, susceptibility data with very small $\Delta h$ is overshadowed by the fluctuations due to ensemble averaging.
The spin susceptibility in Fig.~\ref{fig:Sus-N} does exhibit a strong enhancement compared to that of the non-interacting system, and a non-diverging peak near 
the MIT. The peak becomes sharper with increasing systems size. 
However, the fluctuations of the spin susceptibility (error bars in Fig.~\ref{fig:Sus-N}) across disorder realizations become very large for $\alpha \gtrsim 1$,
consistent with the absence of true ferromagnetic order. 

\begin{figure}[b]
  \centering
  \includegraphics[width=0.4\textwidth]{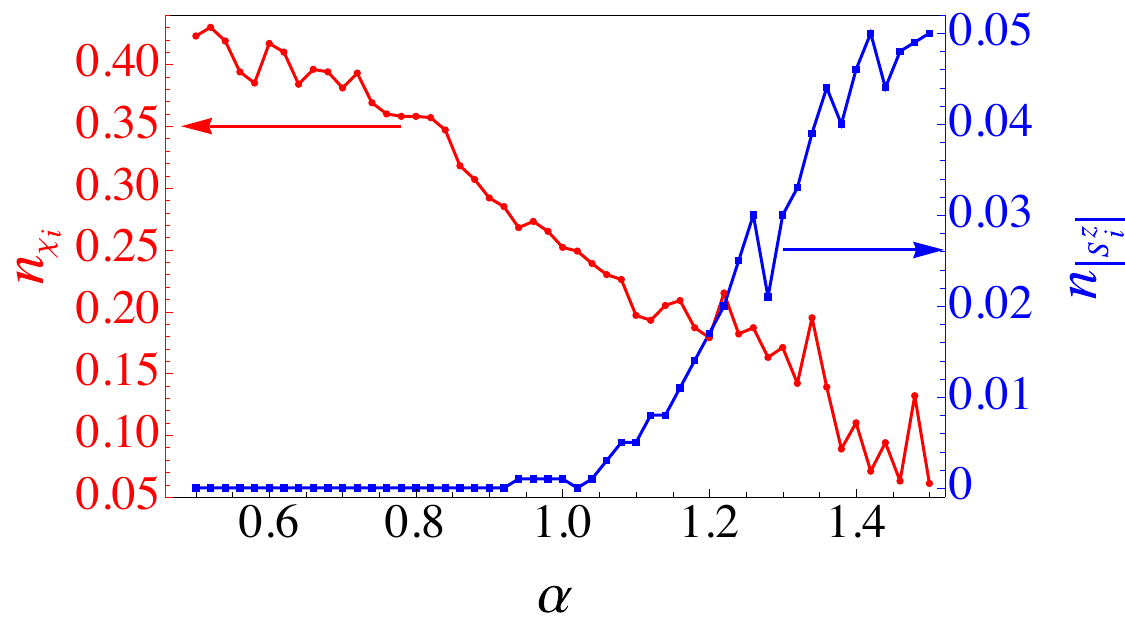}
  \caption{
	The fraction of sites $n_{\scriptscriptstyle{\chi_i}}$ [Eq.~\eqref{eq:nchi}] wherein the local spin susceptibility exceeds the average one $\chi_i> \chi_{\scriptscriptstyle \mathsf{FM}}$ (red),
	and the fraction of sites $n_{\scriptscriptstyle{\left|S_i^z\right|}}$ [Eq.~\eqref{eq:nSiz}] in which the local magnetization exceeds a threshold $\left|S_i^z\right|>0.1$ (blue). 
	In the paramagnetic phase ($\alpha<1$), there are no local moments ($n_{\scriptscriptstyle{\left|S_i^z\right|}} \sim 0$) and every site contributes approximately equally to the spin susceptibility. 
	In the localized phase ($\alpha>1$), the density of local moments increases with $\alpha$ and the spin susceptibility is mainly contributed by sites with local moments. Here $L=1000$ and $U=1$.
  }
  \label{fig:Sichi_count}
\end{figure}

We also evaluate the local spin susceptibility $\chi_i$
\begin{equation}
  \chi_i \equiv 
  \lim_{\Delta h\to 0}
  \frac{S_i^z(\Delta h) - S_i^z(-\Delta h)}{2\Delta h}
  \label{eq:chi-i}
\end{equation}
The spin configuration with $-\Delta h$ is obtained via Hartree-Fock numerics with a random initial state. 
We then evaluate the spin configuration with $\Delta h$ using the $-\Delta h$ one as the initial state. 
Physically, this corresponds to the measurement of one sample subject to a slow time-varying weak field when $\Delta h$ is small.
Fig.~\ref{fig:Sichi_count} shows the fraction of sites with $\chi_i$ greater than the overall spin susceptibility $\chi_{\scriptscriptstyle \mathsf{FM}}$ (red) and the fraction of sites with 
large 
local moments 
(with the criteria $|S_i^z|>0.1$),

\begin{subequations}
\begin{align}
  n_{\scriptscriptstyle\chi_i} 
  & = 
  \frac{1}{L} \sum_i \Theta(\chi_i - \chi_{\scriptscriptstyle \mathsf{FM}})\,,\label{eq:nchi}\\
  n_{\scriptscriptstyle \left|S_i^z\right|} 
  & = 
  \frac{1}{L} \sum_i \Theta(S_i^z - 0.1)\,. \label{eq:nSiz}
\end{align}
\end{subequations}
Here $\Theta(x)$ is the Heaviside step function.

The number of sites with contribution to $\chi_{\scriptscriptstyle \mathsf{FM}}$ above the mean decreases with $\alpha$, whilst the number of large  local moments increases with $\alpha$. In the paramagnetic phase ($\alpha\lesssim1$), almost every electron responds to the magnetic field and nearly half of the sites have $\chi_i$ greater than the average value. In the localized phase ($\alpha > 1$), electrons near the local moments exhibit a stronger contribution to the spin susceptibility than the other electrons. However, Fig.~\ref{fig:Sichi_count}  indicates that the number of sites contributing to the excess susceptibility  $n_{\scriptscriptstyle{\chi_i}}$ remains consistently larger than the number of sites hosting a large magnetic moment with $\left|S_i^z\right|>0.1$, until $\alpha \gtrsim 1.4$ (limit of strong Anderson localization). 
As shown in Fig.~\ref{fig:chi-i-alpha} of Appendix~\ref{app:spinprofiles}, for $\alpha = 1.4$ 
the local spin susceptibility for a \emph{particular} HF ground state near a subset of the local moments takes a much larger value than the other sites.

\begin{figure}
  \centering
  \includegraphics[width=0.48\textwidth]{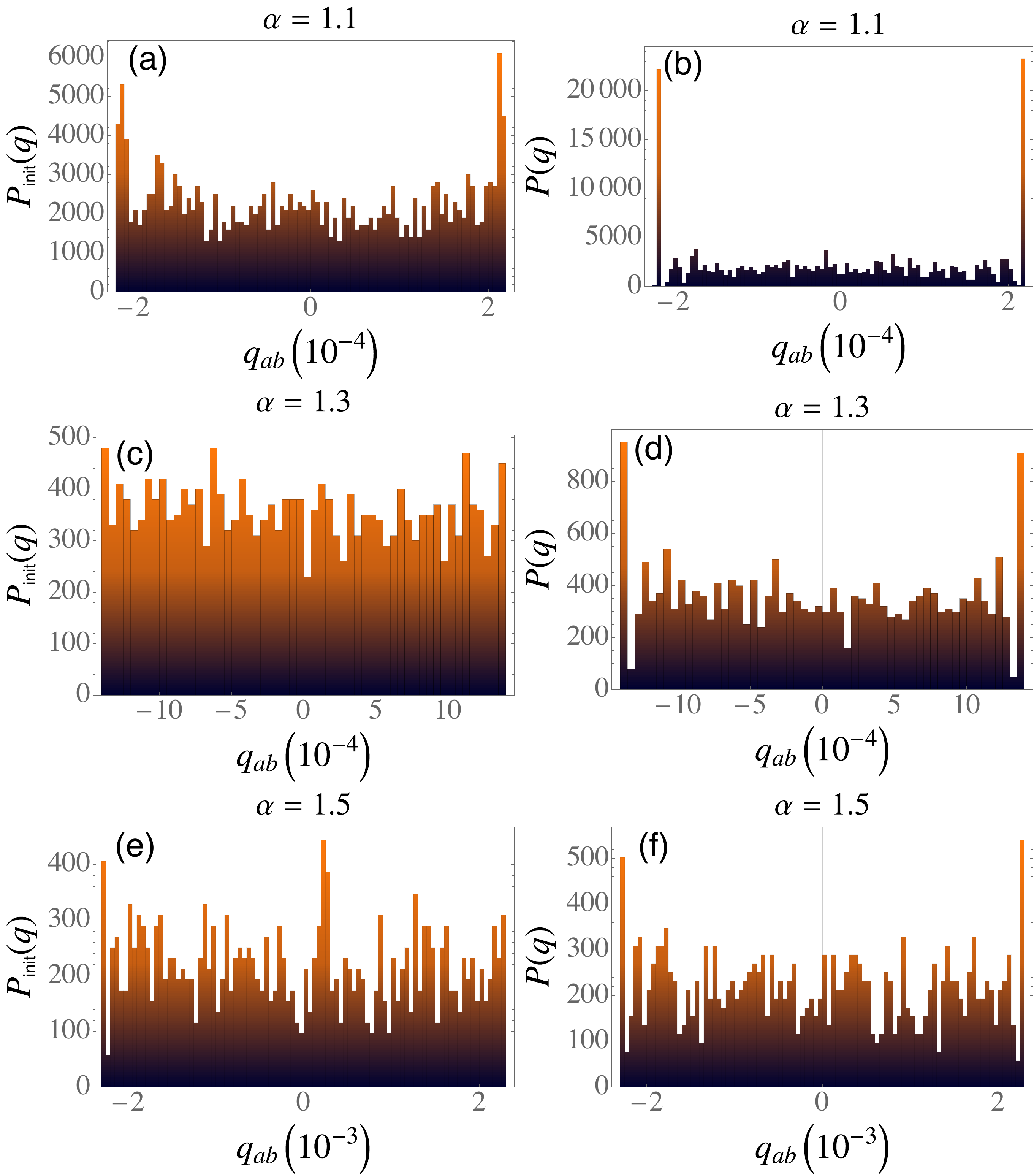}
  \caption{The distribution $P(q)$ [Eq.~\eqref{eq:Pq-a}] of the replica-symmetry-breaking order parameter $q_{ab}$ with $a$ 
  	fixed for a particular reference state [converged solution to the self-consistent Hartree-Fock (HF) numerics].
	The initial (pre-HF-convergence) condition for each of the replica $b$'s is obtained by randomly flipping a portion 
	of the spin configuration of the reference state $a$.
	We stochastically generate several thousand such modified initial conditions.  
	The resulting initial overlap distributions $P_{\scriptscriptstyle{\mathsf{init}}}(q)$ between these and the reference state $a$ 
	for three different values of $\alpha$ are shown in panels (a), (c), and (e). 
	The overlap distributions $P(q)$ for the corresponding HF-converged replica-$b$ states are shown (b), (d), and (f).
	In the localized phase, $q_{ab}$ exhibits a bimodal distribution with two peaks at $\pm q_{\scriptstyle\mathsf{max}}$, 
	and the distribution between the two peaks is approximately uniform and nonzero. 
	The probability density of the two bimodal peaks decreases and that between the two peaks increases with increasing $\alpha$, 
	implying weaker correlations between the local moments with decreasing localization length. 
	The Hubbard interaction is $U = 1$ and the ensembles of states are obtained for a fixed disorder realization. 
  }
  \label{fig:Pq-a}
\end{figure}

%\begin{figure}
%  \centering
%  \includegraphics[width=0.42\textwidth]{fig7.pdf}
%  \caption{ The distribution $P(q)$ [Eq.~\eqref{eq:Pq-a}] of the replica symmetry breaking order parameter $q_{ab}$ with $a$ fixed for a particular reference state
%	(converged solution to the self-consistent Hartree-Fock numerics).
%	The initial conditions for each of the replica $b$'s are obtained by flipping the spin configurations of the reference state $a$.
%	In the localized phase, $q_{ab}$ exhibits a bimodal distribution with two peaks at $\pm q_{\scriptstyle\mathsf{max}}$, and the distribution between the two peaks is approximately uniform and nonzero. 
%	The probability density of the two bimodal peaks decreases and that between the two peaks increases with increasing $\alpha$, implying weaker correlations between the local moments with decreasing
%         localization length. The Hubbard interaction is $U = 1$ and the system size is $L = 1000$. 
%  }
%  \label{fig:Pq-a}
%\end{figure}

\subsubsection{Spin-glass order \label{sec:SGO}}

A classical equilibrium spin glass exhibits a very large number of nearly-degenerate metastable ``pure'' states with different spin configurations \cite{MezardBook}.
These nearly-degenerate states are separated by large free-energy barriers in a complex configuration landscape.  
Below the spin-glass transition temperature, the system will converge (in dissipative dynamics or simulated annealing) into one of these 
ergodicity-breaking metastable states.

The mean-field states obtained in self-consistent numerics depend upon the choice of initial conditions in the spin-glass phase.
By choosing different initial conditions, we obtain an ensemble of ``replicas'' (solutions).
Since we restrict our attention to collinear solutions of the mean-field equations, we can characterize these
in terms of the local magnetization $S^z_{i,a}$ profile, where $i$ labels the lattice site and $a$ indexes the replica.  
Examples drawn from converged solutions in the insulating phase 
with $\alpha = 1.1$ are depicted in Fig.~\ref{fig:Szi-alpha=1.1}, Appendix~\ref{app:spinprofiles}.
(Convergence criteria and stability for a \emph{particular} state are discussed in Sec.~\ref{sec:MF-Details}.)

A collinear (Ising) spin-glass phase can be characterized by the distribution function for the overlaps $q_{a b}$, defined via
\begin{equation}
  q_{ab} 
  \equiv
  \frac{1}{L}
  \sum_i S_{i,a}^z \, S_{i,b}^z.
  \label{eq:qab}
\end{equation}
Here $a$ and $b$ are indices labeling different replicas and $S^z_{i,a}$ is the spin configuration of 
the converged Hartree-Fock mean-field state in replica $a$. 
In practice, we select the replicas via the following steps:
  \renewcommand{\theenumi}{\roman{enumi}}
  \begin{enumerate}
	\item Choose a random initial state and feed it into the Hartree-Fock numerics to obtain convergence. This is taken to be the reference state $a$.
	\item Sample a random number $r\in[0, 1)$ from the uniform distribution.
	\item Stochastically flip a portion $r$ of the spins in the reference state $a$ to obtain a new initial state $\{S^z_{i,0b}\}$.
	\item Feed the new initial state into the Hartree-Fock numerics to obtain a new Hartree-Fock converged state
	  $b$ with spin configuration  $\{S^z_{i,b}\}$.
	\item Repeat ii, iii, and iv to obtain an ensemble of several thousand Hartree-Fock-converged states.
\end{enumerate}
The overlap between the \emph{initial} spin configuration used to seed the calculation for replica $b$ and that of the 
converged reference state $a$ is approximately uniform (see the left column in Fig.~\ref{fig:Pq-a}). 
The initial distribution $P_{\mathsf{init}}(q)$ is obtained via
  \begin{subequations}
\begin{align}
  P_{\mathsf{init}} (q) 
  &\equiv
  \sum_b \delta(q-q_{ab}^0)\,,\\
  q_{ab}^0 
&= \frac{1}{L} \sum_i S_{i,a}^z S_{i,0b}^z\,.
  \label{eq:Pq_init}
\end{align}
  \end{subequations}
Here $\{S_{i,0b}^z\}$ is the spin configuration obtained in step (iii)  of the above procedure.

With self-consistency, the distribution of the overlaps between the converged solutions $\{b\}$ and that of the reference state $a$ 
deviates substantially from the initial uniform distribution, as shown in  
the right column in Fig.~\ref{fig:Pq-a}.

\begin{figure}
  \centering
  \includegraphics[width=0.38\textwidth]{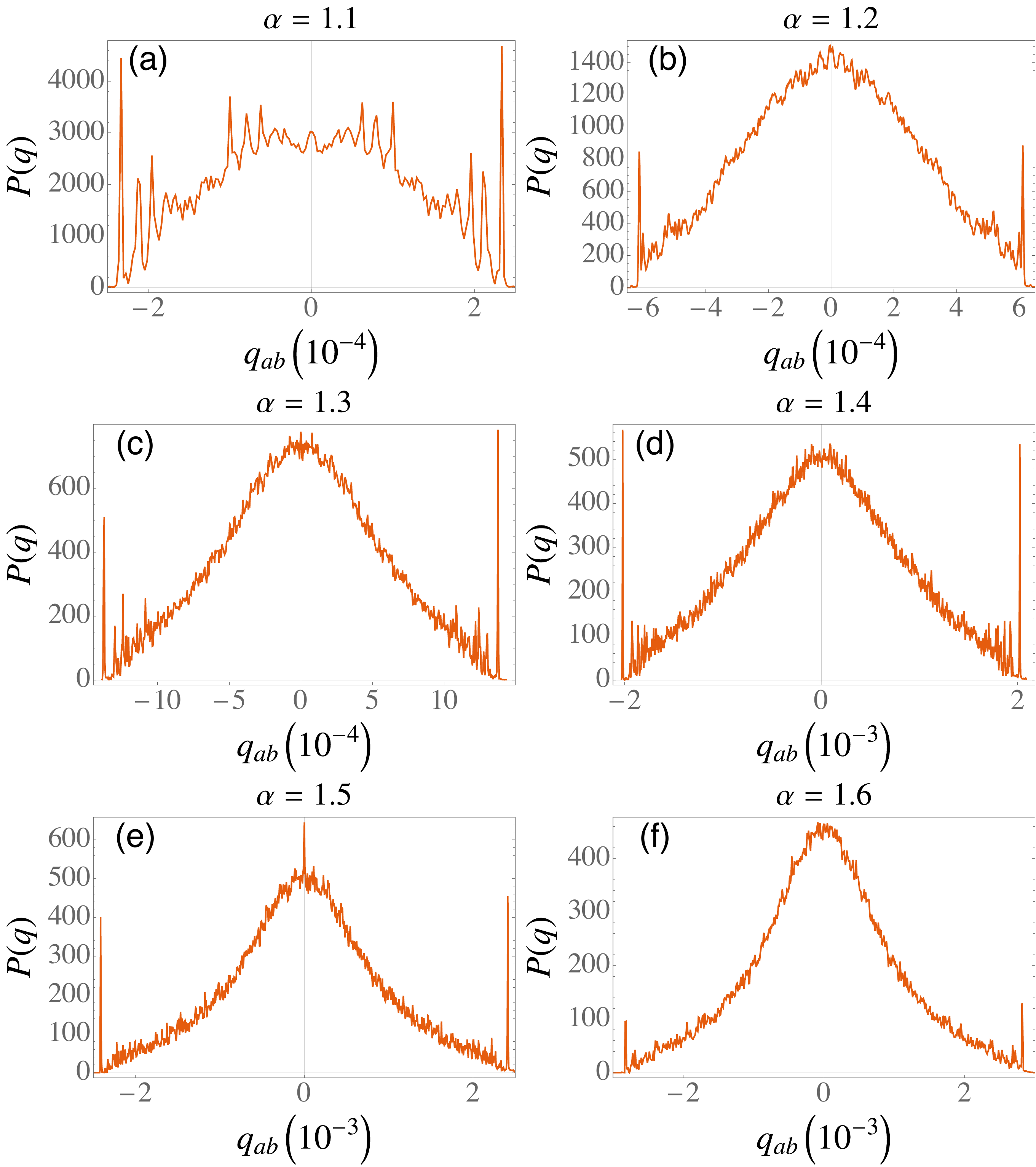}
  \caption{ The distribution $P(q)$ [Eq.~\eqref{eq:Pq-Avg}] of the replica symmetry breaking order parameter $q_{ab}$ between all the replicas. 
	The ensembles of the replicas are chosen in the same way as that in Fig.~\ref{fig:Pq-a}.
	The distribution functions show two sharp peaks at $\pm q_{\scriptstyle \mathsf{max}}$ and another smooth peak at $q=0$.
	The peaks at $\pm q_{\scriptstyle \mathsf{max}}$ characterize the spin glass order and their values decrease with increasing $\alpha$.
	The smooth peak at $q=0$ results from the choice of the replica ensemble (set of initial conditions for the self-consistent numerics).
  }
  \label{fig:Pq-Avg}
\end{figure}

The spin-glass order is characterized by the distribution of the overlaps $q_{ab}$. 
Figs.~\ref{fig:Pq-a}(b,d,f) 
show the distribution function of $q_{ab}$ defined via
\begin{equation}
  P(q) 
  \equiv
  \sum_b
  \delta(q-q_{ab})
\,.
  \label{eq:Pq-a}
\end{equation}
Here we fix the replica $a$ as the reference state and sum over all the replica $b$'s,
and  $\{S_{i,b}^z\}$ is the spin configuration obtained in step (iv) of the above procedure.
The $q_{ab}$ in Fig.~\ref{fig:Pq-a} with fixed $a$ shows a bimodal distribution in the localized phase ($\alpha>1$).
Two peaks appear in the distribution at $\pm q_{\scriptstyle \mathsf{max}}$, 
reflecting the spin $\text{SU}(2)$ symmetry of the PRBM-Hubbard model.
Between the two peaks, the distribution is  approximately uniform,
but with a probability density smaller than that 
in the distribution of overlaps 
between the reference state $a$ and the \emph{initial conditions} seeding the calculation of the replica $b$'s. 
With increasing $\alpha$, $q_{\scriptstyle \mathsf{max}}$ increases and the peak value at $\pm q_{\scriptstyle \mathsf{max}}$ of the distribution function decreases.
At large $\alpha$ ($\alpha>1.5$), the localization length of the single-particle wave functions is reduced to the order of the lattice constant.
As a result, the correlations between the local moments become much weaker and the spin-glass order is strongly diminished.  
The distribution of overlaps between the converged replica solutions becomes closer 
to the uniform distribution of the initial conditions, with much lower peaks at $\pm q_{\scriptstyle \mathsf{max}}$.

Fig.~\ref{fig:Pq-Avg} shows the replica ensemble-averaged distribution function,
\begin{equation}
  P(q) 
  =
  \sum_{a,b}
  \delta(q-q_{ab})\,.
  \label{eq:Pq-Avg}
\end{equation}
Here the replicas are still chosen in the same way as previously stated, 
but we consider the distribution of the overlap between all of the replicas, rather than the overlap with a single reference state.
The bimodal distribution of the spin glass order is still clear with two peaks at $\pm q_{\scriptstyle \mathsf{max}}$.

The distribution of overlaps exhibited in Fig.~\ref{fig:Pq-a} is qualitatively similar to the continuous Parisi replica-symmetry breaking (RSB) solution
for the classical Ising Sherrington-Kirkpatrick model \cite{MezardBook}; a similar ansatz obtains for the classical
isotropic Heisenberg glass \cite{Binder1986}. Combined with the direct inspection of converged Hartree-Fock solutions 
shown in Figs.~\ref{fig:Szi-alpha} and \ref{fig:Szi-alpha=1.1}, we conclude that the interacting Anderson insulator exhibits spin-glass order. 
Although we do not attempt to discern whether true RSB occurs in the thermodynamic limit, we note that the 
field-theory calculations presented in Sec.~\ref{sec:Keldysh-summary} that fail to identify the incipient glass order expand 
around a replica-symmetric saddle point.

We note that classical spin-glass models with power-law interactions were investigated long ago \cite{Khanin1980,Cassandro1982,Kotliar1983,Binder1986}.
The Hamiltonian takes the form
\begin{align}
	H = \sum_{i < j} \frac{\epsilon_{i j} \, \sigma_i \, \sigma_j}{|i - j|^{\Upsilon}}.
\end{align}
Here $\sigma_i \in \pm 1$ is an Ising variable, and $\{\epsilon_{i j}\}$ are taken to be identically distributed, independent Gaussian random variables
with zero mean. This model exhibits a finite-temperature spin-glass transition only for $\Upsilon < 1$, with non-mean-field exponents in the window
$2/3 < \Upsilon < 1$. However, we stress that the effective spin exchange induced in the self-consistently determined, interacting Anderson insulator
studied in this paper has both long-ranged power-law tails 
and
strong short-ranged correlations within the localization volume. The latter originate from the critical character of the 
wave functions at the self-consistently determined MIT near $\alpha = 1$.

\subsubsection{Interaction-dressed MIT}

\begin{figure}
  \centering
  \includegraphics[width=0.49\textwidth]{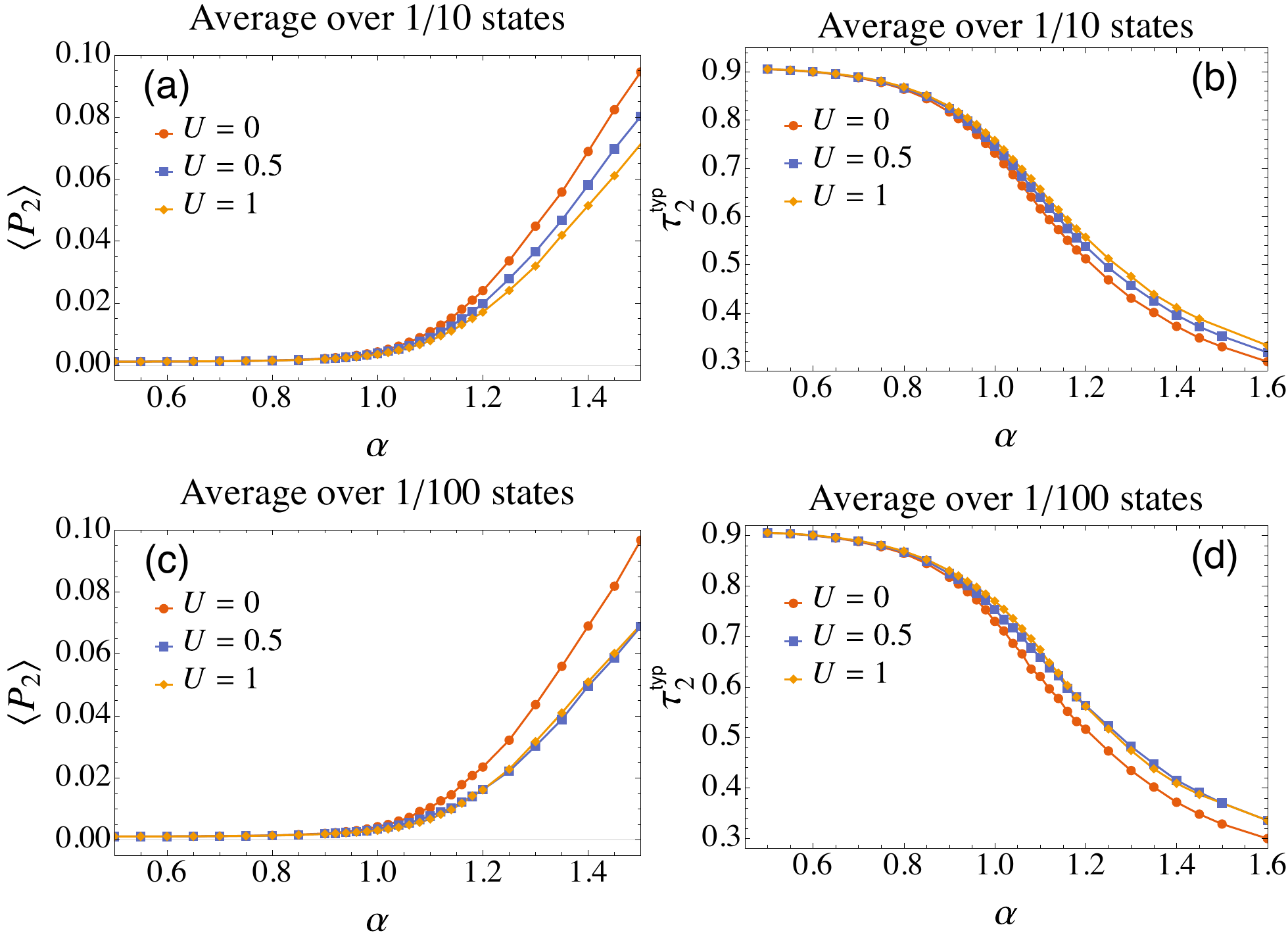}
  \caption{The inverse participation ratio $P_2$ [(a) and (c)] and typical multifractal dimension 
  $\tau_2^{\scriptscriptstyle \mathsf{typ}}$ [(b) and (d)]
  [see Eq.~(\ref{eq:P2tau2}) and text following] 
  in the PRBM-Hubbard model for different interaction strengths. 
  Results are ensemble averaged over disorder realizations and portions of the 
  Hartree-Fock spectrum near the Fermi energy, 
  as indicated.
  The system size is $L=2000$ and all data are averaged over $100$ disorder realizations. 
  The MIT is illustrated by the rising of $P_2$ and drop of $\tau_2^{\scriptscriptstyle \mathsf{typ}}$ near $\alpha=1$. 
  Repulsive Hubbard-$U$ produces a delocalizing correction to the system, leading to smaller $P_2$ for stronger interaction.
  This is consistent with the enhanced DOS due to Altshuler-Aronov effects in the spin-triplet channel,
  Fig.~\ref{fig:DOS}.}
  \label{fig:P2tau2}
\end{figure}

As reviewed in Sec.~\ref{sec:PRBM}, the noninteracting MIT can be identified via the scaling behavior of the IPR 
[Eq.~(\ref{IPRDef})] with system size. For our Hartree-Fock numerics in the spinful PRBM-Hubbard model, 
we define the IPR and the effective multifractal dimension $\tau_2$ 
of a Hartree-Fock eigenstate $\psi_E$
via
\begin{equation}
  P_2(E) = \sum_i \left|\sum_{\sigma = \uparrow,\downarrow} \left|\psi_{E \sigma}(x_i)\right|^2\right|^2\,,\quad
  \tau_2 = -\ln{P_2}/\ln{L}.
  \label{eq:P2tau2}
\end{equation}
 
\begin{figure}
  \centering
  \includegraphics[width=0.49\textwidth]{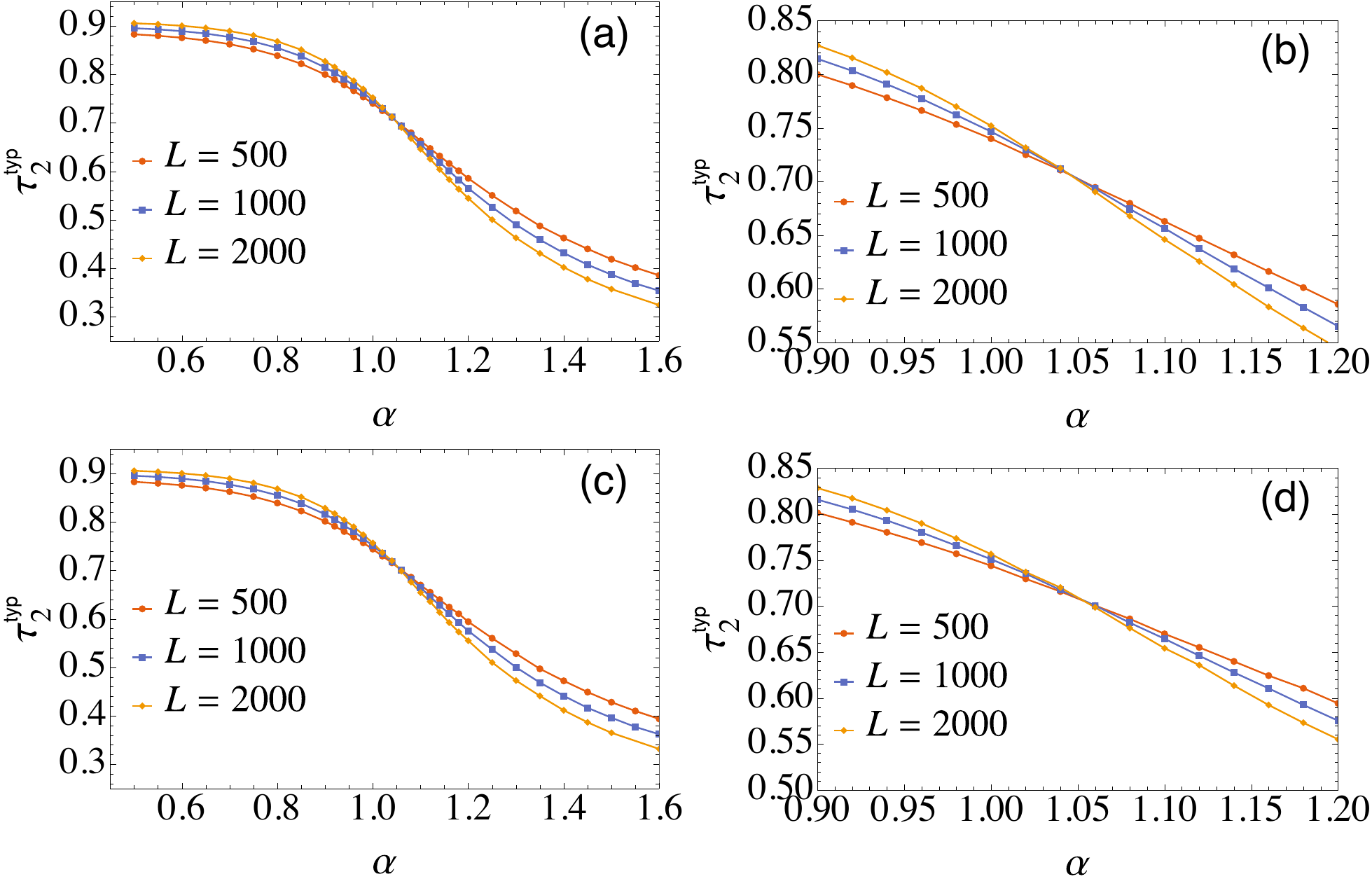}
  \caption{The typical fractal dimension $\tau_2^{\scriptscriptstyle \mathsf{typ}}$ for the PRBM-Hubbard model with $U=1$ and different system sizes. 
  (a) and (b): data that is averaged over $1/5$ of the energy spectrum around Fermi energy; 
  (c) and (d): data averaged over $1/10$ of the states around Fermi energy. 
  The $\tau_2^{\scriptscriptstyle \mathsf{typ}}$ curves for different system sizes intersect at the same point, 
  indicating that an estimate of the transition point can be extracted from finite-size scaling.}
  \label{fig:tau2-L}
\end{figure}

Fig.~\ref{fig:P2tau2} shows the averaged IPR $\left\langle P_2 \right\rangle$ and typical $\tau_2^{\scriptscriptstyle\mathsf{typ}}$ for different interaction strengths. The data shown in Fig.~\ref{fig:P2tau2} are averaged over disorder realizations and states around the Fermi energy. 
The typical $\tau_2^{\scriptscriptstyle \mathsf{typ}}$ is defined as the ensemble and state average of the exponent $\tau_2$ (via the average of the logarithm 
$\left\langle \ln P_2 \right\rangle$ rather than the log of the average $\ln\left\langle P_2 \right\rangle$).
The spectrum-wide Anderson transition approximately survives in the presence of the interactions. 
The wave functions receive antilocalizing corrections from the interactions (consistent with magnetic fluctuations and the enhanced DOS), which is shown by the smaller $P_2$ (correspondingly, larger $\tau_2^{\scriptscriptstyle\mathsf{typ}}$) for the interacting system compared to the non-interacting one.

\begin{figure}[b]
  \centering
  \includegraphics[width=0.35\textwidth]{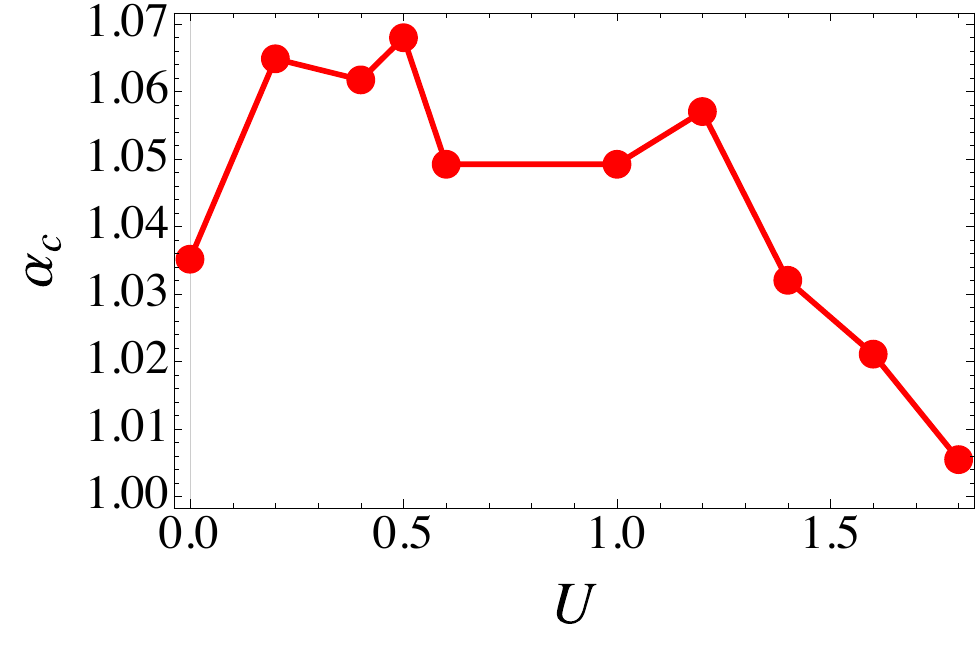}
  \caption{The interacting Anderson-Mott transition point $\alpha_c$ in the PRBM-Hubbard model. 
  	The transition point $\alpha_c$ is obtained as the crossing point of the 
	typical multifractal dimension $\tau_2^{\scriptscriptstyle \mathsf{typ}}$
	of $L=2000$ and $L=500$. 
  	The $L=2000$ and $L=500$ systems are averaged over $100$ and $400$ disorder realizations, respectively. 
  	Weak repulsive interaction results in anti-localizing correction to the Anderson transition and increases $\alpha_c$, 
  	while Mott localization starts to play a role for strong interactions.}
  \label{fig:transition}
\end{figure}

Fig.~\ref{fig:tau2-L} shows the system-size scaling of the typical multifractal dimension $\tau_2^{\scriptscriptstyle \mathsf{typ}}$. 
The $\tau_2^{\scriptscriptstyle \mathsf{typ}}$ at different system sizes intersect at the same point; the latter estimates location of the interaction-dressed MIT. 
The estimated critical point of the interacting Anderson-Mott transition is shown in Fig.~\ref{fig:transition}. Near the critical point, the spin-triplet interaction dominates and moderate bare interaction strengths produce an anti-localizing correction to the single-particle wave functions, driving the transition point to larger $\alpha$. When the interaction is strong, Mott physics steps in and results in localizing correction due to the correlation effects.

%%%%%%%%%%%%%%%%%%%%%%%%%%%%%%%%%%%%%%%%%%%%%%%%%%%%%%%%%%%%%%%%%%%%%%%%%%%%%%%%%%%%%%%
%%%%%%%%%%%%%%%%%%%%%%%%%%%%%%%%%%%%%%%%%%%%%%%%%%%%%%%%%%%%%%%%%%%%%%%%%%%%%%%%%%%%%%%
%%%%%%%%%%%%%%%%%%%%%%%%%%%%%%%%%%%%%%%%%%%%%%%%%%%%%%%%%%%%%%%%%%%%%%%%%%%%%%%%%%%%%%%
%%%%%%%%%%%%%%%%%%%%%%%%%%%%%%%%%%%%%%%%%%%%%%%%%%%%%%%%%%%%%%%%%%%%%%%%%%%%%%%%%%%%%%%
%%%%%%%%%%%%%%%%%%%%%%%%%%%%%%%%%%%%%%%%%%%%%%%%%%%%%%%%%%%%%%%%%%%%%%%%%%%%%%%%%%%%%%%
%%%%%%%%%%%%%%%%%%%%%%%%%%%%%%%%%%%%%%%%%%%%%%%%%%%%%%%%%%%%%%%%%%%%%%%%%%%%%%%%%%%%%%%
%%%%%%%%%%%%%%%%%%%%%%%%%%%%%%%%%%%%%%%%%%%%%%%%%%%%%%%%%%%%%%%%%%%%%%%%%%%%%%%%%%%%%%%
%%%%%%%%%%%%%%%%%%%%%%%%%%%%%%%%%%%%%%%%%%%%%%%%%%%%%%%%%%%%%%%%%%%%%%%%%%%%%%%%%%%%%%%
%%%%%%%%%%%%%%%%%%%%%%%%%%%%%%%%%%%%%%%%%%%%%%%%%%%%%%%%%%%%%%%%%%%%%%%%%%%%%%%%%%%%%%%
%%%%%%%%%%%%%%%%%%%%%%%%%%%%%%%%%%%%%%%%%%%%%%%%%%%%%%%%%%%%%%%%%%%%%%%%%%%%%%%%%%%%%%%
%%%%%%%%%%%%%%%%%%%%%%%%%%%%%%%%%%%%%%%%%%%%%%%%%%%%%%%%%%%%%%%%%%%%%%%%%%%%%%%%%%%%%%%

\section{Technical details for mean-field numerics}
\label{sec:MF-Details}

We employ the self-consistent Hartree-Fock approximation to solve the PRBM-Hubbard model in Eq~\eqref{eq:HPRBM+U},
\begin{align}
 	U n_{i\uparrow} n_{i\downarrow} 
  	=&\, 
  	U 
  	n_{i\uparrow} \langle n_{i\downarrow}\rangle 
  	+
  	U
  	\langle n_{i\uparrow}\rangle n_{i\downarrow} 
  	-
  	U 
  	\langle n_{i\uparrow}\rangle \langle n_{i\downarrow}\rangle 
\nonumber\\
	&\,
	-
	U 
	\langle c_{i\uparrow}^{\dagger}c_{i\downarrow}\rangle c_{i\downarrow}^{\dagger}c_{i\uparrow}
	-
	U 
	c_{i\uparrow}^{\dagger}c_{i\downarrow}\langle c_{i\downarrow}^{\dagger}c_{i\uparrow}\rangle 
\nonumber\\
	&\,
	+ 
	U 
	\langle c_{i\uparrow}^{\dagger}c_{i\downarrow} \rangle \langle c_{i\downarrow}^{\dagger}c_{i\uparrow}\rangle  \,.  
	\label{eq:MF}
\end{align}
The first line gives the local Hartree shift and the second line corresponds to the Fock terms. 
By a priori choosing the $z$-axis as the possible symmetry-breaking direction of spin polarization, we can ignore 
the Fock terms and the Hartree-Fock and Hartree approximations are equivalent \cite{Fazekas1999}. 
Thus, we set the initial value of the Fock terms to $0$ and they are not generated in self-consistent mean-field theory. 
Then we numerically solve the mean-field Hamiltonian self-consistently (including the inhomogeneous Hartree shifts) and ensemble average over
disorder realizations to obtain the physical quantities.

In the self-consistent Hartree-Fock numerics, the local density of spin up and down electrons are evaluated iteratively according to the mean-field equation in Eq.~\eqref{eq:MF}. The difference of ${n_{i\sigma}}$ between two subsequent iterations are defined as,
\begin{equation}
  \Delta n = \frac{1}{L}\sqrt{\sum_{i\sigma} \left(n_{i\sigma}^{m+1} - n_{i\sigma}^m\right)^2} \,.
  \label{eq:deltan}
\end{equation}
Here $n_{i\sigma}^m$ is the density of spin $\sigma$ electrons at site $i$ in the $m$-th iteration. 
We use $\Delta n<10^{-6}$ as the criteria for the convergence of the self-consistent mean-field theory. 

In disordered systems, it is known that the straightforward iteration procedure in Eq.~\eqref{eq:MF} suffers from convergence issues \cite{Sinova2000}. The iterations can get stuck and oscillate between metastable states at the local extrema of the mean-field functional.
The oscillation is especially serious and the convergence condition is very hard to be reached with finite iterations for generic initial conditions
in the Anderson insulating phase, where single-particle wave functions are strongly localized. In our numerics, we employ Broyden's 
method \cite{Sriv1984, Johnson1988} to accelerate and ensure the convergence of the iterative numerics. The iterations converge well even deep 
in the localized phase when the method is applied.

%%%%%%%%%%%%%%%%%%%%%%%%%%%%%%%%%%%%%%%%%%%%%%%%%%%%%%%%%%%%%%%%%%%%%%%%%%%%%%%%%%%%%%%
%%%%%%%%%%%%%%%%%%%%%%%%%%%%%%%%%%%%%%%%%%%%%%%%%%%%%%%%%%%%%%%%%%%%%%%%%%%%%%%%%%%%%%%
%%%%%%%%%%%%%%%%%%%%%%%%%%%%%%%%%%%%%%%%%%%%%%%%%%%%%%%%%%%%%%%%%%%%%%%%%%%%%%%%%%%%%%%
%%%%%%%%%%%%%%%%%%%%%%%%%%%%%%%%%%%%%%%%%%%%%%%%%%%%%%%%%%%%%%%%%%%%%%%%%%%%%%%%%%%%%%%
%%%%%%%%%%%%%%%%%%%%%%%%%%%%%%%%%%%%%%%%%%%%%%%%%%%%%%%%%%%%%%%%%%%%%%%%%%%%%%%%%%%%%%%
%%%%%%%%%%%%%%%%%%%%%%%%%%%%%%%%%%%%%%%%%%%%%%%%%%%%%%%%%%%%%%%%%%%%%%%%%%%%%%%%%%%%%%%
%%%%%%%%%%%%%%%%%%%%%%%%%%%%%%%%%%%%%%%%%%%%%%%%%%%%%%%%%%%%%%%%%%%%%%%%%%%%%%%%%%%%%%%
%%%%%%%%%%%%%%%%%%%%%%%%%%%%%%%%%%%%%%%%%%%%%%%%%%%%%%%%%%%%%%%%%%%%%%%%%%%%%%%%%%%%%%%
%%%%%%%%%%%%%%%%%%%%%%%%%%%%%%%%%%%%%%%%%%%%%%%%%%%%%%%%%%%%%%%%%%%%%%%%%%%%%%%%%%%%%%%
%%%%%%%%%%%%%%%%%%%%%%%%%%%%%%%%%%%%%%%%%%%%%%%%%%%%%%%%%%%%%%%%%%%%%%%%%%%%%%%%%%%%%%%
%%%%%%%%%%%%%%%%%%%%%%%%%%%%%%%%%%%%%%%%%%%%%%%%%%%%%%%%%%%%%%%%%%%%%%%%%%%%%%%%%%%%%%%

\section{Keldysh response theory}
\label{sec:Keldysh}

\subsection{Non-linear sigma model}

The NLsM for the non-interacting PRBM was first studied in Ref.~\cite{Mirlin1996}. 
In Keldysh field theory \cite{Kamenev2009, Kamenev2011}, the action can be written as
\begin{equation}
  	S_0 
	= 
	\frac{1}{2\lambda}
	\intop_{k}
	\mathsf{Tr}\left[
				|k|^{\sigma}
				\,
				\hat{Q}(-k)
				\,
				\hat{Q}(k)
		\right]
	+
	i h
	\intop_{x}
	\mathsf{Tr}\left[
			\hat{\omega} \, \hat{Q}(x)
		\right].
  \label{eq:SQ0}
\end{equation}
Here the exponent $\sigma = 2 \alpha - 1$, and $h=\pi \nu_0$ characterizes the DOS at the Fermi energy. 
The coupling $\lambda$ is the inverse of the superdiffusive ``conductivity'' [Eq.~(\ref{eq:super-cond})].
For the unitary class studied here, the matrix field $\hat{Q}(x) \rightarrow Q_{\omega,\omega'}^{s \tau,s' \tau'} \sim c_{s,\tau}(\omega,x) \bar{c}_{s',\tau'}(\omega',x)$ 
describes hydrodynamic diffuson modes of the disordered system, and 
carries indices in 
frequency ($\omega,\omega'$) 
$\otimes$ 
spin ($s,s'$) 
$\otimes$ 
Keldysh 
($\tau,\tau'$)
spaces. $\hat{Q}(k)$ denotes the Fourier transform of the position-space field $\hat{Q}(x)$.
The latter satisfies everywhere the local constraint $\hat{Q}^2(x) = \hat{1}$.
In Eq.~(\ref{eq:SQ0}), $\hat{\omega}$ is the diagonal matrix of frequencies that span the real line.

Compared to the NLsM describing the dirty electron gas near two dimensions, 
Eq.~(\ref{eq:SQ0})
is nonlocal due to the non-analytic kinetic term when $\sigma<2$, which arises from the long-ranged hopping in the PRBM model. 
For the PRBM with $a(r)$ given by Eq.~\eqref{eq:aij} and $1/2 < \alpha < 3/2$, 
the coupling strength $\lambda$ can be evaluated as
\begin{align}
  	\lambda = \frac{\pi J_0}{b^{2\alpha} c_{\alpha}} \,, 
	\quad
	 c_\alpha = \int_{0}^\infty d x \, x^{-2\alpha} (1-\cos{x})\,.
\label{eq:lambda}
\end{align}
Here $J_0$ is given by Eq.~\eqref{eq:J0}, and $c_\alpha$ is an order-one constant.
The parameter $\lambda \sim b^{1 - 2 \alpha}$, so that the NLsM is in the weak coupling regime when $b \gg 1$. 
The NLsM is renormalizable at $\sigma = \alpha = 1$, and exhibits a spectrum-wide Anderson MIT for this parameter value \cite{Mirlin1996, Mirlin2000a}. 

A repulsive Hubbard $U > 0$ gives rise to repulsive density-density (spin-singlet) $U_s > 0$ and attractive spin-triplet interactions $U_t < 0$ near the Fermi surface \cite{BK1994,Coleman2015}.
These can be incorporated into the interacting version of the Keldysh NLsM. Following conventions in Ref.~\cite{Liao2017}, 
the action reads
\begin{align}
  	S 
	=&
	S_0
	+
	ih
	\intop_{x}
	\mathsf{Tr}\left[
			\left(\tilde{V}_{\mathsf{cl}}+\tilde{V}_{\mathsf{q}}\tauh^{1}\right)
			\hat{M}_{F}\left(\hat{\omega}\right)
			\hat{Q}
			\,
			\hat{M}_{F}\left(\hat{\omega}\right)
		\right] 
\nonumber\\
 &
	-
	i\frac{4h}{\pi}
	\intop_{t,x}
	\tilde{V}_{\mathsf{cl}}\tilde{V}_{\mathsf{q}}
	-
	i\frac{2}{U_{s}}
	\intop_{t,x}
	\rho_{\mathsf{cl}}\rho_{\mathsf{q}}
\nonumber \\
 & 
	-
	ih
	\intop_{x}
	\mathsf{Tr}\left[
			\left(\tilde{\boldsymbol{B}}_{\mathsf{cl}}+\tilde{\boldsymbol{B}}_{\mathsf{q}}\tauh^{1}\right)
			\cdot
			\shb
			\,
			\hat{M}_{F}\left(\hat{\omega}\right)
			\hat{Q}
			\,
			\hat{M}_{F}\left(\hat{\omega}\right)
		\right]
\nonumber\\
 &
	-
	i\frac{4h}{\pi}
	\intop_{t,x}
	\tilde{B}_{\mathsf{cl}}\cdot\tilde{B}_{\mathsf{q}}
	-
	i\frac{2}{U_{t}}
	\intop_{t,{x}}
	\boldsymbol{S}_{\mathsf{cl}}\cdot\boldsymbol{S}_{\mathsf{q}}\,.
 \label{eq:Sq}
\end{align}
The Pauli matrices $\{\tauh^{a}\}$ act on Keldysh space, 
while $\{\sh^a\}$ act on spin-1/2 components; $\shb$ is the vector of spin matrices.
The operator $\hat{M}_F$ encodes the distribution function of the electrons,
\begin{equation}
 	 \hat{M}_{F}\left(\omega\right)
	=
	\left[
	\begin{array}{cc}
		1 & F(\omega)\\
		0 & -1
	\end{array}
	\right]\,,
  \label{eq:M_F}
\end{equation}
with $F(\omega)=\tanh(\beta \omega/2) = 1 - 2 f(\omega)$; $f$ is the equilibrium Fermi-Dirac distribution.
The local fields $\rho(t,x)$ and $\boldsymbol{S}(t,x)$ 
encode the dynamical components of the electron particle and spin densities, and the subscripts $\mathsf{cl}$ and $\mathsf{q}$ 
denote the classical and quantum components in Keldysh field theory. 
In addition, $V$ and $\boldsymbol{B}$ represent external electric and magnetic fields used to define response functions. 
In Eq.~(\ref{eq:Sq}), we employ the notations
\begin{equation}
	\tilde{V}_{\mathsf{cl/q}}=V_{\mathsf{cl/q}}+\rho_{\mathsf{cl/q}}\,,
	\qquad
	\tilde{\boldsymbol{B}}_{\mathsf{cl/q}}=\boldsymbol{B}_{\mathsf{cl/q}}-\boldsymbol{S}_{\mathsf{cl/q}}\,.
\label{eq:tildeV}
\end{equation}

%%%%%%%%%%%%%%%%%%%%%%%%%%%%%%%%%%%%%%%%%%%%%%%%%%%%%%%%%%%%%%%%%%%%%%%%%%%%%%%%%%%%%%%
%%%%%%%%%%%%%%%%%%%%%%%%%%%%%%%%%%%%%%%%%%%%%%%%%%%%%%%%%%%%%%%%%%%%%%%%%%%%%%%%%%%%%%%
%%%%%%%%%%%%%%%%%%%%%%%%%%%%%%%%%%%%%%%%%%%%%%%%%%%%%%%%%%%%%%%%%%%%%%%%%%%%%%%%%%%%%%%
%%%%%%%%%%%%%%%%%%%%%%%%%%%%%%%%%%%%%%%%%%%%%%%%%%%%%%%%%%%%%%%%%%%%%%%%%%%%%%%%%%%%%%%
%%%%%%%%%%%%%%%%%%%%%%%%%%%%%%%%%%%%%%%%%%%%%%%%%%%%%%%%%%%%%%%%%%%%%%%%%%%%%%%%%%%%%%%
%%%%%%%%%%%%%%%%%%%%%%%%%%%%%%%%%%%%%%%%%%%%%%%%%%%%%%%%%%%%%%%%%%%%%%%%%%%%%%%%%%%%%%%

\subsection{Density of states \label{sec:DosTech}}

In the Keldysh NLsM, the density of states can be obtained via
\begin{equation}
  \nu\left( \omega \right) = \frac{\nu_0}{2}\mathsf{Tr}\left[ \tauh^3 \hat{Q}_{\omega, \omega} (x) \right]\,
  \label{eq:dos}
\end{equation}
Here $\nu_0$ is the density of states per spin at zero energy and $\tauh^3$ is the Pauli matrix in Keldysh space.

\begin{figure}[t!]
  \centering
  \includegraphics[width=0.45\textwidth]{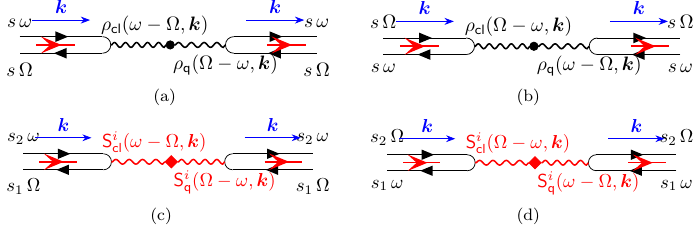}
  \caption{The Feynman diagrams contributing to the Altshuler-Aronov correction to the density of states with singlet [(a) and (b)] and triplet channel [(c) and (d)] interactions.}
  \label{fig:DOS-Diagram}
\end{figure}

We employ the ``$\pi$-$\sigma$'' parameterization of the $Q$ matrix field \cite{Liao2017},
\begin{align}
	\hat{Q}
	\equiv
	\begin{bmatrix}
	\sqrt{\hat{1} - \hat{W} \, \hat{W}^\dagger} & \hat{W} \\
	\hat{W}^\dagger & - \sqrt{\hat{1} - \hat{W}^\dagger \hat{W}}
	\end{bmatrix}_{\tau},
\end{align}
where $\hat{W} \rightarrow W_{\omega,\omega'}^{s,s'}(x)$ is an unconstrained, complex-valued matrix field with indices in frequency and spin spaces. 
Then
\begin{widetext}
\begin{align}
  \tilde{\nu}\left(\omega\right) = &
	4
	-
	\frac{1}{2}
	\sum_{s_{1}s_{2}}
	\,
	\intop_{\Omega,k}\left\langle W_{\omega\Omega}^{s_{2}s_{1}}\left(k\right)W_{\Omega\omega}^{\dagger s_{1}s_{2}}\left(k\right)\right\rangle 
 	-
	\frac{1}{2}
	\sum_{s_{1}s_{2}}
	\,
	\intop_{\Omega,k}\left\langle W_{\Omega\omega}^{s_{2}s_{1}}\left(k\right)W_{\omega\Omega}^{\dagger s_{1}s_{2}}\left(k\right)\right\rangle +\mathcal{O}\left(W^{4}\right) \,.
\label{eq:nu-WW}
\end{align}
Here $\tilde{\nu} = 2\nu(\omega)/\nu_0$.
The diagrams with bare propagators cancel 
because the DOS is noncritical without interactions.
Non-vanishing diagrams arise due to the interactions (Altshuler-Aronov corrections \cite{Lee1985}). 
At one-loop order these are shown in Fig.~\ref{fig:DOS-Diagram}. 
The correction in the singlet channel in Figs.~\ref{fig:DOS-Diagram}(a) and (b) is given by
\begin{equation}
  	\delta\tilde{\nu}_s 
	= 
	-\left(ih\right)^{2}\intop_{\Omega, k}
	\left[\begin{array}{r}\Delta_{0}^{2}\left(\Omega-\omega,k\right)i\Delta_{\rho}^{R}\left(\omega-\Omega,k\right)\left(F_{\omega}-F_{\Omega}\right) \\
  	+\Delta_{0}^{2}\left(\omega-\Omega,k\right)i\Delta_{\rho}^{R}\left(\Omega-\omega,k\right)\left(F_{\Omega}-F_{\omega}\right)\end{array}\right].
  \label{eq:dnu-s}
\end{equation}
\end{widetext}
Here $\Delta_\rho^R(\omega, k)$ is the retarded density propagator, 
\begin{equation}
  	i \Delta_\rho^R(\omega, k) 
	= 
	\frac{i\pi\gamma_s}{4h} 
	\frac{\Delta_{u,s}(-\omega,k)}{\Delta_0(-\omega, k)}\,,
  \label{eq:Deltarho}
\end{equation}
with 
\bsub
\begin{align}
	\Delta_{0}(\omega, k) 
	&= 
	\frac{1}{\frac{\left|k\right|^\sigma}{\lambda} + i\frac{h\omega}{2}},\,
\\
	\Delta_{u,s}(\omega, k) 
	&= 
	\frac{1}{\frac{\left|k\right|^\sigma}{\lambda} + i\frac{h\omega}{2} (1-\gamma_s)}\,.
  \label{eq:Deltaus}
\end{align}
\esub
The parameter $\gamma_{s}$ is the dimensionless interaction strength in the spin-singlet channel, which incorporates the Fermi-liquid renormalization of the compressibility
\begin{equation}
	  \gamma_s = \frac{2U_s h/\pi}{1+2 U_s h/\pi}\,.
  	\label{eq:gamma-s}
\end{equation}
Ignoring the irrelevant terms at order of $\omega^2$, we have 
\begin{equation}
  	\delta\tilde{\nu}_s 
  	= 
  	\frac{i\pi h\gamma_{s}}{2}\intop_{\Omega, k}
  	F_{\Omega}
  	\,
  	\Delta_{0}\left(-\Omega,k\right)
  	\Delta_{u,s}\left(-\Omega,k\right) 
  	\,.
  	\label{eq:denu-s1}
\end{equation}
Similarly, the correction due to the triplet channel interaction is given by
\begin{equation}
  	\delta\tilde{\nu}_t 
  	= 
  	\frac{i\pi h\gamma_{t}}{2}\intop_{\Omega, k}
  	F_{\Omega}
  	\,
  	\Delta_{0}\left(-\Omega,k\right)
  	\Delta_{u,t}\left(-\Omega,k\right) \,.
  	\label{eq:denu-t}
\end{equation}
Here we have
\begin{equation}
	\Delta_{u,t}(\omega, k) 
	= 
	\frac{1}{\frac{\left|k\right|^\sigma}{\lambda} + i\frac{h\omega}{2} (1-\gamma_t)}\,,
  \label{eq:Deltaut}
\end{equation}
\begin{equation}
  \gamma_t = \frac{2U_t h/\pi}{1+ 2U_t h/\pi}\,.
  \label{eq:gamma-t}
\end{equation}

Define 
\begin{equation}
	I_{1}
	=
	\frac{i\pi h\gamma}{2}
	\intop_{\Omega,k}
	F_{\Omega}
	\,
	\Delta_{0}\left(-\Omega,k\right)
	\Delta_{u}\left(-\Omega,k\right)\,.
\label{eq:I1}
\end{equation}
The integral is UV divergent when $\sigma \leq 1$. Integrating out $\Omega$ first from $-\infty$ to $+\infty$, we
have 
\begin{equation}
	I_{1}=\lambda\ln\left(1-\gamma\right)\intop_{l}\frac{1}{l^{\sigma}}\,.
\end{equation}
In the extended phase ($\sigma<1$), we obtain 
\begin{equation}
	I_{1}=\frac{\lambda\Lambda_{k}^{1-\sigma}}{\pi\left(1-\sigma\right)}\ln\left(1-\gamma\right)\,.
\end{equation}
Here $\Lambda_{k}$ is the UV cutoff in momentum. 
The corresponding correction to the DOS in the ergodic phase is
\begin{equation}
  	\delta{\nu}(\omega)
	= 
	\frac{\nu_0 \lambda\Lambda_{k}^{1-\sigma}}{2\pi\left(1-\sigma\right)}
	\left[
	\ln\left(1-\gamma_{s}\right) 
	+ 
	3 \ln\left(1-\gamma_{t}\right)
	\right].
\end{equation}
At the noninteracting critical point $\sigma = 1$, we have 
\begin{equation}
	I_{1}=\frac{\lambda}{\pi}\ln\left(1-\gamma\right)\ln\left(\frac{\Lambda}{T}\right)\,,
\end{equation}
which leads to the DOS correction in Eq.~(\ref{eq:dnu-MIT}).

We can also calculate the perturbative interaction correction in the localized phase. 
The integral $I_{1}$ is UV and IR convergent and we cannot simply extract the UV behavior of $\Omega$ first. 
We have 
\begin{widetext}
\begin{align}
I_{1}=&-\frac{\lambda\left(2-\gamma\right)\gamma D^{1-\frac{1}{\sigma}}T^{\frac{1}{\sigma}-1}}{4\pi\sigma}
\int_{0}^{\infty}dx\int_{-\infty}^{\infty}dy\frac{x^{\frac{1}{\sigma}}y\tanh\frac{y}{2}}{\left(x^{2}+y^{2}\right)\left[x^{2}+\left(1-\gamma\right)^{2}y^{2}\right]}\,.
\end{align}
\end{widetext}
The overall correction to the DOS in the localized phase diverges as a power-law with decreasing temperature,
\begin{equation}
  \delta\tilde{\nu}\left(\omega\right)=-\lambda D^{1-\frac{1}{\sigma}}T^{\frac{1}{\sigma}-1}\left[\mathfrak{a}_s\mathfrak{A}_{s}+3\mathfrak{a}_t\mathfrak{A}_{t}\right]
\end{equation}
Here $\mathfrak{a}_{s/t}$ are defined via 
\begin{equation}
	  \mathfrak{a}_{s/t} = (2-\gamma_{s/t}) \gamma_{s/t}\,,
\label{eq:a_st}
\end{equation}
and
\begin{align}
&\,
		\mathfrak{A}_{s/t}
\nonumber\\
&\,\,
		=
		\frac{1}{4\pi\sigma}
		\int_{0}^{\infty}dx
		\int_{-\infty}^{\infty}dy
		\frac{x^{\frac{1}{\sigma}}y\tanh\frac{y}{2}}{\left(x^{2}+y^{2}\right)\left[x^{2}+\left(1-\gamma_{s/t}\right)^{2}y^{2}\right]}\,.
\end{align}
We have introduced the dimensionless coordinates $x=\frac{D\left|k\right|^{\sigma}}{T}$
and $y=\frac{\Omega}{T}$.

%%%%%%%%%%%%%%%%%%%%%%%%%%%%%%%%%%%%%%%%%%%%%%%%%%%%%%%%%%%%%%%%%%%%%%%%%%%%%%%%%%%%%%%
%%%%%%%%%%%%%%%%%%%%%%%%%%%%%%%%%%%%%%%%%%%%%%%%%%%%%%%%%%%%%%%%%%%%%%%%%%%%%%%%%%%%%%%
%%%%%%%%%%%%%%%%%%%%%%%%%%%%%%%%%%%%%%%%%%%%%%%%%%%%%%%%%%%%%%%%%%%%%%%%%%%%%%%%%%%%%%%
%%%%%%%%%%%%%%%%%%%%%%%%%%%%%%%%%%%%%%%%%%%%%%%%%%%%%%%%%%%%%%%%%%%%%%%%%%%%%%%%%%%%%%%
%%%%%%%%%%%%%%%%%%%%%%%%%%%%%%%%%%%%%%%%%%%%%%%%%%%%%%%%%%%%%%%%%%%%%%%%%%%%%%%%%%%%%%%
%%%%%%%%%%%%%%%%%%%%%%%%%%%%%%%%%%%%%%%%%%%%%%%%%%%%%%%%%%%%%%%%%%%%%%%%%%%%%%%%%%%%%%%

\begin{figure}
  \centering
  \includegraphics[width=0.48\textwidth]{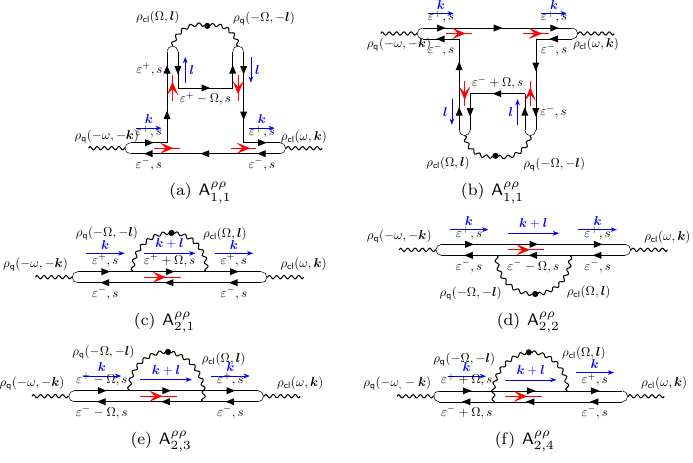}
  \caption{Diagrams $\mathsf{A}^{\rho\rho}$ contributing to the Altshuler-Aronov correction to the density response function 
  in the spin-singlet interaction channel.}
  \label{fig:rho-rho-A}
\end{figure}

\subsection{Density response \label{sec:densresptech}}

\begin{figure}
  \centering
  \includegraphics[width=0.42\textwidth]{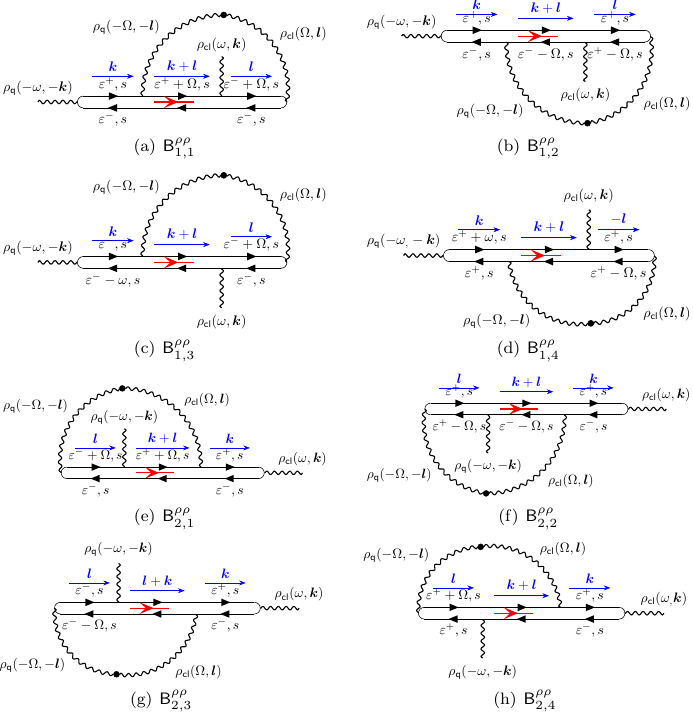}
  \caption{``Wave function renormalization'' diagrams $\mathsf{B}^{\rho\rho}$ contributing to the Altshuler-Aronov correction to the
  density response function in the spin-singlet interaction channel.}
  \label{fig:rho-rho-B}
\end{figure}

The retarded self-energy for the dynamical density field $\rho(t,x)$ due to singlet-channel interactions can be obtained as
\begin{equation}
  	-i\Sigma_{\rho}^{R}
  	=
  	-
  	\frac{h^{2}}{2}
  	\Delta_{0}^{2}\left(-\omega,k\right)
  	\intop_{\varepsilon}
  	\left(F_{\varepsilon^{-}}-F_{\varepsilon^{+}}\right)
  	\left[
  		\Sigma_{X}^{\mathsf{A}^{\rho\rho}}+\Sigma_{X}^{\mathsf{B}^{\rho\rho}}
	\right]\,,
  	\label{eq:SigmaR-rhorho}
\end{equation}
Here $\varepsilon_{\pm} = \varepsilon \pm \omega/2$ and the $\mathsf{A}^{\rho\rho}$ and $\mathsf{B}^{\rho\rho}$ diagrams are shown in Figs.~\ref{fig:rho-rho-A} and \ref{fig:rho-rho-B}, respectively. In the theory of the 2D dirty electron gas, the $\mathsf{A}$ diagrams in Fig.~\ref{fig:rho-rho-A} give the direct correction to the conductivity and the $\mathsf{B}$ diagrams in Fig.~\ref{fig:rho-rho-B} determine the wave function renormalization of the $Q$ field;
both are Altshuler-Aronov corrections in the unitary class. 
In the NLsM for the PRBM-Hubbard model, the kinetic term is nonlocal and has fractional dependence on the momentum. 
As a result, the $\mathsf{A}$ diagrams do not yield a term proportional to $|k|^\sigma$ and the correction to the superdiffusion constant 
comes only from the wave function renormalization in Fig.~\ref{fig:rho-rho-B}. 
The $\mathsf{A}$ diagrams give a term proportional to frequency that is necessary to satisfy the Ward identity for charge conservation.

We obtain
\begin{align}
	\Sigma_{X}^{\rho}
	=&\,
	\frac{i\pi h\gamma_{s} \left|k\right|^{\sigma}}{2\lambda}
\nonumber\\
	&\, 
	\times
	\intop_{\Omega,l}
	\Delta_{0}\left(-\Omega,l\right)
	\Delta_{u,s}\left(-\Omega,l\right)
	\left(F_{\varepsilon+\Omega}-F_{\varepsilon-\Omega}\right).
\label{eq:Sigma-X-rhorho}
\end{align}
Here $\Sigma_X^{\rho} = \Sigma_X^{A^{\rho\rho}} + \Sigma_X^{B^{\rho\rho}}$ and we discard the higher-order terms in $\omega$ 
and the term $F_\varepsilon^+ - F_\varepsilon^-$, which vanishes after integrating out $\Omega$ and $l$.
In this model, the Ward identity requires that the superdiffusive relation $\omega \sim c \left|k\right|^\sigma$ is unaltered by the interaction and disorder scattering. As a result $\Sigma_X$ must be proportional to $\left|k\right|^\sigma$, which is exactly what we have obtained in Eq.~\eqref{eq:Sigma-X-rhorho}. 
The correction to the density response function in the extended phase and at critical point can be obtained by using the integral in 
Eq.~\eqref{eq:I1} and the identity
$
  \intop_{\varepsilon} \left(F_{\varepsilon^-} - F_{\varepsilon^+}\right) = -\frac{\omega}{\pi} \,.
$
In the extended phase ($\sigma<1$), the correction to the density response function is
\begin{align}
  	\delta\Pi\left(\omega,k\right) 
  	=& 
  	\frac{i\left(1-\gamma_{s}\right)^{2}h^{2}\omega\left|k\right|^{\sigma}\Lambda_{k}^{1-\sigma}}{2\pi^{2}\left(1-\sigma\right)} 
  	\nonumber\\
  	&\times\Delta_{u,s}^{2}\left(-\omega,k\right)\ln\left(1-\gamma_{s}\right)\,.
\end{align}
The quantum correction to the superdiffusive ``conductivity'' 
[Eq.~(\ref{eq:super-cond})]
in the extended phase $\sigma<1$ is then given by,
\begin{align}
	\delta\mathfrak{g} & = \frac{2\Lambda_{k}^{1-\sigma}}{\pi^{2}\left(1-\sigma\right)}\ln\left(1-\gamma_{s}\right).
\end{align}
At the critical point ($\sigma=1$), 
\begin{equation}
  	\delta\mathfrak{g}=\frac{2}{\pi^2}\ln\left(1-\gamma_{s}\right)\ln\left(\frac{\Lambda}{T}\right)\,.
\end{equation}
Here $\Lambda$ is the UV cutoff in energy and it is related to the momentum cutoff via the superdiffusive relation $\Lambda = {D} \Lambda_k$. 

On the localized side $\sigma>1$, the integral for the Altshuler-Aronov correction has no UV or IR divergence and we have
\begin{widetext}
\begin{align}
 \delta\Pi\left(\omega,k\right) 
  =& -i\frac{\lambda h^{3}\left(2-\gamma_{s}\right)\gamma_{s}\left(1-\gamma_{s}\right)^{2}D^{2-\frac{1}{\sigma}}T^{\frac{1}{\sigma}}}{8\pi^{2}\sigma}\left|k\right|^{\sigma}\Delta_{u,s}^{2}\left(-\omega,k\right)\mathfrak{B}_s\left(\frac{\omega}{T}\right)\,.
\end{align}
Here $\mathfrak{B}_s\left(\frac{\omega}{T}\right)$ is defined as
\begin{equation}
\mathfrak{B}_s\left(\frac{\omega}{T}\right)=4\int_{0}^{\infty}dx\int_{0}^{\infty}dy\frac{x^{\frac{1}{\sigma}}y\left[4\frac{\omega}{T}\sinh y-4y\sinh\left(\frac{\omega}{T}\right)\right]}{\left(x^{2}+y^{2}\right)\left[x^{2}+\left(1-\gamma_{s}\right)^{2}y^{2}\right]\left[\cosh\left(y\right)-\cosh\left(\frac{\omega}{T}\right)\right]}\,.
\end{equation}
At the leading order in $\omega/T$, we have
\begin{equation}
\label{eq:B}
\mathfrak{B}_s\left(\frac{\omega}{T}\right)  =\mathfrak{B}_{0}\left( \gamma_s \right)\frac{\omega}{T} + \mathcal{O}\left(\left[\frac{\omega}{T}\right]^{3}\right)\,,
\end{equation}
with
\begin{equation}
  \mathfrak{B}_{0}\left( \gamma_s \right)=2\int_{0}^{\infty}dx\int_{0}^{\infty}dy\frac{x^{\frac{1}{\sigma}}y\left(\sinh y-y\right)}{\left(x^{2}+y^{2}\right)\left[x^{2}+\left(1-\gamma_{s}\right)^{2}y^{2}\right]\sinh^{2}\frac{y}{2}} \,.
  \label{eq:B0}
\end{equation}
Thus, the correction to the density response function is
\begin{equation}
  \delta\Pi\left(\omega,k\right)=-i\frac{\lambda h^{3}\left(2-\gamma_{s}\right)\gamma_{s}\left(1-\gamma_{s}\right)^{2}D^{2-\frac{1}{\sigma}}T^{\frac{1}{\sigma}-1}\omega}{8\pi^{2}\sigma}\left|k\right|^{\sigma}\Delta_{u,s}^{2}\left(-\omega,k\right)\mathfrak{B}_{0}\left( \gamma_s \right)\,.
\end{equation}
\end{widetext}
The quantum correction to the superdiffusive ``conductivity'' is given by
\begin{equation}
  \delta\mathfrak{g} =-\frac{\lambda h\mathfrak{a}_sD^{2-\frac{1}{\sigma}}T^{\frac{1}{\sigma}-1}}{2\pi^{2}\sigma}\mathfrak{B}_{0} \left( \gamma_s \right)\,.
\end{equation}
The quantum correction diverges as $T\to0$ with power-law $T^{\frac{1}{\sigma}-1}$
in the localized side ($\sigma>1$), and it decreases the superdiffusive ``conductivity'' as $\gamma_s>0$ and $\mathfrak{B}_0\left( \gamma_s \right)>0$. Therefore, the singlet channel interactions result in localizing corrections. This is similar to the situation of the 2D electron gas \cite{Finkelstein1983, Castellani1984, BK1994}.

For the interactions in the spin-triplet channel, the correction to the density response is given by diagrams similar to those in Figs.~\ref{fig:rho-rho-A} and \ref{fig:rho-rho-B}. 
The total correction to $\delta\mathfrak{g}$ in the ergodic phase ($\sigma < 1$) and at the noninteracting critical point ($\sigma = 1$) is given by 
Eqs.~(\ref{eq:deltasigma}) and (\ref{eq:dsigma-critical}), respectively.
In the localized phase with $\sigma>1$,
\begin{equation}
  \delta\mathfrak{g} %\left(\omega\right)
  =-\frac{3\lambda h\mathfrak{a}_tD^{2-\frac{1}{\sigma}}T^{\frac{1}{\sigma}-1}}{2\pi^{2}\sigma}\mathfrak{B}_{0}\left( \gamma_t \right)\,.
\end{equation}
The natural sign of the spin triplet interaction $\gamma_t$ is negative for repulsive interactions \cite{BK1994,Coleman2015}. 
Thus, the quantum effect in the spin-triplet channel results in an anti-localizing correction to transport.

%%%%%%%%%%%%%%%%%%%%%%%%%%%%%%%%%%%%%%%%%%%%%%%%%%%%%%%%%%%%%%%%%%%%%%%%%%%%%%%%%%%%%%%
%%%%%%%%%%%%%%%%%%%%%%%%%%%%%%%%%%%%%%%%%%%%%%%%%%%%%%%%%%%%%%%%%%%%%%%%%%%%%%%%%%%%%%%
%%%%%%%%%%%%%%%%%%%%%%%%%%%%%%%%%%%%%%%%%%%%%%%%%%%%%%%%%%%%%%%%%%%%%%%%%%%%%%%%%%%%%%%
%%%%%%%%%%%%%%%%%%%%%%%%%%%%%%%%%%%%%%%%%%%%%%%%%%%%%%%%%%%%%%%%%%%%%%%%%%%%%%%%%%%%%%%
%%%%%%%%%%%%%%%%%%%%%%%%%%%%%%%%%%%%%%%%%%%%%%%%%%%%%%%%%%%%%%%%%%%%%%%%%%%%%%%%%%%%%%%
%%%%%%%%%%%%%%%%%%%%%%%%%%%%%%%%%%%%%%%%%%%%%%%%%%%%%%%%%%%%%%%%%%%%%%%%%%%%%%%%%%%%%%%

\begin{figure*}[t!]
  \centering
  \includegraphics[width=0.85\textwidth]{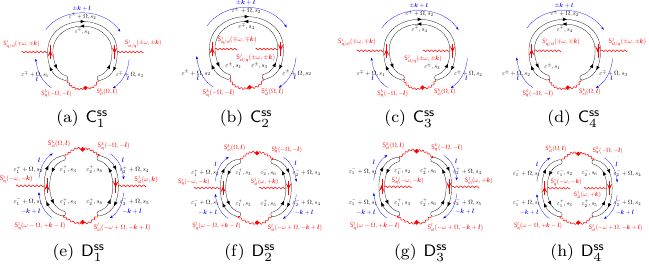}
  \caption{Diagrams $\mathsf{C}$ and $\mathsf{D}$ contributing to the static spin response function.}
  \label{fig:S-S-CD}
\end{figure*}

\subsection{Spin susceptibility} \label{sec:spintech}

The spin response function $\chi_{ij}\left( \omega,k \right)$ encodes both spin transport and the static spin susceptibility. 
Spin and charge transport are renormalized the same way in the absence of magnetic order. 
Different from the charge compressibility that receives no quantum corrections \cite{BK1994}, quantum interference
effects can enhance or suppress the static spin susceptibility. 
The latter is determined by the $\omega\to0$ limit of $\chi_{ij}\left( \omega,k \right)$.
There are two groups of diagrams $\mathsf{C},\mathsf{D}$ contributing to the static magnetic response function 
shown in Fig.~\ref{fig:S-S-CD}.

Up to irrelevant corrections, the self-energy for the dynamical spin field  $\boldsymbol{S}(t,x)$  is 
\begin{align}
	-i\Sigma_{\mathsf{sp}}^{R,\mathsf{C}}
	=&\,
\nonumber\\
	2i\pi
	&\, 
	 h^{3}\gamma_{t}\delta_{ij}\intop_{\Omega l}\Delta_{0}^{2}\left(-\Omega,l\right)\Delta_{u,t}\left(-\Omega,l\right)\left(B_{\Omega}-B_{0}\right)\,.
\label{eq:Sigma-S-S-C1}
\end{align}
Here we have used the identity
$
B_{\omega} \equiv\intop_{\varepsilon}\left(1-F_{\varepsilon}F_{\varepsilon+\omega}\right)
  =\frac{\omega}{\pi}\coth\left(\frac{\omega}{2T}\right)\,.
$
Discarding higher-order terms in $\omega$ and $k$, we have
\begin{widetext}
\begin{align}
-i\Sigma_{\mathsf{sp}}^{R,\mathsf{D}} = & -2h^{6}\delta_{ij}\intop_{\varepsilon_{1}\varepsilon_{2}\Omega l}\Delta_{0}^{4}\left(-\Omega,l\right)\left[i\Delta_{\mathsf{sp}}^{R}\left(\Omega,l\right)\right]^{2}
\left\{ \begin{array}{r}
2\left(F_{\varepsilon_{1}^{+}+\Omega}-F_{\varepsilon_{1}^{+}}\right)\left(F_{\varepsilon_{2}^{-}+\Omega}-F_{\varepsilon_{2}^{+}}\right)F_{\varepsilon_{1}^{-}+\Omega}\\
-2\left(F_{\varepsilon_{1}^{-}+\Omega}-F_{\varepsilon_{1}^{-}}\right)\left(F_{\varepsilon_{2}^{-}+\Omega}-F_{\varepsilon_{2}^{+}}\right)F_{\varepsilon_{1}^{+}}
\end{array}\right\} \,.
\label{eq:Sigma-S-S-D}
\end{align}
Integrating out $\varepsilon_{1}$ and $\varepsilon_{2}$ gives
\begin{equation}
-i\Sigma_{\mathsf{s}}^{R,\mathsf{D}}=\frac{\pi^{2}h^{4}\gamma_{t}^{2}\delta_{ij}}{2}\intop_{\Omega l}\Delta_{0}^{2}\left(-\Omega,l\right)\Delta_{u,t}^{2}\left(-\Omega,l\right)\frac{\Omega}{\pi}\left(B_{\Omega}-B_{0}\right)+\mathcal{O}\left(\omega\right)\,.
\label{eq:Sigma-S-S-D1}
\end{equation}

In the ergodic phase ($\sigma<1$), the integrals in Eq.~\eqref{eq:Sigma-S-S-C1} and \eqref{eq:Sigma-S-S-D1} are UV divergent and IR convergent. We can first integrate out $\Omega$ from $-\infty$ to $+\infty$ and obtain,
\begin{align}
-i\Sigma_{\mathsf{sp}}^{R,\mathsf{C}} & =\frac{16i}{\pi}\delta_{ij}\left[1+\frac{1}{\gamma_{t}}\ln\left(1-\gamma_{t}\right)\right]\intop_{l}\frac{1}{{D}\left|l\right|^{\sigma}}\,,\\
-i\Sigma_{\mathsf{sp}}^{R,\mathsf{D}} & =8\gamma_{t}^{2}\delta_{ij}\intop_{l}\frac{1}{\pi\gamma_{t}^{2}i{D}\left|l\right|^{\sigma}}\left[1+\frac{1}{1-\gamma_{t}}+\frac{2}{\gamma_{t}}\ln\left(1-\gamma_{t}\right)\right]\,.
\end{align}
Here we use the integrals
\bsub
\begin{align}
\intop_{\Omega}\frac{B_{\Omega}-B_{0}}{\left[i{D}\left|l\right|^{\sigma}+\Omega\right]^{2}\left[i{D}\left|l\right|^{\sigma}+\Omega\left(1-\gamma_{t}\right)\right]} & \approx-\frac{1}{\pi^{2}\gamma_{t}}\left[1+\frac{1}{\gamma_{t}}\ln\left(1-\gamma_{t}\right)\right]\frac{i}{{D}\left|l\right|^{\sigma}}\,,\\
\intop_{\Omega}\frac{\Omega\left(B_{\Omega}-B_{0}\right)}{\left[i{D}\left|l\right|^{\sigma}+\Omega\right]^{2}\left[i{D}\left|l\right|^{\sigma}+\Omega\left(1-\gamma_{t}\right)\right]^{2}} & \approx\frac{1}{\pi^{2}\gamma_{t}^{2}}\left[1+\frac{1}{1-\gamma_{t}}+\frac{2}{\gamma_{t}}\ln\left(1-\gamma_{t}\right)\right]\frac{1}{i{D}\left|l\right|^{\sigma}}\,.
\end{align}
\esub
\end{widetext}
Thus we have 
\begin{equation}
-i\Sigma_{\mathsf{sp}}^{R,\mathsf{CD}}=-i\Sigma_{\mathsf{sp}}^{R,\mathsf{C}}-i\Sigma_{\mathsf{sp}}^{R,\mathsf{D}}=-\frac{i4\lambda h\delta_{ij}}{\pi}\frac{\gamma_{t}}{1-\gamma_{t}}\intop_{l}\frac{1}{\left|l\right|^{\sigma}}\,.
 \label{eq:SigmaCD}
\end{equation}
Integrating out $l$ yields
\begin{equation}
-i\Sigma_{\sigma}^{R,\mathsf{CD}}=-\frac{i4\lambda h\delta_{ij}}{\pi^{2}\left(1-\sigma\right)}\frac{\gamma_{t}}{1-\gamma_{t}}\Lambda_{k}^{1-\sigma}\,.
\label{eq:SigmaCD1}
\end{equation}
The corresponding correction to the spin density response function is
\begin{align}
	\delta\chi_{ij}^{\mathsf{CD}}&\left(\omega,k\right) 
\nonumber\\
 	=&
 	-
 	\frac{2}{U_{t}^{2}}
 	\Delta_{\mathsf{\sigma}}^{R,ii}\left(\omega,k\right)
 	\Sigma_{\sigma}^{R,ij}\left(\omega,k\right)
 	\Delta_{\sigma}^{R,jj}\left(\omega,k\right)
\nonumber\\
  	=&
  	-
  	\frac{2\lambda h\delta_{ij}
  	\left(1-\gamma_{t}\right)}{\pi^{2}\left(1-\sigma\right)}
  	\frac{
  		\left(\frac{\left|k\right|^{\sigma}}{\lambda}-\frac{ih\omega}{2}\right)^{2}
	}{
		\left[\frac{\left|k\right|^{\sigma}}{\lambda}-\frac{ih\omega}{2}\left(1-\gamma_{t}\right)\right]^{2}
	}
	\gamma_{t}
	\Lambda_{k}^{1-\sigma}\,.
\end{align}
Here $U_{t}=\frac{\pi}{2h}\frac{\gamma_{t}}{1-\gamma_{t}}$. In the static 
limit, $\omega\to0$, we have 
\begin{equation}
\delta\chi_{ij}\left(\omega\to0\right)=\frac{2\lambda h\delta_{ij}\left(1-\gamma_{t}\right)}{\pi^{2}\left(1-\sigma\right)}\gamma_{t}\Lambda_{k}^{1-\sigma}\,.
\label{eq:dchi}
\end{equation}
Recall that the bare (semiclassical) spin response function is given by Eq.~(\ref{chiSemiClassical})
with $\chi_{0}=\frac{2h}{\pi}\left(1-\gamma_{t}\right)$. 
Then we obtain Eq.~(\ref{eq:chi-ij-0}). 
At the noninteracting critical point there is a logarithmic divergence that can be cut in the infrared by the temperature $T$.
This leads to Eq.~(\ref{eq:chi-MIT}).
The latter is identical to the correction for the dirty 2D electron gas except for the numerical
prefactor \cite{Finkelstein1983}.

\begin{figure}[b!]
  \centering
  \includegraphics[width=0.4\textwidth]{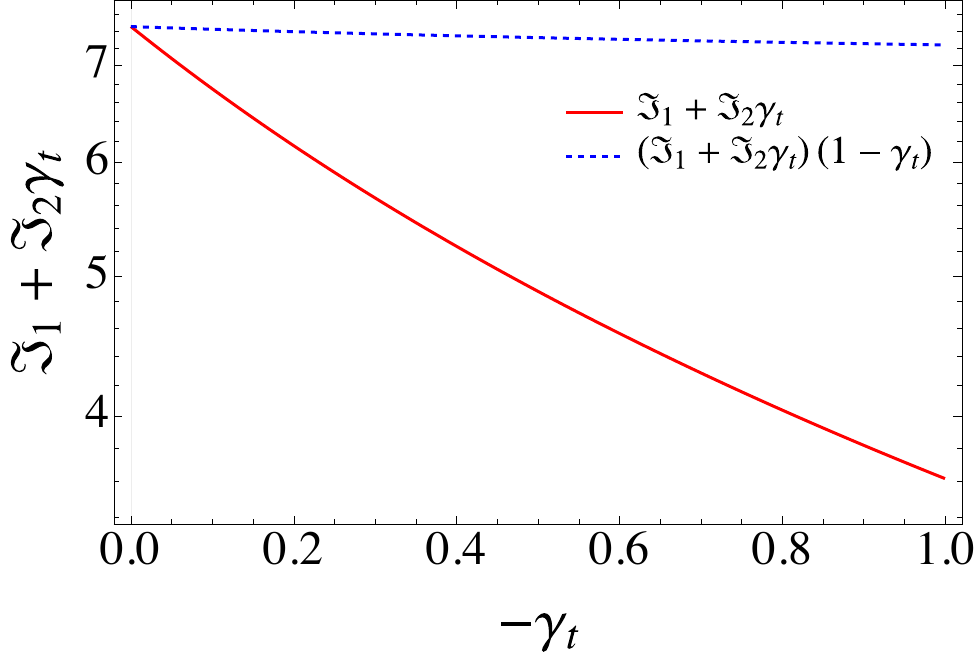}
  \caption{The dependence of the coefficients for the quantum correction to the magnetic susceptibility in the localized phase
  as a function of the spin-triplet interaction strength $\gamma_t$, see Eq.~(\ref{chiLoc}).}
  \label{fig:I1I2}
\end{figure}

Now we turn to the
localized phase with $\sigma>1$. In this case, the integrals are
convergent in both the UV and IR limits, and we can discard the UV cutoffs
$\Lambda$ and $\Lambda_{k}$. The $\mathsf{C}$ diagrams are given by
\begin{align}
  -i\Sigma_{\mathsf{sp}}^{R,\mathsf{C}} & =2i\pi h^{3}\gamma_{t}\delta_{ij}I_{2}\,,
\end{align}
with
\begin{widetext}
\begin{align}
I_{2} & =\frac{1}{4\pi^{2}}\left(\frac{2}{h}\right)^{3}\int_{-\infty}^{\infty}d\Omega\int_{-\infty}^{\infty}dl\frac{1}{\left(D\left|l\right|^{\sigma}-i\Omega\right)^{2}\left[D\left|l\right|^{\sigma}-i\Omega\left(1-\gamma_{t}\right)\right]}\left(B_{\Omega}-B_{0}\right)\nonumber \\
 & =T^{\frac{1}{\sigma}-1}\frac{1}{\sigma}\left(\frac{1}{D}\right)^{\frac{1}{\sigma}}\frac{1}{2\pi^{3}}\left(\frac{2}{h}\right)^{3}\int_{-\infty}^{\infty}dy\int_{0}^{\infty}dx\frac{x^{\frac{1}{\sigma}-1}\left(y\coth\frac{y}{2}-2\right)}{\left(x-iy\right)^{2}\left[x-iy\left(1-\gamma_{t}\right)\right]}\,.
 \end{align}
The $\mathsf{D}$ diagrams are given by
 \begin{align}
-i\Sigma_{\mathsf{sp}}^{R,\mathsf{D}} & =\frac{\pi^{2}h^{4}\gamma_{t}^{2}\delta_{ij}}{2}I_{3}\,,
\end{align}
with
\begin{align}
I_{3} & =\frac{1}{4\pi^{3}}\left(\frac{2}{h}\right)^{4}\int_{-\infty}^{\infty}dl\int_{-\infty}^{\infty}d\Omega\frac{\Omega\left(B_{\Omega}-B_{0}\right)}{\left(D\left|l\right|^{\sigma}-i\Omega\right)^{2}\left[D\left|l\right|^{\sigma}-i\Omega\left(1-\gamma_{t}\right)\right]^{2}}\nonumber \\
 & =T^{\frac{1}{\sigma}-1}\frac{1}{\sigma}\left(\frac{1}{D}\right)^{\frac{1}{\sigma}}\frac{1}{2\pi^{4}}\left(\frac{2}{h}\right)^{4}\int_{0}^{\infty}dx\int_{-\infty}^{\infty}dy\frac{x^{\frac{1}{\sigma}-1}y\left(y\coth\frac{y}{2}-2\right)}{\left(x-iy\right)^{2}\left[x-iy\left(1-\gamma_{t}\right)\right]^{2}}\,.
\end{align}
Thus
 \begin{equation}
   -i\Sigma_{\mathsf{sp}}^{R,\mathsf{CD}}  =-\frac{8i\delta_{ij}}{\pi^{2}}T^{\frac{1}{\sigma}-1}\frac{1}{\sigma}\left(\frac{1}{D}\right)^{\frac{1}{\sigma}}\left[\mathfrak{I}_{1}\gamma_{t}+\mathfrak{I}_{2}\gamma_{t}^{2}\right]\,.
\end{equation}
Here
\begin{subequations}
  \label{eq:I1I2}
\begin{align}
\mathfrak{I}_{1} & =-\int_{0}^{\infty}dx\int_{-\infty}^{\infty}dy\frac{x^{\frac{1}{\sigma}}\left[x^{2}-\left(3-2\gamma_{t}\right)y^{2}\right]\left(y\coth\frac{y}{2}-2\right)}{\left(x^{2}+y^{2}\right)^{2}\left[x^{2}+y^{2}\left(1-\gamma_{t}\right)^{2}\right]}\,,\\
\mathfrak{I}_{2} & =-\int_{0}^{\infty}dx\int_{-\infty}^{\infty}dy\frac{x^{\frac{1}{\sigma}}y^{2}\left(2-\gamma_{t}\right)\left[x^{2}-y^{2}\left(1-\gamma_{t}\right)\right]\left(y\coth\frac{y}{2}-2\right)}{\left(x^{2}+y^{2}\right)^{2}\left[x^{2}+y^{2}\left(1-\gamma_{t}\right)^{2}\right]^{2}}\,.
\end{align}
\end{subequations}
The correction to the spin response function is obtained as,
\begin{equation}
\delta\chi_{ij}^{\mathsf{CD}}\left(\omega,k\right)=-\frac{2\delta_{ij}T^{\frac{1}{\sigma}-1}\left(1-\gamma_{t}\right)^{2}}{\pi^{2}\sigma D^{\frac{1}{\sigma}}}\left(\frac{\frac{\left|k\right|^{\sigma}}{\lambda}-\frac{ih\omega}{2}}{\frac{\left|k\right|^{\sigma}}{\lambda}-\frac{ih\omega}{2}\left(1-\gamma_{t}\right)}\right)^{2}\left(\mathfrak{I}_{1}\gamma_{t}+\mathfrak{I}_{2}\gamma_{t}^{2}\right)\,.
\end{equation}
\end{widetext}
In the static limit ($\omega\to0)$, we have
\begin{equation}
\delta\chi_{ij}=-\frac{2\delta_{ij}T^{\frac{1}{\sigma}-1}\left(1-\gamma_{t}\right)^{2}}{\pi^{2}\sigma D^{\frac{1}{\sigma}}}\left(\mathfrak{I}_{1}\gamma_{t}+\mathfrak{I}_{2}\gamma_{t}^{2}\right)\,.\label{eq:dchi-1}
\end{equation}
Thus, we obtain the static spin susceptibility in terms of the non-interacting one,
\begin{equation}
	\label{chiLoc}
	\chi_{ij}
	=
	\delta_{ij}
	\chi_{0}
	\left[
			1
			-
			\frac{2T^{\frac{1}{\sigma}-1}\left(1-\gamma_{t}\right)}{\pi\sigma D^{\frac{1}{\sigma}}h}
			\gamma_{t}
			\left(\mathfrak{I}_{1}+\mathfrak{I}_{2}\gamma_{t}\right)
		\right]\,.
\end{equation}
Fig.~\ref{fig:I1I2} shows $\mathfrak{I}_1+\mathfrak{I}_2 \, \gamma_t$ as a function of the spin-triplet interaction strength $\gamma_t$ (red line). 
It is always positive for attractive $\gamma_t < 0$, and the static spin susceptibility in the localized phase is therefore enhanced compared to its bare value. The blue dashed line in Fig.~\ref{fig:I1I2} shows the coefficient $\left(\mathfrak{I}_1+\mathfrak{I}_2\gamma_t\right)(1-\gamma_t)$, which is almost a constant over the range plotted. At low temperature, the correction to the spin susceptibility diverges in a power-law fashion $\sim T^{1/\sigma-1}$.

%%%%%%%%%%%%%%%%%%%%%%%%%%%%%%%%%%%%%%%%%%%%%%%%%%%%%%%%%%%%%%%%%%%%%%%%%%%%%%%%%%%%%%%
%%%%%%%%%%%%%%%%%%%%%%%%%%%%%%%%%%%%%%%%%%%%%%%%%%%%%%%%%%%%%%%%%%%%%%%%%%%%%%%%%%%%%%%
%%%%%%%%%%%%%%%%%%%%%%%%%%%%%%%%%%%%%%%%%%%%%%%%%%%%%%%%%%%%%%%%%%%%%%%%%%%%%%%%%%%%%%%
%%%%%%%%%%%%%%%%%%%%%%%%%%%%%%%%%%%%%%%%%%%%%%%%%%%%%%%%%%%%%%%%%%%%%%%%%%%%%%%%%%%%%%%
%%%%%%%%%%%%%%%%%%%%%%%%%%%%%%%%%%%%%%%%%%%%%%%%%%%%%%%%%%%%%%%%%%%%%%%%%%%%%%%%%%%%%%%
%%%%%%%%%%%%%%%%%%%%%%%%%%%%%%%%%%%%%%%%%%%%%%%%%%%%%%%%%%%%%%%%%%%%%%%%%%%%%%%%%%%%%%%

\subsection{RG-improved DOS}

The interaction correction to the DOS obtained above in Sec.~\ref{sec:DosTech}
treats the coupling $\lambda$, interactions $\gamma_{s/t}$ and the bare density of states as constant parameters.
To more accurately account for the interaction corrections, we should turn to the RG equations, 
where these quantities are running parameters.
The RG equations for the coupling $\lambda$, the DOS $\nu$ and the spin susceptibility can be extracted from 
Eqs.~\eqref{eq:dsigma-critical}, ~\eqref{eq:dnu-MIT} and ~\eqref{eq:chi-MIT}, respectively. 
At the critical point, the RG flow equations are
\bsub
\begin{align}
  \frac{d \lambda}{d l} &= -\frac{\lambda^{2}}{2\pi}\left[\ln\left(1-\gamma_{s}\right)+3\ln\left(1-\gamma_{t}\right)\right] \,,\\
  \frac{d\ln\nu}{dl}&=\frac{\lambda}{4\pi}\left[\ln\left(1-\gamma_{s}\right)+3\ln\left(1-\gamma_{t}\right)\right]\,,\\
  \frac{d\chi_{t}}{dl}&=-\frac{\chi_{t}\lambda \gamma_{t}}{\pi}\,.
  \label{eq:RG}
\end{align}
\esub
The RG equation for $h$ takes the similar form as that in 2D unitary class with SU(2) symmetry,
\begin{equation}
  \frac{d\ln h}{dl}=-\frac{\lambda}{4\pi}\left(\gamma_{s}+3\gamma_{t}\right)\,.
  \label{eq:RG-h}
\end{equation}
Thus, we obtain the RG equations for the interactions and the energy (dynamical critical scaling)
\bsub 
\label{eq:RG-gt-E}
\begin{align}
 	\frac{d\gamma_{s}}{dl}&=-\frac{\lambda}{4\pi}\left(1-\gamma_{s}\right)\left(\gamma_s + 3 \gamma_t\right), 
	\\
 	\frac{d\gamma_{t}}{dl}&=\frac{\lambda}{4\pi}\left(1-\gamma_{t}\right)\left(\gamma_t - \gamma_s\right), 
	\label{eq:gammatRG}
	\\
  	\frac{d\ln E}{dl}&= -d-\frac{d\ln h}{dl}=-1+\frac{\lambda}{4\pi}\left(\gamma_{s}+3\gamma_{t}\right)\,.
\end{align}
\esub
The RG coupled equations in general cannot be solved exactly. 
However, the flow of the parameter $\lambda$ is logarithmically slow in the perturbative regime, and can therefore be initially neglected. 
Linearizing the flow equations for the interactions to linear order in $\gamma_{s,t}$ reveals a single relevant direction,
wherein the triplet interaction $\gamma_t \rightarrow - \infty$, indicating the onset of strong magnetic fluctuations near the MIT. 
We therefore specialize to the following simplified flow equations governing the intermediate-scale flow of the DOS, 
\bsub
\begin{align}
  \frac{d \ln{\nu}}{d l} &=   3\lambda\ln{\left( 1-\gamma_t \right)}\,, \\
  \frac{d\gamma_t}{d l} &= \lambda \gamma_t\,, \\
  \frac{d\ln E}{dl}&=-1\,.
  \label{eq:RG-nu-gamma-t}
\end{align}
\esub
Here we rescale the coupling $\lambda\to 4\pi \lambda$. 
We also ignore the spin-singlet interaction $\gamma_s$ and focus on the spin-triplet interaction $\gamma_t$, which dominates near the MIT. 
Solving the simplified RG equations, we obtain
\begin{align}
 	 \gamma_t(l) 
 	 &= 
 	 \gamma_t(0) \, e^{\lambda l} \,,
 \\
  	\ln\left[{\frac{\nu(l)}{\nu(0)}}\right] 
  	&= 
  	-
  	3 \left[ 
  		\mathsf{Li}_2(\gamma_t^0 e^{\lambda l}) - \mathsf{Li}_2(\gamma_t^0) 
	\right]\,,
\\
	\ln\left[{\frac{E(l)}{E(0)}}\right]
	&= 
	-l\,.
\end{align}
Here $\mathsf{Li}_2$ is the dilogarithm function and $\gamma_t^0 = \gamma_t(l=0)$. Identifying $\nu(l) = \nu(E)$ and $E(l)=E$, we obtain Eq.~(\ref{eq:nu-RG}).
The RG-improved expression for the DOS diverges at zero energy, which is qualitatively consistent with the 
numerical results shown in Fig.~\ref{fig:DOS}.

%%%%%%%%%%%%%%%%%%%%%%%%%%%%%%%%%%%%%%%%%%%%%%%%%%%%%%%%%%%%%%%%%%%%%%%%%%%%%%%%%%%%%%%
%%%%%%%%%%%%%%%%%%%%%%%%%%%%%%%%%%%%%%%%%%%%%%%%%%%%%%%%%%%%%%%%%%%%%%%%%%%%%%%%%%%%%%%
%%%%%%%%%%%%%%%%%%%%%%%%%%%%%%%%%%%%%%%%%%%%%%%%%%%%%%%%%%%%%%%%%%%%%%%%%%%%%%%%%%%%%%%
%%%%%%%%%%%%%%%%%%%%%%%%%%%%%%%%%%%%%%%%%%%%%%%%%%%%%%%%%%%%%%%%%%%%%%%%%%%%%%%%%%%%%%%
%%%%%%%%%%%%%%%%%%%%%%%%%%%%%%%%%%%%%%%%%%%%%%%%%%%%%%%%%%%%%%%%%%%%%%%%%%%%%%%%%%%%%%%
%%%%%%%%%%%%%%%%%%%%%%%%%%%%%%%%%%%%%%%%%%%%%%%%%%%%%%%%%%%%%%%%%%%%%%%%%%%%%%%%%%%%%%%
%%%%%%%%%%%%%%%%%%%%%%%%%%%%%%%%%%%%%%%%%%%%%%%%%%%%%%%%%%%%%%%%%%%%%%%%%%%%%%%%%%%%%%%
%%%%%%%%%%%%%%%%%%%%%%%%%%%%%%%%%%%%%%%%%%%%%%%%%%%%%%%%%%%%%%%%%%%%%%%%%%%%%%%%%%%%%%%
%%%%%%%%%%%%%%%%%%%%%%%%%%%%%%%%%%%%%%%%%%%%%%%%%%%%%%%%%%%%%%%%%%%%%%%%%%%%%%%%%%%%%%%
%%%%%%%%%%%%%%%%%%%%%%%%%%%%%%%%%%%%%%%%%%%%%%%%%%%%%%%%%%%%%%%%%%%%%%%%%%%%%%%%%%%%%%%
%%%%%%%%%%%%%%%%%%%%%%%%%%%%%%%%%%%%%%%%%%%%%%%%%%%%%%%%%%%%%%%%%%%%%%%%%%%%%%%%%%%%%%%

\section{Conclusion}\label{sec:Conclusion}

%We have already summarized our main results and conclusions in the Introduction and in Secs.~\ref{sec:Keldysh-summary} and \ref{sec:MF-summary}. 
We have studied the 1D PRBM-Hubbard model with complementary sigma model analytics and self-consistent Hartree-Fock numerics. 
  In the NLsM, the local Hubbard interaction can be decoupled into two effective channels: spin-singlet and spin-triplet ones. 
  The interaction in the spin-singlet channel is shown to suppress the DOS at the Fermi energy, while the one in the spin-triplet channel enhances the DOS. 
  The Hartree-Fock numerics shows that the spin-triplet interaction dominates near the MIT and the DOS is enhanced by the interaction-mediated quantum interference. 
  The dominance of spin-triplet interaction is supplemented with the spin response function and a magnetic instability is predicted via the NLsM calculation. 
  We clarify the nature of the predicted magnetic fluctuation and show that a spin-glass transition appears to occur alongside the interacting Anderson MIT. 
%  The density response calculation predicts that the spin-triplet interaction gives delocalization correction, which is consistent withe numerics. 
%  Our study clarifies the long-standing puzzle of the magnetic instability predicted by the NLsM and identiies the spin-glass local-moment phase in this case study.
  The insights obtained by analyzing our 1D model are potentially general and may shed light on the MIT and associated anomalous magnetic fluctuation phenomena in higher dimensions
   \cite{BK1994,Shashkin2001, Shashkin2006, Ani2006,Pudalov2012,Pudalov2018,Hossain2020}.

We close with open questions. 
We have obtained good qualitative agreement between analytics and numerics in the metallic phase, indicative of a magnetic instability that occurs near the MIT. 
The identification of spin-glass order in the insulator was obtained numerically. Does this result survive the incorporation of correlation effects for the PRBM-Hubbard model? 
A natural alternative would be spin-liquid behavior, which arises in quantum spin-glass models such as the Sachdev-Ye and Sachdev-Ye-Kitaev models \cite{SYK2022}.
One way to go beyond Hartree-Fock would be to diagonalize exactly the full Hamiltonian in a restricted basis of HF states \cite{Vojta1998}.

Another question concerns the nature of the spin glass. What are the statistics of the self-consistently determined magnetic exchange constants, and does
the resulting effective spin model exhibit continuous replica symmetry breaking? It would also be interesting to consider magnonic fluctuations on top 
of the frozen background, incorporating magnon-magnon scattering in the insulator. 

The results employed here suggest that interaction-mediated interference effects and local-moment formation and frustration are two sides of the same coin that merge 
continuously at the MIT. Is it possible to capture the onset of spin-glass order within the interacting sigma model formalism employed here, by allowing for replica-symmetry-broken
saddle-points? Replica symmetry breaking has appeared before in sigma model calculations, but only as a formal device to recover random matrix correlations in the zero-dimensional
($\alpha = 0$), non-interacting version of the PRBM model studied here \cite{Kamenev1999}.

Finally, we note that in real materials, Coulomb interactions play a crucial role \cite{DS2005,DS2013}.
The potential interplay between spin and charge glassiness \cite{Epperlein1997}
in the insulating phase is a key problem for future work.

\begin{acknowledgments}

We thank Sarang Gopalakrishnan, Vadim Oganesyan, and Peter Wolynes for useful discussions. 
This work was supported by the Welch Foundation Grant No. C-1809.
This work was also supported in part by the Big-Data Private-Cloud Research Cyberinfrastructure MRI-award funded by NSF under grant CNS-1338099 and by Rice University's Center for Research Computing (CRC).

\end{acknowledgments}

%%%%%%%%%%%%%%%%%%%%%%%%%%%%%%%%%%%%%%%%%%%%%%%%%%%%%%%%%%%%%%%%%%%%%%%%%%%%%%%%%%%%%%%
%%%%%%%%%%%%%%%%%%%%%%%%%%%%%%%%%%%%%%%%%%%%%%%%%%%%%%%%%%%%%%%%%%%%%%%%%%%%%%%%%%%%%%%
%%%%%%%%%%%%%%%%%%%%%%%%%%%%%%%%%%%%%%%%%%%%%%%%%%%%%%%%%%%%%%%%%%%%%%%%%%%%%%%%%%%%%%%
%%%%%%%%%%%%%%%%%%%%%%%%%%%%%%%%%%%%%%%%%%%%%%%%%%%%%%%%%%%%%%%%%%%%%%%%%%%%%%%%%%%%%%%
%%%%%%%%%%%%%%%%%%%%%%%%%%%%%%%%%%%%%%%%%%%%%%%%%%%%%%%%%%%%%%%%%%%%%%%%%%%%%%%%%%%%%%%
%%%%%%%%%%%%%%%%%%%%%%%%%%%%%%%%%%%%%%%%%%%%%%%%%%%%%%%%%%%%%%%%%%%%%%%%%%%%%%%%%%%%%%%
%%%%%%%%%%%%%%%%%%%%%%%%%%%%%%%%%%%%%%%%%%%%%%%%%%%%%%%%%%%%%%%%%%%%%%%%%%%%%%%%%%%%%%%
%%%%%%%%%%%%%%%%%%%%%%%%%%%%%%%%%%%%%%%%%%%%%%%%%%%%%%%%%%%%%%%%%%%%%%%%%%%%%%%%%%%%%%%
%%%%%%%%%%%%%%%%%%%%%%%%%%%%%%%%%%%%%%%%%%%%%%%%%%%%%%%%%%%%%%%%%%%%%%%%%%%%%%%%%%%%%%%
%%%%%%%%%%%%%%%%%%%%%%%%%%%%%%%%%%%%%%%%%%%%%%%%%%%%%%%%%%%%%%%%%%%%%%%%%%%%%%%%%%%%%%%
%%%%%%%%%%%%%%%%%%%%%%%%%%%%%%%%%%%%%%%%%%%%%%%%%%%%%%%%%%%%%%%%%%%%%%%%%%%%%%%%%%%%%%%

\appendix

$\phantom{0}$\\

\section{Spin configurations in the self-consistent numerics \label{app:spinprofiles}}

\begin{figure*}[t!]
  \centering
  \includegraphics[width=0.7\textwidth]{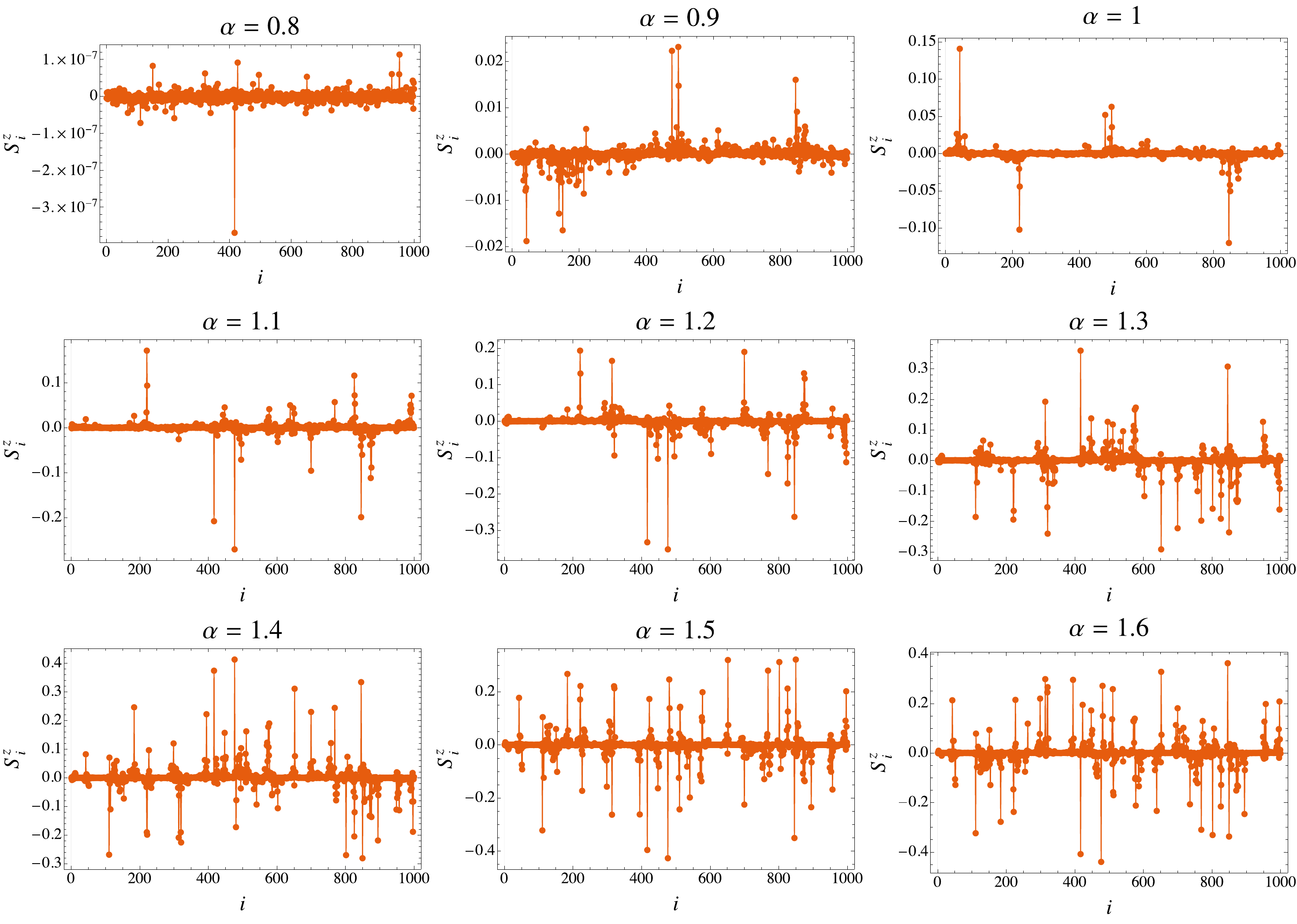}
  \caption{Mean-field converged spin configuration of the PRBM-Hubbard model with $L=1000$ and $U=1$ obtained in self-consistent numerics with different $\alpha$.
  In the extended phase, the local spin density is $0$ and the ground state is spin unpolarized.
  Near the MIT ($\alpha=1$), local magnetic moments start to form and peaks arise in the local spin densities. 
  The density and the strength of the local moments both increases with $\alpha$. 
  The local spin densities can be either positive or negative and the overall ferromagnetic order is absent.
}
  \label{fig:Szi-alpha}
\end{figure*}

\begin{figure*}[h!]
  \centering
  \includegraphics[width=0.7\textwidth]{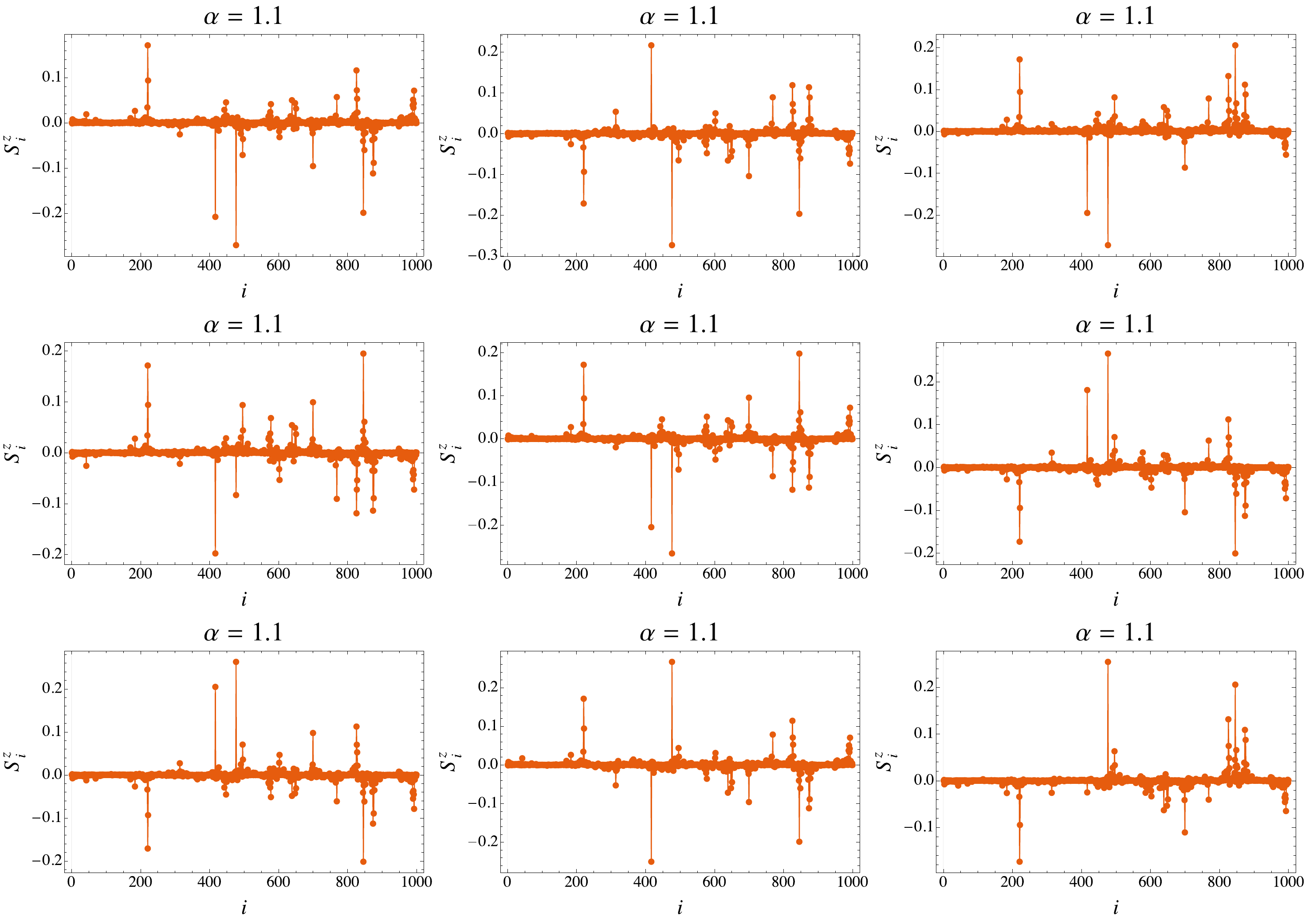}
  \caption{Mean-field converged spin configuration of the PRBM-Hubbard model with $L=1000$ and $U=1$ obtained in self-consistent numerics for $\alpha=1.1$ with different initial conditions. 
  The positions of the local magnetic moments are almost identical for all the ground states, while the sign of their spin polarization depends on the initial conditions.
  The energies of all these states are nearly the same, with relative differences at the order of $10^{-5}$.
}
	\label{fig:Szi-alpha=1.1}
\end{figure*}

Fig.~\ref{fig:Szi-alpha} shows converged spin configurations of the PRBM-Hubbard model obtained by self-consistent Hartree-Fock numerics. 
In the extended phase ($\alpha\lesssim 1$), the local spin density is almost $0$ and there are no magnetic moments.
Near the MIT ($\alpha\approx 1$), dilute magnetic moments start to form, and the density of local moments increases with $\alpha$ in the localized phase ($\alpha\gtrsim 1$). 
The spin polarization of the local moments is not ferromagnetic correlated, and the net magnetization is much smaller than the density of local moments, indicating the absence of ferromagnetic ordering.
Fig.~\ref{fig:Szi-alpha=1.1} shows the local density of spins for $\alpha=1.1$ with different initial conditions for the self-consistent calculations. 
The \emph{positions} of the local moments are almost independent of the initial conditions. 
However, the sign of the local spin density is extremely sensitive to the initial conditions. The converged solutions of the PRBM-Hubbard model in the interacting, localized phase consist of a large number of 
nearly-degenerate states and we obtain one of them in each run of the self-consistent numerics. Each state obtained with different initial conditions can be identified as one ``replica,'' 
and the initial condition-dependence of the ground states implies the replica symmetry breaking in the system, as discussed in Sec.~\ref{sec:SGO}.

\begin{figure*}[h!]
  \centering
  \includegraphics[width=0.7\textwidth]{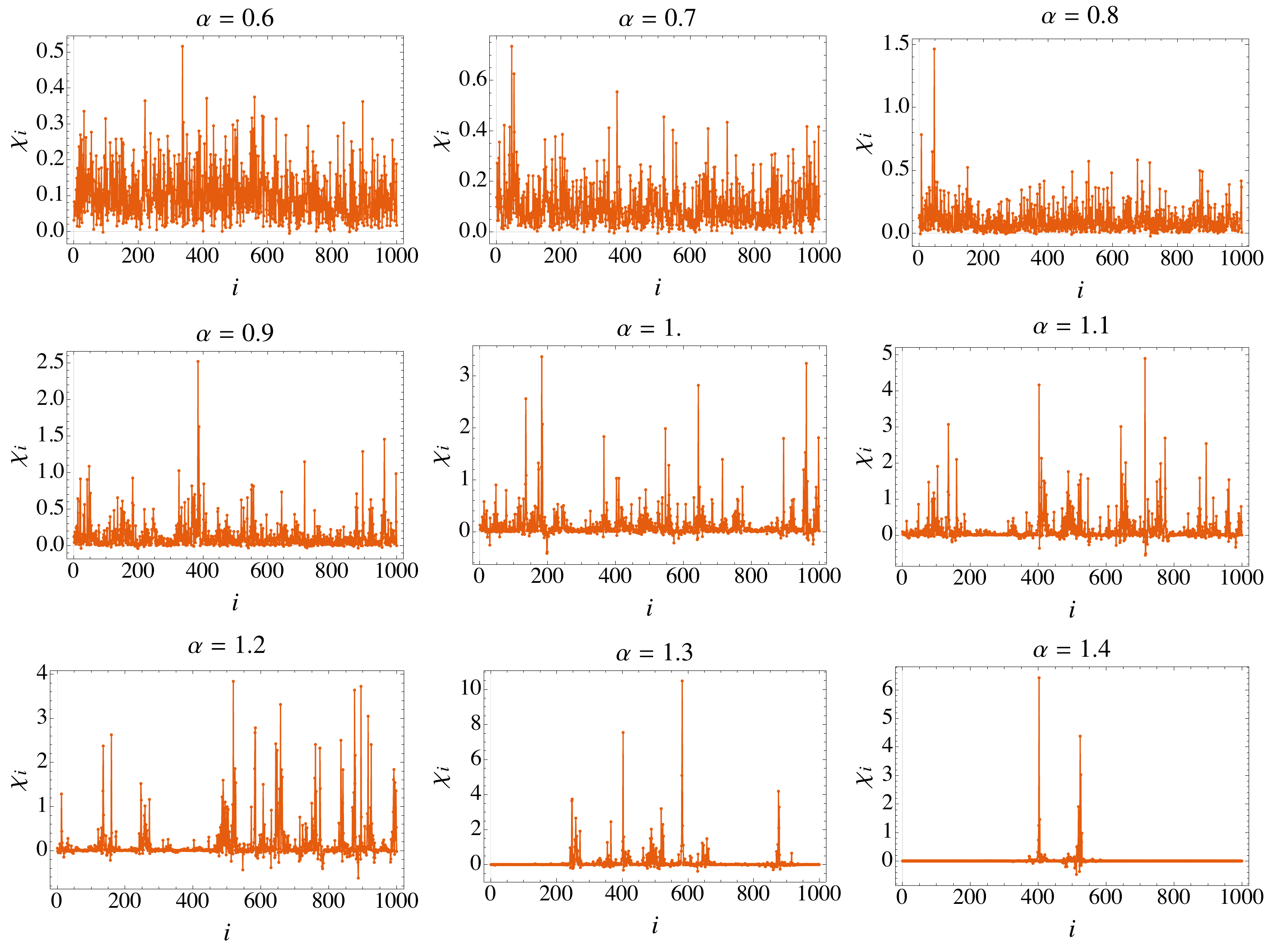}
  \caption{ 
The spatially resolved spin susceptibility [Eq.~\eqref{eq:chi-i}]. In the metallic phase $\alpha < 1$, $\chi_i$ is positive at almost every site and shows a roughly uniform paramagnetic response. 
Deep in the localized phase $\alpha \gtrsim 1.4$, $\chi_i$ is mostly contributed by the subset of the largest local moments (see also Fig.~\ref{fig:Sichi_count}).
The fraction of sites contributing most to the local susceptibility \emph{diminishes} with increasing $\alpha$ 
(compare to Fig.~\ref{fig:Szi-alpha}, which shows the local magnetization profile in typical HF-converged ground states for various $\alpha$). }
	\label{fig:chi-i-alpha}
\end{figure*}

In Fig.~\ref{fig:chi-i-alpha}, we show the spatially resolved spin susceptibility $\chi_i$ defined via Eq.~\eqref{eq:chi-i}. 
In the extended phase ($\alpha < 1)$, almost every site exhibits a positive contribution to the spin susceptibility, consistent with uniform Pauli paramagnetism.
On the other hand, deep in the interacting Anderson insulator ($\alpha \gtrsim 1.4$), the local spin susceptibility shows excessively large peaks at a small subset of the sites hosting the largest local moments
(see also Fig.~\ref{fig:Sichi_count}). 


\begin{thebibliography}{99}
\bibitem{Paalanen1988}
	M. A. Paalanen, J. E. Graebner, R. N. Bhatt, and S. Sachdev,
	\emph{Thermodynamic Behavior near a Metal-Insulator Transition},
	Phys. Rev. Lett. {\bf 61}, 597 (1988).
\bibitem{Milo1989} 
	M. Milovanović, S. Sachdev, and R. N. Bhatt, 
	\emph{Effective-field theory of local-moment formation in disordered metals},
	\href{https://doi.org/10.1103/PhysRevLett.63.82}{Phys. Rev. Lett. \textbf{63}, 82 (1989).}
\bibitem{Bhatt1992} 
	R. N. Bhatt and Daniel S. Fisher, 
	\emph{Absence of spin diffusion in most random lattices}, 
	\href{https://doi.org/10.1103/PhysRevLett.68.3072}{Phys. Rev. Lett. \textbf{68}, 3072 (1992).}
\bibitem{BK1994} 
	D. Belitz and T. R. Kirkpatrick, 
	\emph{The Anderson-Mott transition},
	 \href{https://doi.org/10.1103/RevModPhys.66.261}{Rev. Mod. Phys. \textbf{66}, 261 (1994).}
\bibitem{Feng1999} 
	X. G. Feng, D. Popović, and S. Washburn, 
	\emph{Effect of Local Magnetic Moments on the Metallic Behavior in Two Dimensions}, 
	\href{https://doi.org/10.1103/PhysRevLett.83.368}{Phys. Rev. Lett. \textbf{83}, 368 (1999).}
\bibitem{Miranda2012}
	E. Miranda and V. Dobrosavljevi'c,
	\emph{Dynamical Mean-field Theories of Correlation and Disorder}, in 
	\emph{Conductor-Insulator Quantum Phase Transitions}
	(Oxford University Press, Oxford, England, 2012).
\bibitem{Lee1985} 
	P. A. Lee and T. V. Ramakrishnan, 
	\emph{Disordered electronic systems}, 
	\href{https://doi.org/10.1103/RevModPhys.57.287}{Rev. Mod. Phys. \textbf{57}, 287 (1985).}
\bibitem{Coleman2015}
	P. Coleman,
	\emph{Introduction to Many-Body Physics}
	(Cambridge University Press, Cambridge, England, 2015).
\bibitem{Finkelstein1983}
	A. M. Finkel’stein, 
	\emph{Influence of Coulomb interaction on the properties of disordered metals}, 
	Sov. Phys. JETP \textbf{57}, 97 (1983).
\bibitem{Castellani1984} 
	C. Castellani, C. Di Castro, P. A. Lee, and M. Ma, 
	\emph{Interaction-driven metal-insulator transitions in disordered fermion systems}, 
	\href{https://doi.org/10.1103/PhysRevB.30.527}{Phys. Rev. B \textbf{30}, 527 (1984).}
\bibitem{Finkelstein2010} 
	A. M. Finkel’stein, 
	\emph{Disordered electron liquid with interactions}, 
	\href{https://doi.org/10.1142/S0217979210064642}{Int. J. Mod. Phys. B \textbf{24}, 1855(2010).}
\bibitem{Finkelstein2023} 
	A. M. Finkel'stein and G. Schwiete, 
	\emph{Scale-dependent theory of the disordered electron liquid}, 
	\href{https://doi.org/10.48550/arXiv.2303.00655}{arXiv:2303.00655.}
\bibitem{Feigelman2007} 
	M. V. Feigel'man, L. B. Ioffe, V. E. Kravtsov, and E. A. Yuzbashyan, 
	\emph{Eigenfunction Fractality and Pseudogap State near the Superconductor-Insulator Transition}, 
	\href{https://doi.org/10.1103/PhysRevLett.98.027001}{Phys. Rev. Lett. \textbf{98}, 027001 (2007).}
\bibitem{Feigelman2010} 
	M. V. Feigel'man, L. B. Ioffe, V. E. Kravtsov, and E. Cuevas, 
	\emph{Fractal superconductivity near localization threshold}, 
	\href{https://doi.org/10.1016/j.aop.2010.04.001}{Ann. Phys. \textbf{325}, 1390 (2010).}
\bibitem{Burmistrov2012} 
	I. S. Burmistrov, I. V. Gornyi, and A. D. Mirlin, 
	\emph{Enhancement of the Critical Temperature of Superconductors by Anderson Localization}, 
	\href{https://doi.org/10.1103/PhysRevLett.108.017002}{Phys. Rev. Lett. \textbf{108}, 017002 (2012).}
\bibitem{Evers2008} 
	F. Evers and A. D. Mirlin, 
	\emph{Anderson transitions}, 
	\href{https://doi.org/10.1103/RevModPhys.80.1355}{Rev. Mod. Phys. \textbf{80}, 1355 (2008).}
\bibitem{Chalker1988}
	J. T. Chalker and G. J. Daniell, 
	\emph{Scaling, Diffusion, and the Integer Quantized Hall Effect}, 
	\href{https://doi.org/10.1103/PhysRevLett.61.593}{Phys. Rev. Lett. \textbf{61}, 593 (1988).}
\bibitem{Chalker1990} 
	J. T. Chalker, 
	\emph{Scaling and eigenfunction correlations near a mobility edge}, 
	\href{https://doi.org/10.1016/0378-4371(90)90056-X}{Physica A: Stat. Mech. Appl. \textbf{167}, 253 (1990).}
\bibitem{Cuevas2007} 
	E. Cuevas and V. E. Kravtsov, 
	\emph{Two-eigenfunction correlation in a multifractal metal and insulator}, 
	\href{https://doi.org/10.1103/PhysRevB.76.235119}{Phys. Rev. B \textbf{76}, 235119 (2007).}
\bibitem{KB1996} 
	T. R. Kirkpatrick and D. Belitz, 
	\emph{Quantum critical behavior of disordered itinerant ferromagnets}, 
	\href{https://doi.org/10.1103/PhysRevB.53.14364}{Phys. Rev. B \textbf{53}, 14364 (1996).}
\bibitem{Andreev1998} 
	A. V. Andreev and A. Kamenev, 
	\emph{Itinerant Ferromagnetism in Disordered Metals: A Mean-Field Theory}, 
	\href{https://doi.org/10.1103/PhysRevLett.81.3199}{Phys. Rev. Lett. \textbf{81}, 3199 (1998).}
\bibitem{Chamon2000}
	C. Chamon and E. R. Mucciolo,
	\emph{Nonperturbative Saddle Point for the Effective Action of Disordered and Interacting Electrons in 2D},
	\href{https://doi.org/10.1103/PhysRevLett.85.5607}{Phys. Rev. Lett. \textbf{85}, 5607 (2000).}
\bibitem{NAL2000} 
	B. N. Narozhny, I. L. Aleiner, and A. I. Larkin, 
	\emph{Magnetic fluctuations in two-dimensional metals close to the Stoner instability}, 
	\href{https://doi.org/10.1103/PhysRevB.62.14898}{Phys. Rev. B \textbf{62}, 14898 (2000).}
\bibitem{Kravchenko1994} 
	S. V. Kravchenko, G. V. Kravchenko, J. E. Furneaux, V. M. Pudalov, and M. D’Iorio, 
	\emph{Possible metal-insulator transition at $B=0$ in two dimensions},
	 \href{https://doi.org/10.1103/PhysRevB.50.8039}{Phys. Rev. B \textbf{50}, 8039 (1994).}
\bibitem{Kravchenko1995} 
	S. V. Kravchenko, Whitney E. Mason, G. E. Bowker, J. E. Furneaux, V. M. Pudalov, and M. D’Iorio, 
	\emph{Scaling of an anomalous metal-insulator transition in a two-dimensional system in silicon at $B=0$}, 
	\href{https://doi.org/10.1103/PhysRevB.51.7038}{Phys. Rev. B \textbf{51}, 7038 (1995).}
\bibitem{Abrahams2001} 
	E. Abrahams, S. V. Kravchenko, and M. P. Sarachik, 
	\emph{Metallic behavior and related phenomena in two dimensions},
	 \href{https://doi.org/10.1103/RevModPhys.73.251}{Rev. Mod. Phys. \textbf{73}, 251(2001).}
\bibitem{Kravchenko2010} 
	S. V. Kravchenko and M. P. Sarachik, 
	\emph{A metal-insulator transition in 2d: established facts and open questions}, 
	\href{https://doi.org/10.1142/S021797921006454X}{Int. J. Mod. Phys. B \textbf{24}, 1640 (2010).}
\bibitem{Shashkin2017} 
	A. A. Shashkin and S. V. Kravchenko, 
	\emph{Metal-insulator transition in a strongly-correlated two-dimensional electron system}, 
	in \href{https://doi.org/10.4324/9781315364575}{\textit{Strongly correlated electrons in two dimensions}}, edited by S. V. Kravchenko (Pan Stanford Publishing, 2017).
\bibitem{DS2005} 
	S. Das Sarma and E. H. Hwang, 
	\emph{The so-called two dimensional metal–insulator transition}, 
	\href{https://doi.org/10.1016/j.ssc.2005.04.035}{Solid State Commun. \textbf{135}, 579 (2005).}
\bibitem{Punn2001} 
	A. Punnoose and A. M. Finkel'stein, 
	\emph{Dilute Electron Gas near the Metal-Insulator Transition: Role of Valleys in Silicon Inversion Layers}, 
	\href{https://doi.org/10.1103/PhysRevLett.88.016802}{Phys. Rev. Lett. 88, 016802 (2001).}
\bibitem{Punn2005} 
	A. Punnoose and A. M. Finkel’stein, 
	\emph{Metal-Insulator Transition in Disordered Two-Dimensional Electron Systems}, 
	\href{https://doi.org/10.1126/science.1115660}{Science \textbf{310}, 289 (2005).}
\bibitem{Ani2007} 
	S. Anissimova, S. V. Kravchenko, A Punnoose, A. M. Finkel’stein, and T. M. Klapwijk, 
	\emph{Flow diagram of the metal-insulator transition in two dimensions}, 
	\href{https://doi.org/10.1038/nphys685}{Nat. Phys. \textbf{3}, 707 (2007).}
\bibitem{DS2013} 
	S. Das Sarma, E. H. Hwang, and Q. Li, 
	\emph{Two-dimensional metal-insulator transition as a potential fluctuation driven semiclassical transport phenomenon}, 
	\href{https://doi.org/10.1103/PhysRevB.88.155310}{Phys. Rev. B \textbf{88}, 155310 (2013).}
\bibitem{DS2014} 
	S. Das Sarma and E. H. Hwang, 
	\emph{Two-dimensional metal-insulator transition as a strong localization induced crossover phenomenon}, 
	\href{https://doi.org/10.1103/PhysRevB.89.235423}{Phys. Rev. B \textbf{89}, 235423 (2014).}
\bibitem{Shashkin2001} 
	A. A. Shashkin, S. V. Kravchenko, V. T. Dolgopolov, and T. M. Klapwijk,  
	\emph{Indication of the Ferromagnetic Instability in a Dilute Two-Dimensional Electron System}, 
	\href{https://doi.org/10.1103/PhysRevLett.87.086801}{Phys. Rev. Lett. \textbf{87}, 086801 (2001).}
\bibitem{Shashkin2006}  
	A. A. Shashkin, S. Anissimova, M. R. Sakr, S. V. Kravchenko, V. T. Dolgopolov, and T. M. Klapwijk,  
	\emph{Pauli Spin Susceptibility of a Strongly Correlated Two-Dimensional Electron Liquid}, 
	\href{https://doi.org/10.1103/PhysRevLett.96.036403}{Phys. Rev. Lett. \textbf{96}, 036403 (2006).}
\bibitem{Ani2006}  
	S. Anissimova, A. Venkatesan, A. A. Shashkin, M. R. Sakr, S. V. Kravchenko, and T. M. Klapwijk,  
	\emph{Magnetization of a Strongly Interacting Two-Dimensional Electron System in Perpendicular Magnetic Fields}, 
	\href{https://doi.org/10.1103/PhysRevLett.96.046409}{Phys. Rev. Lett. \textbf{96}, 046409 (2006).}
\bibitem{Pudalov2012}
	N. Tench,  A. Yu. Kuntsevich, V. M. Pudalov, and M. Reznikov, 
	\emph{Spin-Droplet State of an Interacting 2D Electron System},
	\href{https://doi.org/10.1103/PhysRevLett.109.226403}{Phys. Rev. Lett. \textbf{109}, 226403 (2012).}
\bibitem{Pudalov2018}
	V. M. Pudalov, A. Yu. Kuntsevich, M. E. Gershenson, I. S. Burmistrov, and M. Reznikov, 
	\emph{Probing spin susceptibility of a correlated two-dimensional electron system by transport and magnetization measurements},
	\href{https://doi.org/10.1103/PhysRevB.98.155109}{Phys. Rev. B \textbf{98}, 155109 (2018).}
\bibitem{Hossain2020}
	M. S. Hossain, M. K. Ma, K. A. Villegas Rosales, Y. J. Chung, L. N. Pfieffer, K. W. West, K. W. Baldwin, and M. Shayegan, 
	\emph{Observation of spontaneous ferromagnetism in a two-dimensional electron system},
	\href{https://doi.org/10.1073/pnas.2018248117}{Proc. Natl. Acad. Sci. (USA) {\bf 117}, 32244 (2020).}
\bibitem{Mirlin1996}
	A. D. Mirlin, Y. V. Fyodorov, F.-M. Dittes, J. Quezada, and T. H. Seligman,
	\emph{Transition from localized to extended eigenstates in the ensemble of power-law random banded matrices}, 
	\href{https://doi.org/10.1103/PhysRevE.54.3221}{Phys. Rev. E \textbf{54}, 3221 (1996).}
\bibitem{Mirlin2000b}
  	A. D. Mirlin and F. Evers,
  	\emph{Multifractality and critical fluctuations at the Anderson transition},
  	\href{https://doi.org/10.1103/PhysRevB.62.7920}{Phys. Rev. B {\bf 62}, 7920 (2000).}
\bibitem{Mirlin2000a} 	
	A. D. Mirlin, 
	\emph{Statistics of energy levels and eigenfunctions in disordered systems}, 
	\href{https://doi.org/10.1016/S0370-1573(99)00091-5}{Phys. Rep. \textbf{326}, 259 (2000).}
\bibitem{Kamenev2009} 
	A. Kamenev and A. Levchenko, 
	\emph{Keldysh technique and non-linear sigma-model: basic principles and applications}, 
	Adv. Phys. \textbf{58}, 197 (2009).
\bibitem{Kamenev2011} 
	A. Kamenev, 
	\textit{Field Theory of Non-Equilibrium Systems} 2nd Edition, (Cambridge University Press, Cambridge, England, 2023).
\bibitem{Liao2017} 
	Y. Liao, A. Levchenko, and M. S. Foster, 
	\emph{Response theory of the ergodic many-body delocalized phase: Keldysh Finkel’stein sigma models and the 10-fold way},
	 \href{https://doi.org/10.1016/j.aop.2017.08.020}{Ann. Phys. \textbf{386}, 97 (2017).}
\bibitem{DH1993} 
	C. Dasgupta and J. W. Halley, 
	\emph{Phase diagram of the two-dimensional disordered Hubbard model in the Hartree-Fock approximation}, 
	\href{https://doi.org/10.1103/PhysRevB.47.1126}{Phys. Rev. B \textbf{47}, 1126(R) (1993).}
\bibitem{Tusch1993}
	M. A. Tusch and D. E. Logan,
	\emph{Interplay between disorder and electron interactions in a $d = 3$ site-disordered Anderson-Hubbard model: A numerical mean-field study},
	\href{https://doi.org/10.1103/PhysRevB.48.14843}{Phys. Rev. B \textbf{48}, 14843 (1993).}	
\bibitem{Epperlein1997}
	F. Epperlein, M. Schreiber, and T. Vojta,
	\emph{Quantum Coulomb glass within a Hartree-Fock approximation}, 
	\href{https://doi.org/10.1103/PhysRevB.56.5890}{Phys. Rev. B \textbf{56}, 5890 (1997).}
\bibitem{Vojta1998}
	F. Epperlein, M. Schreiber, and T. Vojta,
	\emph{Do Interactions Increase or Reduce the Conductance of Disordered Electrons? It Depends!}, 
	\href{https://doi.org/10.1103/PhysRevLett.81.4212}{Phys. Rev. Lett. \textbf{81}, 4212 (1998).}
\bibitem{Amini2014}
	M. Amini, V. E. Kravtsov, and M. M\"uller,
	\emph{Multifractality and quantum-to-classical crossover in the Coulomb anomaly at the Mott–Anderson metal–insulator transition}, 
	\href{https://doi.org/10.1088/1367-2630/16/1/015022}{New. J. Phys. {\bf 16}, 015022 (2014).}
\bibitem{Binder1986} 
	K. Binder, and A. P. Young, 
	\emph{Spin glasses: Experimental facts, theoretical concepts, and open questions}, 
	\href{https://doi.org/10.1103/RevModPhys.58.801}{Rev. Mod. Phys. \textbf{58}, 801 (1986).}
\bibitem{MezardBook}
	M. Mezard, G. Parisi, and M. A. Virasoro,
	\emph{Spin Glass Theory and Beyond}
	(World Scientific, Singapore, 1987).
\bibitem{bibnote1}
	For a noteable exception in the context of random matrix theory, see Ref.~\cite{Kamenev1999}.
\bibitem{Kamenev1999}	
	A. Kamenev and M. M\'ezard,
	\emph{Wigner-Dyson statistics from the replica method},
	\href{https://doi.org/10.1088/0305-4470/32/24/304}{J. Phys. A \textbf{32}, 4373 (1999).}
\bibitem{Livan2017}
	G. Livan, M. Novaes, and P. Vivo,
	\href{https://doi.org/10.1007/978-3-319-70885-0}{\emph{Introduction to Random Matrices: Theory and Practice}}
	(Springer, Cham, 2018).
\bibitem{Mehta}
	M. L. Mehta,
	\emph{Random Matrices}, Third Edition
	(Elsevier Academic Press, Amsterdam, 2004).
\bibitem{Kravtsov2012}
	V. E. Kravtsov, O. M. Yevtushenko, P. Snajberk, and E. Cuevas, 
	\emph{L\'evy flights and multifractality in quantum critical diffusion and in classical random walks on fractals},
	Phys. Rev. B {\bf 86}, 021136 (2012).
	\href{https://doi.org/10.1103/PhysRevB.86.021136}{Phys. Rev. B \textbf{86}, 021136 (2012).}
\bibitem{Foster2012}
	M. S. Foster and E. A. Yuzbashyan,
	\emph{Interaction-mediated surface state instability in disordered three-dimensional topological superconductors with spin SU(2) symmetry},
	\href{https://doi.org/10.1103/PhysRevLett.109.246801}{Phys. Rev. Lett. {\bf 109}, 246801 (2012).}
\bibitem{Foster2014} 
	M. S. Foster, H.-Y. Xie, and Y.-Z. Chou,
	\emph{Topological protection, disorder, and interactions: Survival at the surface of 3D topological superconductors},
	\href{https://doi.org/10.1103/PhysRevB.89.155140}{Phys. Rev. B {\bf 89}, 155140 (2014).}  
\bibitem{Zhang2022}
	X. Zhang and M. S. Foster,
	\emph{Enhanced amplitude for superconductivity due to spectrum-wide wave function criticality in quasiperiodic and power-law random hopping models},
	\href{https://doi.org/10.1103/PhysRevB.106.L180503}{Phys. Rev. B {\bf 106}, L180503 (2022).}  
\bibitem{Pezzoli2009} 
	M. E. Pezzoli, F. Becca, M. Fabrizio, and G. Santoro, 
	\emph{Local moments and magnetic order in the two-dimensional Anderson-Mott transition}, 
	\href{https://doi.org/10.1103/PhysRevB.79.033111}{Phys. Rev. B \textbf{79}, 033111 (2009).}
\bibitem{DellAnna2017}
	L. Dell'Anna,
	\emph{Beta-functions of non-linear $\sigma$-models for disordered and interacting electron systems},
	\href{https://doi.org/10.1002/andp.201600317}{Ann. Phys. (Berlin) {\bf 529}, 1600317 (2017).}
\bibitem{Edwards1975}
	S. F. Edwards, and P. W. Anderson, 
	\emph{Theory of spin glasses}, 
	J. Phys. F: Met. Phys. \textbf{5}, 965 (1975).
\bibitem{Khanin1980}
	K. M. Khanin, 
	\emph{Absence of phase transitions in one-dimensional spin systems with random Hamiltonian},
	Teor. J. Mat. Fiz. {\bf 43}, 253 (1980).
\bibitem{Cassandro1982}
	M. Cassandro, E. Olivieri, and B. Tirozzi,
	\emph{Infinite Differentiability for One-Dimensional Spin System with Long Range Random Interaction},
	Commun. Math. Phys. {\bf 87}, 229 (1982).
\bibitem{Kotliar1983}
	G. Kotliar, P. W. Anderson, and D. L. Stein,
	\emph{One-dimensional spin-glass model with long-range random interactions},
	\href{https://doi.org/10.1103/PhysRevB.27.602}{Phys. Rev. B \textbf{27}, 602(R)  (1983).}
\bibitem{Fazekas1999} 
	P. Fazekas, 
	\href{https://doi.org/10.1142/2945}{\textit{Lecture Notes on Electron Correlation and Magnetism}}, (World Scientific, Singapore, 1999).
\bibitem{Sinova2000} 
	J. Sinova, A. H. MacDonald, and S. M. Girvin, 
	\emph{Disorder and interactions in quantum Hall ferromagnets near $\nu=1$}, 
	\href{https://doi.org/10.1103/PhysRevB.62.13579}{Phys. Rev. B \textbf{62}, 13579 (2000).}
\bibitem{Sriv1984} 
	G. P. Srivastava, 
	\emph{Broyden's method for self-consistent field convergence acceleration}, 
	\href{https://doi.org/10.1088/0305-4470/17/6/002}{J. Phys. A: Math. Gen. \textbf{17} L317 (1984).}
 \bibitem{Johnson1988} 
	D. D. Johnson, 
	\emph{Modified Broyden’s method for accelerating convergence in self-consistent calculations}, 
	\href{https://doi.org/10.1103/PhysRevB.38.12807}{Phys. Rev. B \textbf{38}, 12807 (1988).}
\bibitem{SYK2022}
	For a review, see e.g.\ 
	D. Chowdhury, A. Georges, O. Parcollet, and S. Sachdev,
	\emph{Sachdev-Ye-Kitaev models and beyond: Window into non-Fermi liquids},
	\href{https://doi.org/10.1103/RevModPhys.94.035004}{Rev. Mod. Phys. \textbf{94}, 035004 (2022)},
	and references therein.
\end{thebibliography}
\end{document}